\documentclass[12pt,a4paper]{report}
\usepackage{amsmath}
\usepackage{amsfonts}
\usepackage{amssymb}
\usepackage{graphicx}

\usepackage{graphicx,epsf}
\usepackage{xcolor}

\usepackage{setspace}

\def\be{\begin{equation}}
\def\ee{\end{equation}}
\def\bea{\begin{eqnarray}}
\def\eea{\end{eqnarray}}
\def\p{\partial}

\def\cs2{c_{\rm{s}}^2}
\def\wt{\widetilde}
\def\frw{{\rm{FRW}}}
\def\SMTP{{\zeta}_{\rm{SMTP}}}
\def\DOTSMTP{{\dot{\zeta}}_{\rm{SMTP}}}
\def\PY{{\sc{Pyessence}}}
\def\LT{{\rm{LTB}}}

\usepackage{float}
\usepackage{mathtools}

\usepackage[english]{babel}

\newcommand{\sfx}{X}
\newcommand{\sfy}{Y}
\newcommand{\dpn}{\delta P_{\mathrm{nad}}}
\newcommand{\C}{\mathbb{C}}
\newcommand{\ph}{\varphi}
\newcommand\eq[1]{Eq.~(\ref{#1})}
\def\beal{\begin{align}}
\def\eeal{\end{align}}
\def\p{\partial}

\begin{document}

\title{Perturbations in Lema{\^\i}tre-Tolman-Bondi and Assisted Coupled Quintessence Cosmologies}
\author{Alexander Leithes}
\date{April 2017}

{\bf{Declaration:}}
\newline
\newline
{\bf{I, Alexander Leithes, confirm that the research included within this thesis is my own work or that where it has been carried out in collaboration with, or supported by others, that this is duly acknowledged below and my contribution indicated. Previously published material is also acknowledged below.
\newline
\newline
I attest that I have exercised reasonable care to ensure that the work is original, and does not to the best of my knowledge break any UK law, infringe any third party's copyright or other Intellectual Property Right, or contain any confidential material.
\newline
\newline
I accept that the College has the right to use plagiarism detection software to check the electronic version of the thesis.
\newline
\newline
I confirm that this thesis has not been previously submitted for the award of a degree by this or any other university.
\newline
\newline
The copyright of this thesis rests with the author and no quotation from it or information derived from it may be published without the prior written consent of the author.
\newline
\newline
Signature: Alexander Leithes
\newline
Date: 19$^{\rm{th}}$ August 2016
\newline
\newline
Details of collaboration and publications: Some work in this thesis is based upon the paper  written in collaboration with Karim A. Malik - `Conserved quantities in Lema{\^\i}tre-Tolman-Bondi cosmology' - DOI: 10.1088/0264-9381/32/1/015010. Other work is based upon the paper written in collaboration with David J. Mulryne, Nelson Nunes and Karim A. Malik - `Linear Density Perturbations in Multifield Coupled Quintessence' - arXiv:1608.00908 DOI: TBC, and the paper `{\sc{Pyessence}} - Generalised Coupled Quintessence Linear Perturbation Python Code - A Guide' - arXiv:1608.00910 DOI: TBC}}

\maketitle

\chapter*{Abstract}
\label{ch:abstract}
\addcontentsline{toc}{chapter}{Abstract}
\section*{}
\singlespacing
In this thesis we present research into linear perturbations in Lema{\^\i}tre-Tolman-Bondi (LTB) and Assisted Coupled Quintessence (ACQ) Cosmologies. First we give a brief overview of the standard model of cosmology. We then introduce Cosmological Perturbation Theory (CPT) at linear order for a flat Friedmann-Robertson-Walker (FRW) cosmology. Next we study linear perturbations to a Lema{\^\i}tre-Tolman-Bondi (LTB) background spacetime. Studying the transformation behaviour of the perturbations under gauge transformations, we construct gauge invariant quantities in LTB. We show, using the perturbed energy conservation equation, that there is a conserved quantity in LTB which is conserved on all scales. We then briefly extend our discussion to the Lema{\^\i}tre spacetime, and construct gauge-invariant perturbations in this extension of LTB spacetime.
We also study the behaviour of linear perturbations in assisted coupled quintessence models in a FRW background. We provide the full set of governing equations for this class of models, and solve the system numerically. The code written for this purpose is then used to evolve growth functions for various models and parameter values, and we compare these both to the standard $\Lambda$CDM model and to current and future observational bounds. We also examine the applicability of the ``small scale approximation", often used to calculate growth functions in quintessence models, in light of upcoming experiments such as SKA and Euclid. We find the results of the full equations deviates from the approximation by more than the experimental uncertainty for these future surveys. The construction of the numerical code,~\PY, written in Python to solve the system of background and perturbed evolution equations for assisted coupled quintessence, is also discussed.
%
%
\chapter*{Acknowledgements}
\label{ch:acknowledgements}
\addcontentsline{toc}{chapter}{Acknowledgements}
This is dedicated to my parents David and Christina Leithes, who gave me everything I needed to make a life, and to Karim Malik, who gave me back my life to make anew.\\ 
\vfill
I would also like to thank Ellie Nalson, David Mulryne, Joe Elliston, Adam Christopherson, Tim Clifton and Ian Huston who were there from the very start and helped me in so many ways. And all my colleagues, friends and family who have supported me.
\vfill
This work was supported by the Science and Technology Facilities Council (STFC) studentship ST/K50225X/1.

\setcounter{tocdepth}{4}
\setcounter{secnumdepth}{4}
\tableofcontents

\onehalfspacing

%


\chapter{Introduction}
\label{ch:1}

\section{Introduction}
\label{sec:BeyondL}

The \emph{Cosmological Constant + Cold Dark Matter} ($\Lambda$CDM) model of cosmology has become our gold standard in explaining the evolution of the universe. In this model, the dark sector of the universe is modelled by a cosmological constant, which is responsible for the acceleration of the universe in the present epoch, and a pressureless fluid that constitutes dark matter. The model is completed by assuming the presence of baryonic matter and a radiation component. 
Remarkably, this simple picture is sufficient to explain every observational probe to date. These include high precision measurements of the \emph{Cosmic Microwave Background} (CMB) \cite{Adam:2015rua,ACT,SPT}, supernovae observations \cite{Perlmutter:1998np,Riess:1998cb,Kowalski:2008ez}, and large scale structure surveys \cite{Anderson:2013zyy,Bonnett:2015pww,Dawson:2015wdb}.

Despite its success, the model raises many unanswered questions such as: Why does the cosmological constant takes such an unnaturally small value? What is the fundamental nature of \emph{Dark Energy} (DE)? These, in addition to other questions such as why the energy density associated with $\Lambda$ is of the same order as that of dark matter -- the coincidence problem -- have led the community to investigate more complex scenarios.\\ 

One possible scenario is inhomogeneous cosmologies. Research into \emph{Lema{\^\i}tre-Tolman-Bondi} (LTB) cosmology had in the past been motivated by seeking an alternative explanation for the late time accelerated expansion of the universe, as indicated by e.g.~SNIa
observations~\cite{Perlmutter:1998np}. Inhomogeneous cosmologies,
including LTB, had been suggested as such an alternative explanation
of these observations (see e.g.~Ref.~\cite{celerier}). Other
observations such as galaxy surveys, large scale structure surveys,
the CMB and indeed any redshift dependent observations
(see for example Refs.~\cite{Moresco:2012jh,Anderson:2013zyy,Ade:2013ktc})
are usually interpreted assuming a flat \emph{Friedmann-Robertson-Walker} (FRW) cosmology - isotropic and
homogeneous on large scales. In order to test the validity of this
assumption other, inhomogeneous, cosmologies such as LTB should also
be considered. There is however some difficulty making LTB match all observations (see e.g. Refs.~\cite{Clarkson:2012bg,Vargas:2015ctw}). However there are environments, such as large voids or overdensities where LTB may prove a more appropriate cosmological model (see e.g. Refs.~\cite{Meyer:2014qla,Sussman:2015bea}), where such overdensities or voids may be approximately spherical in nature, and LTB may then prove a better background model. If such structures are sufficiently large then perturbed LTB may then be more appropriate for studying structure growth within such environments. Consequently there is much active research into LTB and
other inhomogeneous spherically symmetric cosmologies, both at
background order and with perturbations (see
e.g.~Refs.~\cite{Iribarrem:2014dta,Lim:2013rra,Biswas,Bolejko1,Bolejko3,
  CFL,GBH2,MZS10,ZMS,goodman,Sussman1,Zumalacarregui:2012pq,Clarkson:2012bg} for theory and comparison with
observation in general, see e.g.~Refs.~\cite{YNS,Alnes,CFZ,ACT,SPT}
for research relating to CMB and see
e.g.~Refs.~\cite{Bull:2011wi,ZS,MZ11,YNS2,SZ80,kSZobs1,kSZobs2,kSZobs3,GBH}
for research more specific to the kinetic Sunyaev-Zeldovich effect, see e.g.~Refs.~\cite{Finelli:2014yha,Alonso:2010zv,Alonso:2012ds} for structure formation in LTB, including N-body simulations).\\

Within homogeneous and inhomogeneous cosmologies, conserved quantities are useful tools with a wide range of
applications. In particular, they allow us to relate
early and late times in a cosmological model, without explicitly
having to solve the evolution equations, either exactly or taking
advantage of some limiting behaviour. These quantities have been
studied extensively within the context of \emph{Cosmological Perturbation
Theory} (CPT), and usually applied to a FRW
background spacetime.

Using metric based cosmological perturbation theory
\cite{Bardeen80,KS}, we can readily construct gauge-invariant
quantities which are also conserved, that is constant in time (see
e.g.~Ref.~\cite{Lyth85} for early work on this topic).
In a FRW background spacetime, $\zeta$, the curvature perturbation on
uniform density hypersurfaces, is conserved on large scales for
adiabatic fluids.  To show that $\zeta$ is conserved and under what
conditions, we only need the conservation of energy \cite{WMLL}. This
was first shown to work for fluids at linear order, but it holds also
at second order in the perturbations, and in the fully non-linear
case, usually referred to as the $\delta N$ formalism 
\cite{WMLL,MW2003,LMS}.

Instead of, or in addition to, cosmological perturbation theory, we can
also use other approximation schemes to deal with the non-linearity of
the Einstein equations. In particular gradient expansion schemes have
proven to be useful in the context of conserved quantities, again with the
main focus on FRW spacetimes \cite{Salopek:1990jq, Rigopoulos03, LMS, Langlois:2005qp}.
But conserved quantities have also been studied for spacetimes other
than FRW, such as braneworld models (see
e.g.~Ref.~\cite{Bridgman:2001mc}, and anisotropic spacetime
(e.g.~Ref.~\cite{Abolhasani:2013zya}).

The LTB spacetime \cite{Bondi} is a more
general solution to Einstein's field equations than the
Friedmann-Robertson-Walker (FRW) model. While LTB is invariant under
rotations, FRW is rotation and translation invariant, and hence has
homogeneous and isotropic, maximally symmetric spatial sections
\cite{Ellis}.

Gauge-invariant perturbations in general spherically symmetric
spacetimes have been studied already in the 1970s by Gerlach and
Sengupta \cite{Gerlach:1979rw,Gerlach:1980tx}, using a 2+2 split on
the background spacetime. Recent works studying perturbed LTB
spacetimes performs a 1+1+2 split (see e.g.~Refs.~\cite{Tim1,February:2012fp,February:2013qza}). These
splits allow for a decomposition of the tensorial quantities on the
submanifolds into axial and polar scalars and vectors, similar to the
scalar-vector-tensor decomposition in FRW \cite{Bardeen80,KS}. Later in this thesis we perform a 1+3 split of spacetime, without further decomposing
the spatial submanifold. This prevents us from decomposing tensorial
quantities on the spatial submanifold further into axial and polar scalars and
vectors, but provides us with much simpler expressions, well suited
for the construction of conserved quantities.
We therefore study systematically how to construct gauge-invariant quantities in
perturbed LTB spacetimes. To this end we derive the transformation
rules for matter and metric variables under small coordinate - or
gauge - transformations and use these to construct gauge-invariant
quantities. We also derive the perturbed energy density evolution
equation, which allows us to derive a very simple evolution equation
for the spatial metric perturbation on uniform density and comoving
hypersurfaces.\\

Another possible scenario is coupled quintessence. In this model a scalar field, which makes up the DE component of the universe and produces acceleration, is coupled to a pressureless dark matter fluid \cite{Amendola:1999dr,Holden:1999hm,Amendola:1999er,Koivisto:2005nr,Gonzalez:2006cj,Valiviita:2008iv,Amendola:2014kwa,Farrar:2003uw,Copeland:2003cv,Brookfield:2007au,Baldi:2012kt,Piloyan:2013mla,AmenTsuji,Koivisto:2015qua}. 
Recent extensions which have been investigated include Multi-coupled Dark Energy (McDE) (see e.g.~Ref.~\cite{Piloyan:2014gta}), in which the dark matter component of the universe is formed from two fluids that couple differently to a single scalar field. 

In a series of recent papers \cite{Baldi:2012kt,Piloyan:2013mla,Piloyan:2014gta}, perturbations in the McDE model have been calculated numerically and compared with present and future large scale structure experiments. Taking this line of investigation, 
one can model the dark sector of the universe as being made up of $N$ fluids interacting with $M$ scalar fields. This model is known as assisted coupled quintessence (ACQ) \cite{Amendola:2014kwa}. The name derives from the idea that the many fields can act together to generate acceleration, in a similar manner to assisted inflation models of the early universe (see for example~Refs.~\cite{Liddle:1998jc,Malik:1998gy,Kanti:1999vt}). ACQ is a more general model than single field and single fluid models (or McDE) and a natural extension to the existing work in this area. It is also a reasonable assumption to make given the multiple particle species already known from the standard model of particle physics, as well as models beyond the standard model, and is the same assumption as that made in the aforementioned assisted inflation models.

ACQ is the focus of the later parts of this thesis. Our aims are two-fold. First we will calculate the equations of motion for linear perturbations in this rather general model, and incorporate these into a fast numerical code, \PY. In principal, this code can be used to generate quantities such as the growth factor of large scale structure for any coupled quintessence model with an arbitrary number of fields and fluids and arbitrary couplings. We intend to make this code publicly available. Secondly, we will apply this code, initially to revisit the McDE model, and then to consider specific models in which two scalar fields are present. Ongoing and future large scale surveys (see for example Refs.~\cite{Kitching:2015fra,Raccanelli:2015qqa}) offer a chance to distinguish between a cosmological constant and dynamical DE models, and it is important therefore to understand at what level the predictions of assisted models will differ from those of $\Lambda$CDM and those of other quintessence models.

For scales which are small compared to the horizon size today, an approximation to the full perturbed equations of motion has often been used in previous literature, and in particular in the previous study of McDE. A final aim of our work is to evaluate whether this approximation is sufficiently accurate, especially in the light of upcoming surveys.\\

The thesis is structured as follows. The remainder of this chapter will detail the standard $\Lambda$CDM FRW background model of cosmology. We shall move from the Hot Big Bang model, through inflation and finally late time accelerated expansion driven by $\Lambda$ as a form of DE. We shall also describe generalised background governing equations. Chapter 2 will explain cosmological perturbation theory in general and then applied to the standard FRW model. Chapter 3 details using CPT in LTB cosmology in order to construct gauge invariant conserved quantities. It also briefly discusses possible uses for the Spatial Metric Trace Perturbation in for example numerical simulations of structure formation. Chapter 4 returns to FRW cosmology but now models DE as interacting with Cold Dark Matter (CDM) as scalar fields in ACQ models. We describe the growth of structure in ACQ models, conducted using a Python code written specifically for the task. The results are compared with current and future observational bounds. Chapter 5 details the construction of the Python code, \PY, as well as its final structure and use. Finally, in Chapter 6 we discuss the overall conclusions drawn from our research and the possible avenues for further research in the field of cosmological perturbation theory applied to LTB and ACQ cosmologies.

\subsection{Notation Conventions}
\label{sec:Note}
Through out we use the positive metric signature, $-,+,+,+$. We also use natural units where $c=\hbar=1$. With these units the Planck Mass is $M_{\rm{pl}}=G^{-\frac{1}{2}}$.

\section{The Background Cosmology of the Standard $\Lambda$CDM Model}
\label{sec:LCDM}

\subsection{The Background Cosmology}
\label{sec:Background}
In the following sections we shall briefly outline the history and development of the standard $\Lambda$CDM cosmological model, in this chapter at the unperturbed background level only. We shall move from the early motivations for a Hot Big Bang model, through the problems of that model to their resolution in inflationary cosmology and finally to the observations of apparent late time acceleration and the need for an additional component, DE usually as a cosmological constant, $\Lambda$.\\
The discovery by Edwin Hubble \cite{Hubble} of the recession of nearby galaxies gave the first strong evidence for an expanding universe. This discovery that the galaxy recession velocities increased with redshift, coupled with the assumptions of the Cosmological Principle - namely that the universe is homogeneous and isotropic - implied that the universe was expanding. This expansion in turn implied a super-dense, high temperature, high pressure point or singularity at the very earliest time from which the universe expanded in a Hot Big Bang. Further evidence of a Hot Big Bang was provided through the discovery of the CMB~\cite{PenziasWilson}.\\
However, there are problems with the Hot Big Bang model - Flatness, Horizon and Relic problems - which cannot be explained by a simple unmodified model. An additional mechanism, inflation (see e.g.~Ref.~\cite{Guth}), is required in order to counter these problems. Inflation is most simply described using a canonical scalar field, the inflaton, $\ph$, with a kinetic and potential term, which provides the energy driving the process of inflation. Inflationary models are useful in explaining observations including Large Scale Structure surveys (e.g. 2df Galaxy Redshift Survey~\cite{2df}, 6df Galaxy Survey~\cite{6df}, Sloan Digital Sky Survey~\cite{SDSS}), DES~\cite{Bonnett:2015pww}, Euclid and SKA~\cite{Kitching:2015fra}) to small amplitude anisotropies in the CMB i.e. 1 part in $10^5$ fluctuations around a background temperature of 2.725 K \cite{COBE}(e.g. COBE~\cite{COBE}, WMAP~\cite{WMAP}, PLANCK~\cite{PLANCK}).\\

\subsection{The Governing Equations}
\label{sec:equationsofexpansion}
\subsubsection{General Background Equations}
\label{sec: Gen Back}

\emph{General Relativity} GR is defined on pseudo-Riemannian manifolds, where we use the torsion-free metric connection, the Levi-Civita connection, as an affine connection to define differentiation of tangent vectors on such a manifold. In terms of the metric the Levi-Civita connection in Christoffel symbol form is,
\be
\label{christ}
\Gamma^{\mu}_{\hphantom{\mu}\nu\gamma} = \frac{1}{2} g^{\mu \delta} \left( \partial_\nu g_{\delta \gamma} + \partial_\gamma g_{\delta \nu} - \partial_\delta g_{\nu \gamma}  \right) ,
\ee
where $g_{\mu \nu}$ is the spacetime metric and $\partial_\nu$ is the partial derivative with respect to $x^{\nu}$, the spacetime co-ordinates. From the Christoffel symbols we construct the Reimann tensor which describes the intrinsic curvature of our pseudo-Riemannian manifold,

\be
\label{ReiT}
R_{\mu\nu\gamma}^{\hphantom{\mu}\hphantom{\mu}\hphantom{\mu}\delta} = \partial_\nu \Gamma^{\delta}_{\hphantom{\mu}\mu\gamma} - \partial_\mu \Gamma^{\delta}_{\hphantom{\mu}\nu\gamma} + \Gamma^{\alpha}_{\hphantom{\mu}\mu\gamma} \Gamma^{\delta}_{\hphantom{\mu}\nu\alpha} - \Gamma^{\alpha}_{\hphantom{\mu}\nu\gamma} \Gamma^{\delta}_{\hphantom{\mu}\mu\alpha} , 
\ee
where $R_{\mu\nu\gamma}^{\hphantom{\mu}\hphantom{\mu}\hphantom{\mu}\delta}$ is the Reimann tensor. By contracting the Reimann tensor once we get,

\be
\label{RicT}
R_{\mu\delta\gamma}^{\hphantom{\mu}\hphantom{\mu}\hphantom{\mu}\delta} = R_{\mu \gamma} , 
\ee
where $R_{\mu \gamma}$ is the Ricci tensor. Finally, by contracting the Ricci tensor we get,

\be
\label{RicS}
R_{\mu}^{\hphantom{\mu}\mu} = R , 
\ee
where $R$ is the Ricci scalar. We now have all the necessary components to describe the geometry of our spacetime in the Einstein tensor, The Einstein field equations are,

\begin{equation}
\label{Einstein2}
G_{\mu \nu}=8\pi G T_{\mu \nu},
\end{equation}
where $G_{\mu \nu}$ is the Einstein tensor, which describes the geometry of spacetime, $G$ is the universal gravitational constant and $T_{\mu \nu}$ is the energy-momentum tensor, which describes the matter content of the universe. The Einstein tensor, $G_{\mu \nu}$, is defined as,

\be
\label{RicS2}
G_{\mu \nu} = R_{\mu \nu} - \frac{1}{2} g_{\mu \nu} R . 
\ee
The matter content of the universe is described using the energy-momentum tensor, which for a perfect fluid in the absence of anisotropic stress is,

\be
\label{SET}
T_{\mu \nu} = (\rho + P)u_\mu u_\nu + P g_{\mu \nu} , 
\ee
where $T_{\mu \nu}$ is the energy-momentum tensor and $u_\mu$ is the 4-velocity for the fluid defined by,
\begin{equation}
\label{4vel}
u^{\mu} = \frac{dx^{\mu}}{d\tau}\,,
\end{equation}
where $\tau$ is the proper time along the curves to which $u^\mu$ is tangent, related to the line element $ds$ by
\begin{equation}
\label{Propertime}
ds^2 = - d \tau^2 \,.
\end{equation}
The 4-velocity is subject to the constraint,
\begin{equation}
\label{4velcons}
u^{\mu}u_{\mu} = -1 \,.
\end{equation}
The contracted Bianchi identities,

\begin{equation}
\label{contBianchi}
\nabla_ \mu G^{\mu \nu} = 0 
\end{equation}
where $G^{\mu \nu}$ is the Einstein tensor, gives the continuity equation,

\begin{equation}
\label{Bianchi}
\nabla_ \mu T^{\mu \nu} = 0 
\end{equation}
where $T^{\mu \nu}$ is the total energy-momentum tensor. The general expression for the interval is metric form is,
\begin{equation}
\label{intervalmetricform}
ds^2 = g_{\mu \nu} dx^{\mu} dx^{\nu},
\end{equation}
where $ds$ is the interval. The metric tensor is subject to the constraint,
\begin{equation}
\label{MetricConstraint}
g^{\mu \nu}g_{\nu \gamma} = \delta^{\mu}_{\gamma} ,
\end{equation}
where $\delta^{\mu}_{\gamma}$ is the Kronecker delta.
The metric tensor allows us to define a unit time-like vector field
orthogonal to constant-time hypersurfaces,
\be
\label{ndef}
n_\mu \propto \frac{\p t}{\p x^\mu}\,, 
\ee
subject to the constraint
\be
n^\mu n_\mu=-1\,.
\ee
The covariant derivative of any 4-vector can be decomposed as (see for
example~Refs.~\cite{Ellis,Wald84}),
\begin{equation}
\label{4vdecomp}
\nabla_\mu n_\nu = -n_\mu {\dot{n}}_\nu 
+ \frac{1}{3} \Theta_n {\cal{P}}_{\mu \nu} 
+ \sigma_{\mu \nu} + \omega_{\mu \nu} \,,
\end{equation}
where we use the unit normal vector, $n^\mu$, purely as an example, since \eq{4vdecomp} is true for any 4-vector e.g the 4-velocity, $u^\mu$. Here
$\Theta_n$ is the expansion factor, $\sigma_{\mu \nu}$ the shear
tensor, $\omega_{\mu \nu}$ the vorticity tensor, 
and ${\cal{P}}_{\mu \nu}$ is the spatial projection tensor. Note that here, in \eq{4vdecomp} only, ${\dot{n}}_\nu = u^\mu \nabla_\mu n_\nu$, whereas through the rest of this thesis the ``dot'' denotes the derivative with respect to coordinate time. The expansion factor defined with respect to the unit normal vector is, 
\begin{equation}
\label{ExpFacugen}
\Theta_n = \nabla_{\mu} n^{\mu}\,,
\end{equation}
the shear, $\sigma_{\mu \nu}$,
is given by,
\begin{equation}
\label{sheargen}
\sigma_{\mu \nu} 
= \frac{1}{2} {\cal{P}}_{\mu}^{\alpha} {\cal{P}}_{\nu}^{\beta} 
\left( \nabla_{\beta} n_{\alpha} 
+  \nabla_{\alpha} n_{\beta} \right) 
- \frac{1}{3} \Theta_n {\cal{P}}_{\mu \nu} \,,
\end{equation}
where the spatial projection tensor is defined as
\begin{equation}
\label{spatproj}
{\cal{P}}_{\mu \nu} = g_{\mu \nu} + n_{\mu} n_{\nu} \,.
\end{equation} 

\subsubsection{FRW Background}
\label{sec:FRWBack}

A homogeneous, isotropic expanding spacetime is described by the FRW metric,
\begin{equation}
\label{spacetimemetric2}
ds^2 = -dt^2 + a(t)^2d{\rm{\bf{x}}}(x,y,z)^2,
\end{equation}
where $a(t)$ is the scale factor, and shown in Cartesian coordinates. 
The energy-momentum tensor~\eq{SET} in this case is diagonal and because of the homogeneity and isotropy has identical spatial components,

\begin{equation}
\label{FRWstressenergymomentum}
T_{0 0} = -\rho (t) \qquad , \qquad T_{i j} = \delta_{i j} P(t) ,
\end{equation}
where $\rho (t)$ is the density of the universe at time $t$ and $P(t)$ is the pressure at time $t$. When combined with \eq{Einstein2} this gives the exact solutions to the Einstein equations from GR
for the specific conditions for the FRW spacetime. The covariant form of the metric tensor for the background FRW spacetime is,
\begin{equation}
\label{FRWspacetime}
\bar{g}_{\mu \nu}=\begin{pmatrix}
  -1  0  \\
  0  a^2 \delta_{i j}
 \end{pmatrix} ,
\end{equation}
where the `bar' denotes a background quantity and $a$ is the scale factor, whilst the unperturbed contravariant metric is,
\begin{equation}
\label{UnpertgFRWuu}
{\bar{g}^{\mu \nu}}=\begin{pmatrix}
  -1  0  \\
  0  a^{-2} \delta^{i j}
 \end{pmatrix} .
\end{equation}
The background covariant 4-velocity vector, necessarily stationary relative to the unperturbed energy density fluid is given by,
\begin{equation}
\label{Unpert4veld}
u_{\mu}=\left[-1,0,0,0\right] .
\end{equation}
Similarly the contravariant form is given by,
\begin{equation}
\label{Unpert4velu}
u^{\mu}=\left[1,0,0,0\right] .
\end{equation}
From \eq{intervalmetricform} and \eq{FRWspacetime} we can construct \eq{spacetimemetric2} from Subsection~\ref{subsec:GalRecVel}. The Friedmann equation~\cite{Friedmann} is the $0 - 0$ component of the Einstein field equations,

\begin{equation}
\label{Friedmann Equation}
H^2 = \frac{8\pi G}{3} \rho - \frac{k}{a^2},
\end{equation}
where $\rho$ is the overall density of the universe incorporating all matter and radiation, and $k$ is the curvature term which can take be negative, $\mbox{ } 0 \mbox{ }$ or positive. The curvature term is so called because it corresponds to three possible geometries of spacetime, negatively curved (``Saddle" shaped), flat (Planar) and positively curved (Hypersphere) respectively.\\
The acceleration is the $i - j$ component of the Einstein field equations,

\begin{equation}
\label{Acceleration}
\frac{\ddot{a}}{a} = -\frac{4\pi G}{3} \left(\rho + 3P\right) .
\end{equation}
The $\rho$ term, which is the mass content of the universe, causes negative acceleration due to its gravitational attraction. The conservation equation is the time component of the continuity equation,~\eq{Bianchi},

\begin{equation}
\label{Conservation}
\dot{\rho} + 3 \frac{\dot{a}}{a} \left(\rho + P \right)  = 0,
\end{equation}
where $\dot{\rho}$ is the time derivative of the density.\\
For a matter dominated universe it is useful to make a simplification of pressureless matter or ``dust". In the case of pressureless matter the conservation equation,~\eq{Conservation}, becomes,

\begin{equation}
\label{Conservation2}
\dot{\rho} + 3 \frac{\dot{a}}{a} \left(\rho \right)  = 0.
\end{equation}
Therefore,
\begin{equation}
\label{densityscalefactor2}
\rho = \rho_0 \left( \frac{a_0}{a} \right)^3 ,
\end{equation}
where the ``zero" suffix denotes the value today.
Next we need to substitute \eq{densityscalefactor2} into the Friedmann equation, \eq{Friedmann Equation}. The mathematics is simplest if we assume k=0. Equation \eq{Friedmann Equation} becomes,

\begin{equation}
\label{Friedmann Equation3}
\left(\frac{\dot{a}}{a}\right)^2 = \frac{8\pi G}{3} \rho_0 \left( \frac{a_0}{a} \right)^3 .
\end{equation}
The solution of this is,

\begin{equation}
\label{scalefactortimemat}
a  = a_0 \left(\frac{t}{t_0}\right)^{\frac{2}{3}}.
\end{equation}
The derivation of the scale factor - time relation for a radiation dominated universe differs in that $\rho_{\rm{rad}}$ has equation of state $w=\frac{1}{3}$ such that \eq{Friedmann Equation} becomes,

\begin{equation}
\label{Friedmann Equation3b}
\left(\frac{\dot{a}}{a}\right)^2 = \frac{8\pi G}{3} \rho_{\rm{rad}0} \left( \frac{a_0}{a} \right)^4 .
\end{equation}
The solution of this is,

\begin{equation}
\label{Scale Factor Time Rad}
a = a_0 \left(\frac{t}{t_0}\right)^{\frac{1}{2}} .
\end{equation}
Finally, for a cosmological constant we get \eq{FriedmannInf1}, the solution of which is,

\begin{equation}
\label{Scale Factor Time Lambda}
a = a_0 e^{\left(\Lambda\right)^{\frac{1}{2}}[t - t_0]} ,
\end{equation}
giving exponential growth of the scale factor.

\subsection{Observational Evidence for the Hot Big Bang Model}
\label{sec:ObsEvidHBB}
\subsubsection{Galaxy Recession Velocities}
\label{subsec:GalRecVel}
Galaxies at sufficient distances are receding from the observer's position. The Cosmological Principle states that the universe is homogeneous and isotropic. An isotropic universe looks the same in all directions, while a homogeneous universe looks the same from every position, so that from any point, or from any galaxy within the universe, everything must appear to be moving away from these points also. Hubble's Law states~\cite{Hubble},

\textsc{\begin{equation}
\label{hubblelaw}
v = H_0 d
\end{equation}}

where $v$ is the recession velocities of the distant galaxies, $H_0$ is Hubble's Constant and $d$ is the distance to these distant galaxies.\\
For an expanding spacetime we use the Friedmann-Robertson-Walker (FRW) metric, \eq{spacetimemetric2}. The scale factor, $a$, as the name implies scales between the physical and co-moving co-ordinates as follows,

\begin{equation}
\label{co-ordinates}
{\bf{r}} = a{\rm{\bf{x}}} ,
\end{equation}\\
where ${\rm{\bf{x}}}$ is the position vector in comoving coordinates.
This makes the expression $a(t)d{\rm{\bf{x}}}(x,y,z)$ equivalent to $\bf{dr}$ from \eq{co-ordinates}. While $ds^2$ is the square of the line element governed by the spacetime metric, $d{\rm{\bf{x}}}^2$ represents the square of the spatial section only of the line element from the spacetime metric. In short the scale factor is a scaled proper separation or distance between points in space which would vary with the expansion of the universe itself as compared to the co-moving separation or distance which would remain fixed irrespective of any expansion. The Hubble parameter measures the expansion rate and is defined,

\begin{equation}
\label{hubbleparameter}
H = \frac{\dot{a}}{a},
\end{equation}

where $a$ is the scale factor of the universe, and $\dot{a}$ is the first order time derivative of the scale factor (see e.g.~Ref.~\cite{Liddle:2000cg}). The physical co-ordinates can be represented with the position vector $\bf{r}$. 
The most common interpretation of observations is to invoke the Cosmological Principle, implying a uniform expansion of the universe with no particular bias in direction or position.

\subsubsection{The Cosmic Microwave Background}
\label{subsec:CMB}

The second major evidence for an expanding universe came from the discovery of the CMB \cite{PenziasWilson,CMB2}. The conditions at early times implied by an expanding universe were high temperatures and particle densities. At early times, at redshifts, $z = 1089.90 \pm 0.23$~\cite{Ade:2015xua} from current data, 379000 years after the big bang, the universe was much more dense and therefore hotter, around 3000 K. The transition from radiation interacting with matter to not interacting is called decoupling and the time at which it occurred is denoted $t_{\rm{dec}}$. This radiation released at the time of decoupling is of a black body, isotropic and red-shifted due to the expansion of the universe since its time of release to the present. The present day CMB temperature is 2.725 K \cite{COBE}. The CMB is observed to be isotropic to a very small order - 1 part in $10^5$ \cite{COBE}. The detected CMB, at a peak temperature of 2.725 K \cite{COBE}, has a black body radiation curve corresponding to one for a body at a temperature of 3000 K which has undergone cosmological red-shift due to the expansion of the universe since the time of decoupling.

\subsubsection{Primordial Nucleosynthesis}
\label{subsec:nucleosynthesis}
Primordial nucleosynthesis is the formation of the first elements some time after the Hot Big Bang as the universe cools and particle species begin to ``freeze out''. The evidence concerns the relative abundances of the elements formed. The Hot Big Bang model of an expanding universe is readily described back to the very early times at which inflation is taken to have ceased or become insignificant in most models, typically around $t=10^{-34} \mbox{ } s$~\cite{Dodelson}. Since we are dealing with an expanding universe it is useful to recall that these temperatures must also be related to the size of the universe, i.e. the scale factor, $a$. Stefan-Boltzmann's Law gives,

\begin{equation}
\label{ScaleFactorTemperatureRelation}
T \propto \frac{1}{a} .
\end{equation}
The above relation allows very precise predictions to be made for the times at which different fundamental forces and different particle species ``froze out" of the primordial fireball and allowed the formation of matter in the universe today. Table~\ref{Tab:EarlyUniverseTimeline} shows the significant times during the early evolution of the universe.

\begin{table}
\centering
\begin{tabular}{|l|p{3.2cm}|l|p{2.0cm}|l|p{2.0cm}|l|p{2.0cm}|l|}
\hline 
\multicolumn{5}{|c|}{Universe Timeline} \\ 
\hline {\bf TIME} & {\bf DESCRIPTION} & \bf SCALE FACTOR & \bf HORIZON DIST. & \bf REDSHIFT \\
\hline 
$<10^{-34}$ s & $t_{\rm{inf}}$, time at end of inflation & $1.33\times10^{-27}$ & 0.174 m & $7.52\times10^{26}$\\
\hline
$10^{-5}$ s & $t_{\rm{had}}$, time at which hadrons fall out of equilibrium with radiation & $4.21\times10^{-13}$& 34.5 AU& $2.38 \times 10^{12}$\\
\hline
$1$ s & Time after which nuclei could begin to form & $1.33\times10^{-10}$ & 0.563 pc & $7.52\times10^{9}$ \\
\hline 
$\approx 400$ s & $t_{\rm{nuc}}$, time of nucleosynthesis & $2.66\times10^{-9}$ & 11.50 pc & $3.76\times10^{8}$ \\
\hline 
$6570 $ yrs & $t_{\rm{eq}}$, time of equality & $6.00\times10^{-5}$ & 254200 pc & 16700 \\
\hline
$379000$ yrs & $t_{\rm{dec}}$, time of decoupling & $9.00\times10^{-4}$ & 3.812 Mpc & 1090 \\
\hline
$2.997 \times 10^{17}$ s & $t_{\rm{\Lambda dom}}$, start of DE domination & 0.772 & 3.270 Gpc & 2.30\\
\hline
$4.360 \times 10^{17}$ s & $t_0$, current epoch & 1 & 4.236 Gpc & 0 \\
\hline
\end{tabular}
\caption{Timeline highlighting significant times during the early evolution of the universe. This was constructed by evolving the scale factor back in time from the present epoch using Mathematica and taking the initial conditions from the Planck satellite CMB measurement data~\cite{Ade:2015xua}.}
\label{Tab:EarlyUniverseTimeline}
\vspace*{20pt}
\end{table}

The abundances of the various elements, Hydrogen, Helium and traces of metals, primarily Lithium 7, match very closely the abundances as predicted by decreasing temperature with time. The relative abundances of elements are governed by the energies at which particle species are formed and can combine. The particle energies correspond to the temperature of the universe. When the universe was 1 second old the typical particle energies were of the order 1 MeV, which is also the order of nuclear binding energies. Hence, before this time stable nuclei could not form. There is time between hadrons forming and stable nuclei beginning to form, during which the temperature continues to drop with the expansion of the universe and protons and neutrons fall out of thermal equilibrium. Unbound neutrons are unstable have a half life of $\approx 648 \mbox{ } s$. The first nuclei in which neutrons may bind to protons is Deuterium, whose binding energy is $0.1 \mbox{ } MeV$, significantly lower than the temperature at which protons and neutrons fall out of equilibrium. This lower temperature is reached $\approx 400 \mbox{ } s$ after the Big Bang, a time comparable with the half-life of a free neutron. This time is taken as the time of primordial nucleosynthesis, $t_{\rm{nuc}}$, and it is this delay which leads to a ratio of protons to neutrons at this time of $7:1$ (see e.g.~\cite{Hou:2017uap}). The relative abundances of protons and neutrons available to collide and bond leads to the mass fraction of Hydrogen being 0.75 while Helium-4 is 0.25, which agrees very closely with current observed mass fractions. The latest Planck satellite CMB measurement data  \cite{Ade:2015xua} gives a Helium-4 mass fraction of $0.249^{+0.025}_{-0.026}$.

\subsection{Problems of the Hot Big Bang Model}
\label{sec:ProbsHBB}

With the success of the Hot Big Bang Model in explaining galaxy recession velocities, the existence of CMB radiation and the abundances of the various elements found in the universe today it may not appear in need of improvement. However, significant problems remain with the standard Hot Big Bang Model without inflation. The three main problems - the Horizon Problem, the Flatness Problem and Relic Problem - are explored below.\\
Note that in the rest of this chapter wherever the density of the universe, $\rho$, is referred to or the density of matter, $\rho_{\rm{mat}}$, both these terms assume the inclusion of both baryonic and Dark Matter.

\subsubsection{The Horizon Problem}
\label{subsec:HorProb}

This problem (see e.g.~Ref.~\cite{Rindler}) arises from the isotropy observed in the CMB temperature today at $2.725 \mbox{ } K$, uniform to $1$ part in $10^5$ \cite{COBE}, and the horizon distance at different epochs. The observed uniformity in the CMB temperature requires that all parts of the observed universe must be in causal contact at some point in the past. This means that all parts of the observed universe must have been within the horizon distance at some earlier time. If two regions in space observed today are separated by more than the scaled horizon distance at the time the light was emitted, then those two regions were outside each other's horizon distance at that time. Even at the relatively late time of the CMB generation it is possible to see that regions in the CMB are out of contact with each other and yet show all the properties of bodies in thermal equilibrium. The scaled or comoving horizon distance is given by,
\begin{equation}
\label{Comovhor}
d_{(h)} = \int_{0}^{t} \frac{dt}{a} ,
\end{equation}
where $d_{(h)}$ is the comoving horizon distance, and using natural units. Assuming matter domination the angle subtended on the sky by the horizon distance at decoupling may be found from,
\begin{equation}
\label{angle}
\theta = 360 \frac{1}{\pi} \left(\frac{t_{\rm{DEC}}}{t_0} \right)^{\frac{1}{3}} ,
\end{equation}
where $t_{\rm{DEC}}$ is the time of decoupling, $1.2\times 10^{13}s$ and $t_0$ is the time today, $4.3\times10^{17}s$. As such the regions of the CMB on the sky which would be out of causal contact would be separated by only $\approx 1 \mbox{ } ^\circ$. This is in stark contrast to the homogeneity of the CMB temperature over the whole sky. Moving further back in time towards the Big Bang the problem is magnified with regions in causal contact decreasing in size down to microscopic or Planck scales.

\subsubsection{The Flatness Problem}
\label{subsec:flatness}

The Flatness Problem (see e.g.~Ref.~\cite{Dicke}) concerns the density of the universe, $\rho$, as compared with the density of a universe whose expansion lies on the boundary between halting followed by future collapse in a ``Big Crunch" or continuing forever. This density is called the critical density, $\rho_{\rm{crit}}$, and is defined,
\begin{equation}
\label{CriticalDensity}
\rho_{\rm{crit}} (t) = \frac{3 H^2}{8 \pi G}.
\end{equation}
The critical density corresponds to a flat universe. A universe with positive curvature in the absence of a component such as DE would recollapse while a universe with negative curvature would expand forever. It is useful at this stage to introduce the density parameter \cite{Peacock},

\begin{equation}
\label{DensityParameter}
\Omega (t) = \frac{\rho (t)}{\rho_{\rm{crit}} (t)},
\end{equation}

where $\Omega$ is the density parameter. All the terms are time dependent, implying that the critical density at the current epoch will differ from that in the past. The density of the universe will include ordinary matter, Dark Matter and DE.\\
By substituting \eq{CriticalDensity} and \eq{DensityParameter} into \eq{Friedmann Equation} we have,

\begin{equation}
\label{FriedmannDensityParameter}
\Omega (t) - 1 = \frac{k}{a^2 H^2}.
\end{equation}

From this equation we can see that if the universe is at the critical density and therefore $k = 0$ then $\Omega (t) = 1$ for all time. However for any non-zero k,

\begin{equation}
\label{EvolutionofDensityParameter}
\left| \Omega (t) - 1 \right| \propto \frac{1}{\dot{a}^2}.
\end{equation}

Now, \eq{Acceleration} shows that for any universe dominated by matter or radiation with non-zero density and pressure $\ddot{a} < 0$, and therefore $\dot{a}$ must be decreasing. This implies that in both cases the density parameter must diverge away from unity. In a radiation dominated universe $a \propto t^{\frac{1}{2}}$ while in a matter dominated universe $a \propto t^{\frac{2}{3}}$ and in both cases this leads to large deviations from unity at the current time for relatively small deviation in the early history of the universe.\\
Current observations, for example the Planck 2015 results \cite{Ade:2015xua}, put the density parameter at the current time, $\Omega_0$ at $\Omega_0 = 1.0008^{+0.0040}_{-0.0039}$. Given the age of the universe is $t_0 = 10^{17} \mbox{ } s$ and the time at onset of nucleosynthesis is $t = 1 \mbox{ }s$ this implies by expressing \eq{EvolutionofDensityParameter} in terms of values at the current time, $\left| \Omega (t) - 1 \right|_{\rm{nuc}} < 10^{-17}$, giving a value of the density parameter so close to unity at that time that it appears to require a high level of tuning to produce a universe at early times which results in the universe currently observed.

\subsubsection{Relic Problem}
\label{subsec:Relic}

The problem of relics arises as a result of the conditions in the very earliest history of the universe at very high energies and temperatures. At these very high energies particle physics theories suggest that the forces \footnote{The electro-weak force and the strong nuclear force.} are unified i.e. requiring a \emph{Grand Unifying Theory} (GUT \footnote{This GUT is not necessarily a complete one incorporating gravity at this time.}) \cite{Liddle,Giacomelli,Press,Taylor,Asaka}, and the creation of high mass, stable particles are required by particle physics models at these energies. Giacomelli et al.~\cite{Giacomelli} quotes typical energies and masses for one type, magnetic monopoles, as $\approx 10^{16} \mbox{ } - \mbox{ } 10^{17} \mbox{ } GeV$ (as compared to protons at $\approx 1 \mbox{ } GeV$). Other candidates for relic particles include Domain Walls~\cite{Press}, Supersymmetric particles such as the Gravitino~\cite{Taylor} and Moduli~\cite{Asaka} fields from superstring theory.\\
When a particle's thermal or kinetic energy is greater than their mass energy ($k_{B} T \approx mc^2$) we take it to be relativistic in nature. As such the density of radiation and relativistic particles ($\rho_{rad}$) falls much more rapidly than for non-relativistic particles, which scales as matter ($\rho_{mat}$) over the history of the universe. Magnetic monopoles, which are many orders of magnitude more massive than the constituent particles we see in the universe today in ordinary baryonic matter, become non-relativistic at $T \approx 10^{16} \mbox{ } GeV \mbox{ } = 10^{28} \mbox{ } K$. This occurs at $t=10^{-10} \mbox{ }s$ which is also of the order of the time at which they first form. Their density comes to dominate the evolution of the universe almost as soon as they are formed and long before any other particle species form. In a matter dominated universe $a \propto t^{\frac{2}{3}}$ whereas $a \propto t^{\frac{1}{2}}$ for a radiation dominated universe, so the expansion rate will be much greater once the magnetic monopoles start to dominate. By the time baryons have formed they will be spatially separated from each other by too great a distance for proton-neutron collisions to be likely. This would lead in turn to a lack of Helium 4 in the universe in conflict with observational evidence.

\subsection{Inflation - an Elegant Solution to the Problems of the Hot Big Bang}
\label{sec:inflation}

\subsubsection{The Basics of Inflation}
\label{subsec:InfBas}

Inflation provides a solution to the problems of the Hot Big Bang model through a period of accelerated expansion i.e. $\ddot{a} > 0$. Note: the Friedmann equation and acceleration equations are quoted in this section for illustrative purposes. They are covered in more detail in the governing equations section, Section~\ref{sec:equationsofexpansion}. Assuming a cosmological constant is the dominant energy content of the universe at this time we can simplify the Friedmann equation,
\begin{equation*}
\label{Friedmann Equationsimp}
H^2 = \frac{8\pi G}{3} \rho - \frac{k}{a^2} ,
\end{equation*}
to,
\begin{equation}
\label{FriedmannInf1}
\left( \frac{\dot{a}}{a} \right)^2 = \frac{8 \pi G}{3} \Lambda_{\rm{inf}} .
\end{equation}

Equation~\eqref{FriedmannInf1} shows that $\frac{\dot{a}}{a} = constant$ which implies an exponential expansion. We can find the minimum value of the pressure required for accelerated expansion from the acceleration equation,
\begin{equation*}
\label{Accelerationsimp}
\frac{\ddot{a}}{a} = -\frac{4\pi}{3} \left(\rho + 3P\right) .
\end{equation*}
For positive acceleration we require a negative pressure term. From \eq{Acceleration}, given that we know $\ddot{a}$ (and by definition, $a$) must be positive we can see that,

\begin{equation}
\label{PressureInf}
P < - \frac{\rho}{3},
\end{equation}
 
or

\begin{equation}
\label{eostake1}
w < - \frac{1}{3},
\end{equation}

where $w$ is the equation of state, defined as,
\begin{equation}
\label{eostake2}
w = \frac{P}{\rho}.
\end{equation}
If we replace the generalised density, $\rho$, with our inflationary cosmological constant, $\Lambda_{\rm{inf}}$, an equation of state for an inflationary cosmological constant may be obtained from the conservation equation, 
\begin{equation*}
\label{Conservationsimp}
\dot{\rho} + 3 \frac{\dot{a}}{a} \left(\rho + P \right)  = 0,
\end{equation*}
to give,
\begin{equation}
\label{ConservationInf}
3 \frac{\dot{a}}{a} \left(\Lambda_{\rm{inf}} + P \right)  = 0,
\end{equation}
\eq{ConservationInf} leads to an equation of state for $\Lambda_{\rm{inf}}$ of,

\begin{equation}
\label{EqnStateInf}
P  = -\Lambda_{\rm{inf}},
\end{equation}
or $w=-1$. This simple inflationary cosmological constant model, de Sitter~\cite{Sitter}, could not generate the observed universe, however it is sufficient to demonstrate the possibility of inflation and allows us to address the problems of the Hot Big Bang model.

\subsubsection{A Solution to The Horizon Problem}
\label{subsec:solhor}

The predictions of a universe undergoing ordinary non-inflationary expansion disagree with the observed homogeneity in the CMB, and the distribution of matter at late times on the largest scales. To solve the horizon problem light must have been able to travel much further in the universe at some time before both decoupling and the present day. This condition can be expressed in terms of the horizon distance, \eq{Comovhor} as, 

\begin{equation}
\label{ScaledHorizonDistanceInf2}
\int_{t_{\rm{b}}}^{t_{\rm{f}}} \frac{dt}{a(t)} > 2 \int_{t_{\rm{dec}}}^{t_0} \frac{dt}{a(t)},
\end{equation}
where $t_{\rm{b}}$ is the time at the start of inflation, $t_{\rm{f}}$ is the time inflation finishes and $t_{\rm{dec}}$ is the time of decoupling when the CMB was produced and $t_0$ is today. With appropriate values for, $t_{\rm{b}}$, $t_{\rm{f}}$ and $\Lambda_{\rm{inf}}$ it is indeed possible to satisfy the condition in \eq{ScaledHorizonDistanceInf2}. Therefore inflation provides a solution to the horizon problem. Due to the exponential nature of the expansion during inflation and the importance of the length of time for inflation the time for inflation is often given in e-foldings.\\

\subsubsection{A Solution to The Flatness Problem}
\label{subsec:solfla}

The observed value of the density parameter lying very close to unity would require fine tuning in the absence of a mechanism for this to arise naturally. A less finely-tuned model would drive the density parameter very close unity at very early times such that it remains close to this value to the present day.
In \eq{EvolutionofDensityParameter} we saw that $\left| \Omega (t) - 1 \right| \propto {\dot{a}^{-2}}$. For accelerated expansion $\ddot{a}$ is positive and therefore $\dot{a}$ must be increasing pushing $\Omega (t)$ towards unity, in this case exponentially fast. Therefore, it takes a very short time compared to the history of the universe to push the density parameter so close to unity that today it is still unity to within one part in $10^3$~\cite{Ade:2015xua}.

\subsubsection{Explaining the Apparent lack of Relics}
\label{subsec:solrel}

The relic problem is usually taken to be solved by assuming they are generated before or during the period of inflation. Given the exponential rate of expansion during their formation they become separated by large distances due to the rapidly increasing scale factor. From the solution to the horizon problem in we see that these relic particles will also be pushed beyond each other's co-moving horizon distance. Consequently they are likely to be beyond each other's co-moving horizon distance today and their particle density so low there may not be a single magnetic monopole within our current co-moving horizon distance. Even allowing for one, or a few, magnetic monopoles within our co-moving horizon distance the probability of it interacting with a detector on earth would be vanishingly small. Additionally, their density would be subdominant to all other constituents and therefore would not lead to early matter domination, inconsistent with other predictions and observations e.g. primordial nucleosynthesis.\\

\subsubsection{The Details of Inflation}
\label{subsec:Infdet}

\begin{figure}
\centerline{\includegraphics[angle=0,width=80mm]{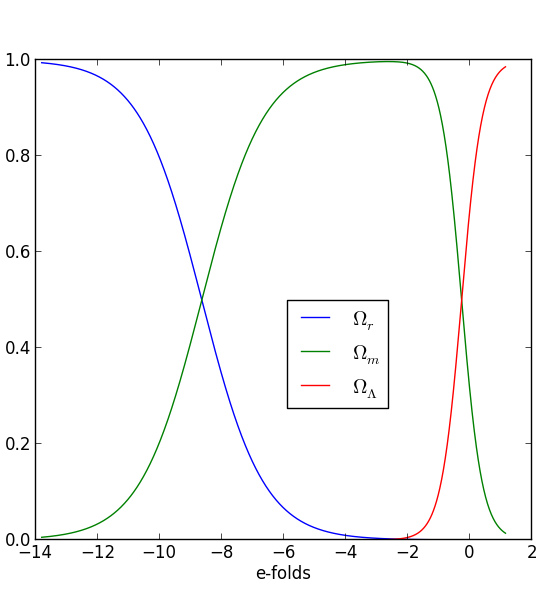}}  
\caption{The evolution of density with scale factor for a cosmological constant, matter and radiation. Once a cosmological constant dominates it does so for all time. This figure was produced using the \PY~code used in Chapter 4 and described in more detail in Chapter 5. The background evolution was plotted from initial conditions derived from values today taken from the Planck satellite CMB measurement data~\cite{Ade:2015xua}, $\Omega_\Lambda = 0.6911 \pm 0.0062 , \Omega_m = 0.3089 \pm 0.0062 , \Omega_r = 8.75893 \pm 0.00003 (\times 10^{-5})$.}
\label{densityevolutions}
\vspace*{-10pt}
\end{figure}

Figure~\ref{densityevolutions} serves to show a qualitatively comparison between the evolution of radiation, matter and a cosmological constant density parameters in a universe with these constituents. The density parameters today as taken from the Planck satellite CMB measurement data~\cite{Ade:2015xua} are,

\bea
\label{•}
\Omega_\Lambda = 0.6911 \pm 0.0062 , \\ \nonumber \Omega_m = 0.3089 \pm 0.0062 , \\ \nonumber \Omega_r = 8.75893 \pm 0.00003 (\times 10^{-5}) . 
\eea
Once a cosmological constant dominates the density of the universe it will do so for all time. Since in this de Sitter model the inflationary cosmological constant dominates from the outset the universe will never reach a period of radiation or matter domination and consequently not match observations. However, a period of constant or near constant energy density would be useful in our models in order to generate a similar inflationary period. In addition this energy density must at some point decay away in order to allow for both the radiation dominated and matter dominated phases at later times. A simple way to satisfy the above conditions is to introduce a scalar field, $\ph$, to describe the energy content of the universe (see e.g.~Ref.~\cite{Zeldovich})\footnote{The governing equations quoted in this section are covered in more detail in Section~\ref{sec:equationsofexpansion}.}. The Lagrangian for such a field is

\be
\label{lagsf}
{\cal{L}_\ph} = -\frac{1}{2} g^{\mu \nu} \partial_\mu \ph \partial_\nu \ph - V(\ph) ,
\ee
where $\ph$ is the scalar field, the first term is a kinetic term, whilst $V(\ph)$ is the potential.\\
Invoking again the cosmological principle as described in Subsection~\ref{subsec:GalRecVel} - that the universe is homogeneous and isotropic - this homogeneity also implies that the inflaton scalar field must be the same everywhere i.e. invariant with position. Hence the scalar field is dependent only on time, $\ph \equiv \ph (t)$. The energy density for such a scalar field is given by,

\begin{equation}
\label{EnergyDensityPhi}
\rho = \frac{1}{2} \dot{\ph}^2 + V(\ph) ,
\end{equation}
where $\dot{\ph}$ is the time derivative of the scalar field, the first term can be thought of as the kinetic term introduced above, and similarly the second term is the potential term. The pressure in the FRW spacetime \cite{Zeldovich} is given by,

\begin{equation}
\label{PressurePhi}
P = \frac{1}{2} \dot{\ph}^2 - V(\ph).
\end{equation}
If $V(\ph)$ is near constant for a period, with only small variation in $\ph$, $V(\ph)$ will dominate producing a negative pressure necessary for inflation as with the de Sitter model. The Einstein field equations give us the Friedmann equation, which for a scalar field is,
\begin{equation}
\label{FriedmannPhi}
H^2 = \frac{8\pi G}{3} \left( \frac{1}{2} \dot{\ph}^2 + V(\ph) \right) ,
\end{equation}
where we have taken the curvature term to be zero. If the scalar field causes inflation this would flatten the universe, making this a reasonable assumption.
By substituting \eq{EnergyDensityPhi} and \eq{PressurePhi} into the conservation equation we obtain the Klein-Gordon equation,

\begin{equation}
\label{KleinGordon}
\ddot{\ph} + 3 H \dot{\ph} + V'(\ph) = 0,
\end{equation}

where a `dash' denotes the derivative with respect to $\ph$. Finally by substituting the scalar field density into the acceleration equation we find,

\begin{equation}
\label{AccelerationPhi}
\frac{\ddot{a}}{a} = -\frac{8\pi G}{3} \left(\dot{\ph}^2 - V(\ph) \right).
\end{equation}

\subsubsection{The Slow Roll Approximation}
\label{subsec:InfSRA}

During standard inflation it is assumed the scalar field ``slowly rolls", meaning that the scalar field, $\ph$, is changing very slowly during the period of inflation. This is called the slow roll approximation (SRA) and allows us to also approximate the governing equations and make them analytically treatable. For the SRA $\dot{\ph}^2 \ll V(\ph)$ \cite{Kinney}, which in \eq{AccelerationPhi} gives the required positive acceleration. It also allows us to re-write the Friedmann equation, \eq{FriedmannPhi}, as,

\begin{equation}
\label{FriedmannSRA}
H^2 \simeq \frac{8\pi}{3 {M_{\rm{pl}}}^2} \left(V(\ph) \right) ,
\end{equation}
Similarly in the SRA we assume that $\ddot{\ph} \ll 3H \dot{\ph} + V'(\ph)$ \cite{Kinney}, so the Klein-Gordon equation, \eq{KleinGordon}, becomes,

\begin{equation}
\label{KleinGordonSRA}
3 H \dot{\ph} + V'(\ph) \simeq 0 .
\end{equation}
We define slow roll parameters, $\epsilon$ and $\eta$ to describe the small changes occurring. The first slow roll parameters is defined (see e.g.~Ref.~\cite{Taylor}),

\begin{equation}
\label{EpsilonV}
\epsilon = \frac{{M_{\rm{pl}}}^2}{16 \pi} \left(\frac{V'(\ph)}{V(\ph)}\right)^2 ,
\end{equation}
where $\epsilon$ is our first slow roll parameter. It may also be expressed using the Friedmann equation in first order form, in terms of $\ph$ (see e.g.~Ref.~\cite{Kinney}),
\begin{equation}
\label{EpsilonPhi}
\epsilon(\ph) = \frac{4 \pi}{{M_{\rm{pl}}}^2} \left(\frac{\dot{\ph}}{H}\right)^2 .
\end{equation}
We can see in \eq{EpsilonPhi} that as long as $\dot{\ph}$ is very small compared to $H$ then $\epsilon \ll 1$. This is one of the necessary conditions for the SRA \cite{Liddle}.\\
Our second slow roll parameter is defined~\cite{Taylor},

\begin{equation}
\label{EtaVapprox}
\eta = \frac{{M_{\rm{pl}}}^2 }{8 \pi} \left( \frac{V''}{V} \right) .
\end{equation}
or expressed in terms of $\ph$ as~\cite{Kinney},

\begin{equation}
\label{EtaPhi}
\eta = \frac{\ddot{\ph}}{H\dot{\ph}} .
\end{equation}
We can see in \eq{EtaPhi} that as long as the magnitude of $\ddot{\ph}$ is very small compared to $H\dot{\ph}$ then $|\eta| \ll 1$. This is a second necessary condition for the SRA \cite{Liddle}.
It can be useful to relate the slow roll parameters to the number of e-foldings occurring during inflation and to each other. The relation between scale factor and time measured in e-foldings is given by,

\begin{equation}
\label{scalefactortimeSRA}
a = a_0 e^{-N} ,
\end{equation}
where in this case $a_0$ is the scale factor today and $N$ is the number of e-foldings. One e-fold is the time it takes for the horizon distance to change by a factor of $e$ and so \eq{scalefactortimeSRA} becomes,
\be
\label{Nloga}
N = \ln \left(\frac{a}{a_0}\right) .
\ee
Consequently we introduce the convention here of counting e-foldings backwards from the end of inflation, or any other relevant end time e.g. today. The number of e-foldings may then be related to the Hubble parameter by differentiating \eq{scalefactortimeSRA} with respect to time and dividing by the scale factor to give,

\begin{equation}
\label{HubblepNSRA2}
dN = -H dt .
\end{equation}
Next we need to link the number of e-foldings to the slow roll parameter, $\epsilon$,

\begin{equation}
\label{EpsilonN}
\epsilon \simeq \frac{1}{H} \frac{dH}{dN} .
\end{equation}

Both slow roll parameters are $\ph$ dependent and both describe characteristics of the potential, $V(\ph)$. \eq{EpsilonV} contains the term $\frac{V'(\ph)}{V(\ph)}$, the normalised slope of the potential. \eq{EtaVapprox} contains the term $\frac{V''(\ph)}{V(\ph)}$, the normalised curvature of the potential.
For the SRA to hold it is necessary that $V'$ and $V''$ be very small, or put more formally in terms of the slow roll parameters, $\epsilon \ll 1$ and $|\eta| \ll 1$. It is worth noting however that this condition alone is not sufficient to ensure the SRA will hold however \cite{Liddle}, since although $V(\ph)$ may be very slowly changing or near flat, $\dot{\ph}$ could be large.\\

\subsection{Dark Energy Driving Late Time Accelerated Expansion}
\label{sec:Lambda}
We now briefly look at the final missing component of the standard $\Lambda$CDM cosmology, namely DE. We shall describe the observations which made an additional component necessary and how DE may be used to explain these observations.

\subsubsection{Observations of Late Time Accelerated Expansion}
\label{subsec:ObsproLambda}
\begin{figure}
\centerline{\includegraphics[angle=0,height=90mm]{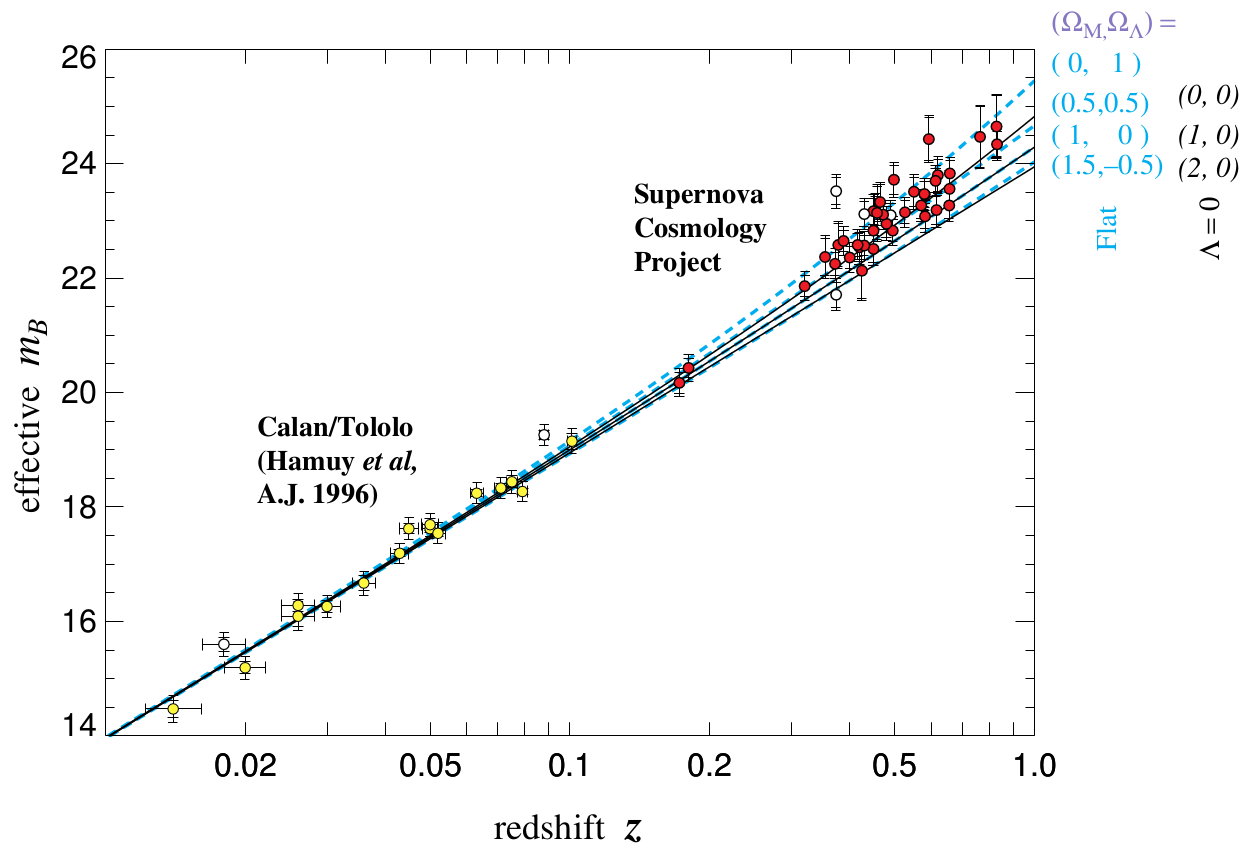}} 
\caption{Hubble diagram from~\cite{Perlmutter:1998np} showing the deviation from the Hubble law for distant type Ia supernovae.}
\label{SNIapic}
\end{figure}

In 1998 Perlmutter et al.~\cite{Perlmutter:1998np} and Reiss et al.~\cite{Riess:1998cb} announced the discovery of the apparent acceleration in the expansion rate of the universe, made through analysis of the Hubble diagram for distant supernovae. Figure~\ref{SNIapic} show the initial results from the Supernova Cosmology Project~\cite{Perlmutter:1998np}. 

These observations are usually attributed to a late time accelerated expansion of the universe. As we shall see in Chapter~\ref{ch:3} this is not the only possible explanation. An inhomogeneous cosmology where the expansion of space is not only time dependent but has some additional spatial dependency could produce a similar phenomenon to accelerated expansion since the expansion rate would be different at different distances from the observer. However, in this initial discussion of $\Lambda$CDM cosmology we shall consider only acceleration driven by a cosmological constant. Further evidence for DE comes from several sources including the CMB constraints on the flatness of the universe \cite{Ade:2015xua} giving $| \Omega_k | < 0.005$ where $\Omega_k$ is the density parameter of the curvature. When coupled with the CMB constraints in total matter at $\Omega_M = 0.308 \pm 0.012$, which includes both CDM and baryons, the remaining energy density required for flatness is attributed to DE. Independently, observation of galaxy clusters (see e.g.~Ref.~\cite{Morandi:2016cet} puts similar constraints on the total matter at around $\Omega_M = 0.311 \pm 0.014$, with similar DE requirements to match the observed flatness. Finally, the Baryon Accoustic Oscillation (BAO) data from galaxy surveys \cite{Anderson:2013zyy,Bonnett:2015pww,Dawson:2015wdb} also favour models with a DE component of around $\Omega_{DE} = 0.75$.

\subsubsection{A Cosmological Constant Driving Late Time Accelerated Expansion}
\label{subsec:LambdaInf}
Since no exit from late time accelerated expansion has been observed the simplest inflationary model, de Sitter, may be employed to drive late time accelerated expansion. Hence the introduction of a cosmological constant, $\Lambda$, in the $\Lambda$CDM model.\\
As such the standard model of cosmology, namely $\Lambda$CDM in flat FRW evolves as follows. From an initial inflationary period the universe passes through radiation domination to a period of CDM domination and finally to a new accelerated expansion epoch at late times due to DE domination in the form of a cosmological constant, $\Lambda$ (see Figure~\ref{densityevolutions}). The latest Planck values for the density parameter for DE is $\Omega_\Lambda=0.6911 \pm 0.0062$.

\chapter{Cosmological Perturbation Theory}
\label{ch:2}

%

\section{Structure in the Universe}
\label{sec:Conc}

\emph{Cosmological Perturbation Theory} (CPT) is a vital tool in the analysis of the universe across all epochs. For a more comprehensive description of this field see e.g.~Ref.~\cite{Liddle:2000cg} but a brief overview follows.\\
Inflation provides the mechanisms whereby the small scale anisotropies in the universe, as seen in both the CMB and galaxy distributions may be generated by the initial conditions in the universe. Quantum fluctuations in the inflaton become perturbations in the density of matter, and the inflation it drives simultaneously freezes in these matter perturbations, and associated gravitational perturbations, from early times such that we can observe them today. CPT is the tool which allows us to model perturbed cosmologies, link primordial perturbations to late time matter distributions and model the evolution of perturbations, including density perturbations, over time. In the standard $\Lambda$CDM model of cosmology we typically assume a flat FRW spacetime.

\section{Cosmological Perturbation Theory in Flat FRW}
\label{PertTheoryFRW}
\subsection{Introduction}
\label{CPTintro}
In this section we look at CPT in flat FRW cosmology in more detail. Ultimately we seek to apply these same techniques, modified as necessary, to LTB cosmology. In both cases we shall be looking for the perturbed forms of cosmologically significant scalars, vectors and tensors and investigating conserved quantities and conservation equations. We do this since these these conserved quantities, such as, for example, the gauge-invariant curvature perturbation, allow us to link early to late times in the formation and evolution of structure in the universe e.g. through the density perturbation on flat hypersurfaces. Consequently, we shall also construct gauge invariant quantities. Since these will contain no gauge or coordinate artefacts they are useful when comparing with other research in CPT which is formulated in a gauge invariant way.

\subsection{The Perturbed Metric and 4-Velocities}
\label{sec4vel}
We perform a $3+1$ decomposition of spacetime into spatial hypersurfaces of constant time, as can be seen in the FRW metric used earlier~\eq{spacetimemetric2}. This allows us to further decompose quantities into scalar, vector and tensor components according to their transformations on spatial 3-hypersurfaces. At linear order scalar, vector and tensor perturbations are decoupled. The metric may be decomposed into a background metric and a perturbed metric as,
\be
\label{metricsplit}
g_{\mu \nu} = \bar{g}_{\mu \nu} + \delta g_{\mu \nu} ,
\ee
then the perturbed portion of the metric is given by,
\begin{equation}
\label{PertonlygFRWdd}
\delta g_{\mu \nu}=\begin{pmatrix}
  -2 \Phi & aB_i \\
  aB_j & a^2 2C_{i j} 
 \end{pmatrix} ,
\end{equation}
where $\Phi$ is the lapse function, or perturbation in the proper time coordinate, $B_i$ is the perturbation in the mixed temporal and spatial components of the metric and $C_{i j}$ is the perturbation in the spatial only components of the metric. $\Phi$ is a scalar perturbation.
The perturbed components of the contravariant form of the metric may be found using the constraint,
\begin{equation}
\label{gconstraint}
g^{\mu \nu} g_{\nu \gamma} = {\delta^{\mu}}_{\gamma} .
\end{equation}
The perturbed metric is,
\begin{equation}
\label{PertonlygFRWuu}
\delta g^{\mu \nu}=\begin{pmatrix}
  +2 \Phi & a^{-1} B^i\\
  a^{-1} B^j & -2 a^{-2} C^{i j}
 \end{pmatrix} .
\end{equation}
The line element derived from the covariant form of the perturbed metric is given by,
\begin{equation}
\label{PertlineFRW}
ds^2=-\left(1+2\Phi\right)dt^2 + 2a B_i dx^i dt + a^2\left(\delta_{i j} + 2C_{i j}\right)dx^i dx^j ,
\end{equation}
The $B_i$ component is a ``true" vector perturbation and may be further decomposed as,
\be
\label{Bdecomp}
B_i = B,_i - S_i ,
\ee
where $B$ is a scalar perturbation and $S_i$ the divergence-free vector perturbation and the `comma' denotes the partial derivative with respect to the coordinates. Similarly $C_{i j}$ may be further decomposed as,
\be
\label{Cdecomp}
C_{i j} = - \psi \delta_{i j} + E,_{i j} + F_{(i,j)} + \frac{1}{2} h_{i j} ,
\ee
where $\psi$ and $E$ are scalar perturbations, $F_i$ is the divergence-free vector perturbation and $h_{i j}$ is divergence-free, trace-free tensor perturbation.\\
The unperturbed form of the 4-velocities using the metric for flat FRW in coordinate time with a negative signature, in natural units is defined as in \eq{4vel}. We define the 3-velocity with respect to conformal time, $\eta$, as,
\begin{equation}
\label{3veldefn}
{{v}}^i = \frac{dx^{i}}{d \eta} ,
\end{equation}
where,
\be
\label{conftocoord}
dt = a d \eta  .
\ee
We use \eq{Propertime} to give $d\tau$, where $\tau$ is the proper time along the curves to which $u^\mu$ is tangent, to linear order as,
\begin{equation}
\label{dtauvsdt}
d\tau = (1+2\Phi)^{\frac{1}{2}} dt .
\end{equation}
From this and \eq{4vel} we find the timelike component of the 4-velocity is,

\begin{equation}
\label{4vel0b}
u^0 = \frac{dx^{0}}{d \tau} = \frac{dt}{(1+2\Phi)^{\frac{1}{2}} dt} = (1-\Phi) .
\end{equation}
Similarly the spatial component of the 4-velocity is found to be,
\begin{equation}
\label{4velia}
u^i = \frac{dx^{i}}{d \tau} = \frac{dx^{i}}{(1+2\Phi)^{\frac{1}{2}} ad\eta} ,
\end{equation}
which when combined with \eq{3veldefn}, and remembering that in the background there is no spatial velocity for the fluid, and therefore any 3-velocity is by definition a perturbation,
\begin{equation}
\label{4velib}
u^i = (1+2\Phi)^{-\frac{1}{2}} \frac{{{v}}^i}{a} = \frac{{{v}}^i}{a} ,
\end{equation}
to linear order.\\
This gives the 4-velocity as,
\begin{equation}
\label{Pert4veluresc}
u^{\mu}=\left[ (1-\Phi) , \frac{v^i}{a} \right] .
\end{equation}
As with the metric, the 4-velocity may be separated into a background and a perturbed metric such that,
\be
\label{4vdecompu}
u^{\mu} = \bar{u}^{\mu} + \delta u^{\mu} .
\ee
In this case the perturbed 4-velocity becomes simply,
\begin{equation}
\label{onlyPert4veluresc}
\delta u^{\mu}=\left[ -\Phi , \frac{v^i}{a} \right] .
\end{equation}
The covariant 4-velocities may be obtained simply by the metric acting upon the contravariant 4-velocities,
\begin{equation}
\label{4veldmet}
u_{\mu}=u^{\nu}g_{\nu \mu} .
\end{equation}
The indices may be split to give the time and spatial components separately as,
\begin{equation}
\label{4vel0dmet}
u_{0}=u^{\nu}g_{\nu 0}\\
= u^{0}g_{0 0} + u^{i}g_{i 0}\\
= -(1-\Phi)(1+2\Phi) + \frac{v^i}{a} a B_i ,
\end{equation}
which to linear order becomes,
\begin{equation}
\label{4vel0d}
u_{0} = -(1+\Phi) .
\end{equation}
Similarly the spatial component of the 4-velocity is found to be,
\begin{equation}
\label{4velidmet}
u_{i}=u^{\nu}g_{\nu i}\\
= u^{0}g_{0 i} + u^{j}g_{i j}\\
= (1-\Phi)aB_i + \frac{v^j}{a} a^2 ({\delta}_{ij} + 2C_{ij})
\end{equation}
which to linear order becomes,
\begin{equation}
\label{4velid}
u_{i} = aB_i + a v_i .
\end{equation}
Therefore we may write the covariant perturbed 4-velocity for flat FRW as,
\begin{equation}
\label{Pert4veld}
u_{\mu}=\left[ -(1+\Phi) , aB_i + a v_i  \right] .
\end{equation}
This may also be decomposed as,
\be
\label{4vdecompd}
u_{\mu} = \bar{u}_{\mu} + \delta u_{\mu} .
\ee
In this case the perturbed 4-velocity becomes simply,
\begin{equation}
\label{OnlyPert4veld}
\delta u_{\mu}=\left[ -\Phi , aB_i + a v_i  \right] .
\end{equation}
The expansion scalar as defined in \eq{ndef} for unit normal vector field in FRW is,
\be
\label{expfFRW}
\Theta_n = 3H \left(1- A \right) - 3\dot\psi +
 \nabla^2 \sigma \,,
\ee
where $\sigma$ is the shear defined,
\be
\label{shear}
\sigma = \dot{E} - B \,.
\ee
\\
\subsection{The Perturbed Energy-Momentum Tensor}
\label{Tmunu}
The unperturbed energy-momentum tensor, $T_{\mu \nu}$, for a perfect fluid in the absence of anisotropic stress is given in~\eq{SET}. We now perturb $T_{\mu \nu}$ as follows,
\begin{equation}
\label{PertTdd}
T_{\mu \nu}=\left(\bar{P} + \delta P + \bar{\rho} + \delta \rho \right)u_{\mu} u_{\mu} + \left( \bar{P} + \delta P \right) g_{\mu \nu} ,
\end{equation}
where $\bar{P}$ and $\bar{\rho}$ are the background pressure and energy density respectively, whilst $\delta P$ and $\delta \rho$ are the perturbations in these same quantities. The energy-momentum tensor may also be decomposed into a background tensor and a perturbed tensor such that,
\be
\label{SETdecomp}
T_{\mu \nu} = {\bar{T}_{\mu \nu}} + \delta T_{\mu \nu} .
\ee
\\
The various components of $T_{\mu \nu}$ may be found by substituting for the appropriate components of the perturbed 4-velocity,~\eq{4vel0b} and~\eq{4velib} and perturbed metric~\eq{PertlineFRW},
\begin{equation}
\label{PertTdd00b}
T_{0 0}=\left(\bar{P} + \delta P + \bar{\rho} + \delta \rho \right)(1+\Phi)^2 - \left( \bar{P} + \delta P \right)(1+2\Phi) ,
\end{equation}
which to linear order becomes,
\begin{equation}
\label{PertTdd00c}
T_{0 0} = \bar{\rho} +2\Phi \bar{\rho} + \delta \rho  ,
\end{equation}
giving the unperturbed portion of the $00$ component of ${T^{\mu}}_{\nu}$ as, ${\bar{T}_{0 0}} = \bar{\rho}$ and the perturbation only as $\delta T_{0 0} = 2\Phi \bar{\rho} + \delta \rho$.
Raising the index gives the $00$ component to linear order more concisely as,
\begin{equation}
\label{PertTud00}
{T^{0}}_{0} = - \bar{\rho} - \delta \rho .
\end{equation}
This may be stated alternatively as the unperturbed portion of the $00$ component of the ${T^{\mu}}_{\nu}$ being, ${{\bar{T}^{0}}_{0}} = -\bar{\rho}$ and the perturbation only being $\delta {T^{0}}_{0} = -\delta \rho$.
The other components of ${T^{\mu}}_{\nu}$  to linear order are,
\begin{equation}
\label{PertTud0i}
{T^{0}}_{i}= \left(\bar{P} + \delta P + \bar{\rho} + \delta \rho \right)u^{0} u_{i} = \left(\bar{P} + \bar{\rho} \right) \left( a B_i + a v_i \right) ,
\end{equation}
or the unperturbed portion of the $0i$ component of the ${T^{\mu}}_{\nu}$ is, ${{\bar{T}^{0}}_{i}} = 0$ and the perturbation only being $\delta {T^{0}}_{i} = \left(\bar{P} + \bar{\rho} \right) \left( a B_i + a v_i \right)$.\\
Finally the spatial only components of ${T^{\mu}}_{\nu}$ we find,
\begin{equation}
\label{PertTudij}
{T^{i}}_{j}= \left(\bar{P} + \delta P \right) {\delta^{i}}_{j} ,
\end{equation}
to linear order, since all the multipliers generated by $u^{i} u_{j}$ are second order, leaving only the right-hand term in the expression. This gives us the unperturbed portion of the $ij$ component of the ${T^{\mu}}_{\nu}$ as, ${{\bar{T}^{i}}_{j}} = \bar{P} {\delta^{i}}_{j} $ and the perturbation only as $\delta {T^{i}}_{j} = \delta P {\delta^{i}}_{j} $.\\
\subsection{Conservation Equations}
\label{consquant}
We find the conservation equation\footnote{Cadabra~\cite{Cadabra}, a tensor manipulation package, was use to aid in many of these derivations} for the perturbed energy momentum tensor $T^{\mu \nu}$ using the continuity equation, \eq{Bianchi}, such that,
\begin{eqnarray}
\label{PertEconsback}
{{\nabla}_{\mu} \bar{T}^{\mu 0}} &=& \dot{\bar{\rho}} + 3 H \left( \bar{\rho} + \bar{P} \right) ,
\end{eqnarray}
which is the fluid equation for the background, where ${{\nabla}_{\mu} \bar{T}^{\mu 0}} = 0$ and,
\begin{equation}
\label{PertEconspert}
\delta {\nabla}_{\mu} {T}^{\mu 0} = {\partial}_{i}{{v}^{i}} a^{-1} (\bar{\rho} +\bar{P} ) + \dot{\delta \rho} + \dot{C}_i^i  (\bar{\rho} + \bar{P}) + 3 H (\delta \rho  + \delta P)  ,
\end{equation}
again where $\delta {\nabla}_{\mu} {T}^{\mu 0} = 0$. We obtain the equivalent momentum conservation equation,
\begin{eqnarray}
\label{PertMomCons}
{\nabla}_{\mu} T^{\mu i} &=& \left( \bar{\rho} + \bar{P} \right) \left( 4 H a^{-1} {v}^{i} + {\dot{v}^{i}} a^{-1} + {\dot{B}^{i}} a^{-1} \right) \\ \nonumber &+& \dot{\bar{\rho}} a^{-1} {v}^{i} + {\partial}^{i}{\delta P} a^{-2} ,
\end{eqnarray}
again, where ${\nabla}_{\mu} T^{\mu i} =0$. This contains only perturbed quantities i.e. ${\nabla}_{\mu} T^{\mu i} = \delta {\nabla}_{\mu} T^{\mu i}$.\\
We derive here only the perturbed conservation equations since they are needed for the following sections on gauge transformations and gauge invariance. We postpone the derivation of the perturbed Einstein field equations in FRW to chapter 3 where they are needed for comparison with LTB and Lema{\^\i}tre cosmologies.
\\
\subsection{Gauge Transformations}
\label{GaugeTrans}
In order to find gauge-invariant perturbations we must first understand the transformation behaviour of the perturbed quantities. There are two approaches to gauge transformations; passive and active. In the passive approach we specify the relation between the two coordinate systems i.e.  the original coordinates and the ``shifted" coordinates. The change in the perturbed quantities under this coordinate transformation is then calculated, but at the same physical point. In the active approach the transformation in the perturbed quantities is induced by a mapping, but is calculated at the same coordinate point. We shall first use the passive approach for the transformation behaviour of the density perturbations for illustrative purposes (throughout the rest of this thesis we use the active approach). We shall assign the manifold in which the original coordinates live unmarked coordinates, e.g. $x^{\mu}$, whilst shifted coordinates shall be marked with a tilde, e.g. ${\tilde{x}}^{\mu}$, such that,
\begin{equation}
\label{gaugetrans1}
{\tilde{x}}^{\mu} = x^{\mu} + \delta x^{\mu} ,
\end{equation}
where $\delta x^{\mu}$ is the coordinate shift. We first look at the energy density, $\rho(x^{\mu})$. The coordinate shift $\delta x^{\mu}$ may be decomposed into, 
\be
\label{coorddecomp}
\delta x^{\mu} = [\delta t , \delta x^{i}] .
\ee
Note that the $\delta x^{i}$ could itself be further decomposed into scalar and vector components,
\be
\label{spcoorddecomp}
\delta x^{i} = \delta^{ij} \delta {x_{,j}} + {\gamma}^i . 
\ee
If we do not decompose the density into a background and perturbation and just apply the change in coordinates we will have simply performed a passive gauge transformation as in \eq{gaugetrans1}, i.e., 
\begin{equation}
\label{gaugetrans2}
\tilde{\rho}({\tilde{x}}^{\mu}) =\tilde{\rho}(x^{\mu} + \delta x^{\mu}) = \tilde{\rho}(x^{\mu}) + \frac{\partial \tilde{\rho}(x^{\mu})}{\partial x^{\mu}} \delta x^{\mu} +\mathcal{O}(\delta {x^{\mu}}^2) .
\end{equation}
To compare perturbed quantities in the background manifold with those in the perturbed manifold we must decompose such a quantity into a background and perturbed portion, e.g.,
\begin{equation}
\label{decomp1}
\tilde{\rho}({\tilde{x}}^{\mu}) = {\tilde{\bar{\rho}}}({\tilde{x}}^{\mu}) + \delta \tilde{\rho}({\tilde{x}}^{\mu}) .
\end{equation}
4-scalar quantities are covariant, i.e. $\tilde{\rho}({\tilde{x}}^{\mu}) = {\rho}({x}^{\mu})$. We assume $\bar{\rho}(x^{\mu}) = {\tilde{\bar{\rho}}}(x^{\mu})$. From these we can find,
\begin{eqnarray}
\label{densitypert1}
\tilde{\rho}({\tilde{x}}^{\mu}) &=& {\tilde{\bar{\rho}}}({\tilde{x}}^{\mu}) + \delta \tilde{\rho}({\tilde{x}}^{\mu}) \\ \nonumber &=& {\tilde{\bar{\rho}}}(x^{\mu} + \delta x^{\mu}) + \delta \tilde{\rho}(x^{\mu} + \delta x^{\mu}) .
\end{eqnarray}
Taylor expanded and linearised gives us the perturbation in the perturbed manifold's relation to that in the background manifold,
\begin{equation}
\label{densitypert3}
\delta \tilde{\rho}({\tilde{x}}^{\mu}) = \delta \rho(x^{\mu}) -  {\dot{\bar{\rho}}} \delta t .
\end{equation}\\
We now use the active approach to examine the transformation behaviour of vector or tensor quantities, using the Lie derivative. For this we take the perturbation in the coordinates as the vector through which we project our vector or tensor quantity of interest. The Lie derivative acting on a tensor is defined,
\be
\label{Lie1}
{\pounds}_{\delta x^{\gamma}} g^{\mu \nu} = \delta x^{\gamma} \partial_\gamma g^{\mu \nu} - g^{\mu \gamma} \partial_\gamma  \delta x^{\nu} - g^{\gamma \nu} \partial_\gamma  \delta x^{\mu}  ,
\ee 
where, in this context, $\delta x^{\gamma}$ is the projection vector acting upon the tensor, $g^{\mu \nu}$. The gauge transformation for a tensor to linear order is,
\be
\label{gauge_gen}
\wt{\delta {\mathbf{T}}} = \delta \mathbf{T} 
+ {\pounds}_{\delta x^{\mu}} \bar{\mathbf{T}} ,
\ee
where $\mathbf{T}$ is generalised tensor.\\ 
Below we apply the Lie derivative to the perturbed contravariant 4-velocities~\cite{din},
\begin{eqnarray}
\label{Liederiveuu1}
{\tilde{u}}^{\mu} &=& \exp \left[ {\pounds}_{\delta x^{\mu}} u^{\mu} \right]    \\ \nonumber &=& \left[ 1 + {\pounds}_{\delta x^{\mu}} + \mathcal{O}({\delta}^2) \right] u^{\mu} \\ \nonumber &=& u^{\mu} + {\pounds}_{\delta x^{\mu}} u^{\mu} + \mathcal{O}({\delta}^2) .
\end{eqnarray}
To linear order this becomes,
\begin{equation}
\label{Liederiveuu2}
{\tilde{u}}^{\mu} = u^{\mu} + \delta x^{\nu} {\partial}_{\nu} u^{\mu} - u^{\nu} {\partial}_{\nu} \delta x^{\mu} .
\end{equation}
The $\mu=0$ equation is as for the lapse function i.e.
\be
\label{transu0}
{\tilde{u}}^{0} = {{u}}^{0} - \dot{\delta t} .
\ee
The $\mu=i$ equation is,
\begin{eqnarray}
\label{Liederiveuu3i}
{\tilde{u}}^{i} &=& u^i  + \delta x^{\nu} \partial_{\nu} u^i  - u^{\nu}  \partial_{\nu} \delta x^i  \\ \nonumber &=& u^i - \dot{\delta x^i} + \mathcal{O}({\delta}^2) .
\end{eqnarray}
to linear order. Since $u^i \equiv \frac{v^i}{a}$ this gives,
\begin{equation}
\label{3velFRW}
\frac{{\tilde{v}}^{i}}{a} = \frac{v^i}{a}   - \dot{\delta x^i} + \mathcal{O}({\delta}^2)
\end{equation}
This same approach may be applied to the perturbed metric tensor in which case the Lie derivative is,
\begin{eqnarray}
\label{Liederiveguu1}
{\wt{\delta g}}^{\mu \nu} &=& \delta g^{\mu \nu} + {\pounds}_{\delta x^{\gamma}} {\bar{g}}^{\mu \nu} + \mathcal{O}({\delta}^2)  \\ \nonumber &=& \delta g^{\mu \nu} + \delta x^{\gamma} \partial_{\gamma} {\bar{g}}^{\mu \nu} - {\bar{g}}^{\gamma \nu} \partial_{\gamma} \delta x^{\mu} - {\bar{g}}^{\mu \gamma} \partial_{\gamma} \delta x^{\nu} + \mathcal{O}({\delta}^2).
\end{eqnarray}
The components of the metric in the perturbed manifold are therefore for the $00$ component,
\begin{equation}
\label{Liederiveguu00}
{\wt{\delta g}}^{0 0} = \delta g^{0 0} + 2 \dot{\delta x}^{0} + \mathcal{O}({\delta}^2) ,
\end{equation}
for the $i0$ component (and by symmetry the $0j$ component),
\begin{equation}
\label{Liederiveguui0}
{\wt{\delta g}}^{i 0} = \delta g^{i 0} + \dot{\delta x}^{i} - a^{-2} \partial^{i} \delta x^{0} + \mathcal{O}({\delta}^2) ,
\end{equation}
and for the $ij$ component,
\begin{equation}
\label{Liederiveguuij}
{\wt{\delta g}}^{i j} = \delta g^{i j} - 2 H a^{-2} \delta^{i j} \delta x^{0} - a^{-2} \left( \partial^{j} \delta x^{i} + \partial^{i} \delta x^{j} \right) + \mathcal{O}({\delta}^2) .
\end{equation}
From these we obtain the transformation behaviour of the scalar metric perturbations as
\bea
\label{GGTphi}
\tilde{\phi} &=& \phi - \dot{\delta t} \,,\\
\label{GGTpsi}
\tilde{\psi} &=& \psi + H \delta t \,,\\
\tilde{B} &=& B - a\dot{\delta x}+ \delta t \,,\\
\tilde{E} &=& E - \delta x \,.
\eea
The active approach may also be applied to the density perturbations to give,
\be
\label{actdenspert}
\delta \tilde{\rho}({\tilde{x}}^{\mu}) = \delta \rho(x^{\mu}) +  {\dot{\bar{\rho}}} \delta t .
\end{equation}\\
Note the sign change between the passive and active approaches.

\subsection{Selecting and Testing Gauge Invariant Quantities}
\label{GaugeInvQuant}
We construct some useful gauge invariant quantities typically found in the literature in the field of CPT (see e.g.~Refs.~\cite{MW2008,kmdrmgi}).\\
We use the perturbed metric~\cite{kmdrmgi} in which the perturbed spatial metric component $C_{i j}$ is decomposed as in~\eq{Cdecomp} but only the scalar perturbations are retained, i.e. $C_{i j} = E,_{i j} - \psi \delta_{i j}$ where the scalar $\psi$ is the curvature perturbation. This is related to the perturbed intrinsic curvature of spatial 3-hypersurfaces through $R=4\nabla^2\left(\frac{\psi}{a^2}\right)$ where $R$ is the Ricci 3-scalar. From \eq{Liederiveguuij} we have already shown the transformation behaviour of $\psi$ is as in \eq{GGTpsi}.
If we take \eq{densitypert3} and rewrite for uniform density hypersurfaces i.e. $\delta \tilde{\rho} = 0$, we obtain,
\begin{equation}
\label{unidenshyp1}
\delta t \bigg|_{\delta \tilde{\rho}=0} = \frac{\delta \rho }{\dot{\bar{\rho}}} .
\end{equation}
By substituting \eq{unidenshyp1} into \eq{GGTpsi} we find,
\begin{equation}
\label{unidenshyp2}
\tilde{\psi} \bigg|_{\delta \tilde{\rho}=0} = \psi \bigg|_{\delta \tilde{\rho}=0} + H \frac{\delta \rho }{\dot{\bar{\rho}}} \bigg|_{\delta \tilde{\rho}=0} .
\end{equation}
This curvature perturbation~\cite{kmdrmgi,MW2008} is conserved on very large scales, in adiabatic systems, of a fluid with a barotropic equation of state. The gauge-invariant curvature perturbation is denoted by $\zeta$ where $\zeta = - \tilde{\psi} \bigg|_{\delta \tilde{\rho}=0}$. Therefore \eq{unidenshyp2} becomes,
\begin{equation}
\label{unidenshyp3}
-\zeta = \psi \bigg|_{\delta \tilde{\rho}=0} + H \frac{\delta \rho }{\dot{\rho_0}} \bigg|_{\delta \tilde{\rho}=0} .
\end{equation}
By performing the gauge transformation upon the RHS of \eq{unidenshyp3} expressed in the perturbed manifold we can show that the curvature perturbation is gauge invariant, or in other words $\zeta$ is equal to the RHS expression both in the perturbed and unperturbed manifolds and therefore is gauge invariant.\\
We may also construct density perturbations on flat hypersurfaces i.e. $\tilde{\psi} = 0$
\eq{GGTpsi} expressed in terms flat hypersurfaces is,
\begin{equation}
\label{ConsCurv1}
\delta \tilde{t} \Big|_{\psi=0}  = -\frac{\psi}{H} ,
\end{equation}
which when combined with \eq{densitypert3} leads to,
\begin{equation}
\label{denspertflat}
\delta \tilde{\rho} \Big|_{\psi=0}  = \delta \rho + \frac{\dot{\bar{\rho}}\psi}{H} ,
\end{equation}
the expression for a gauge invariant density perturbation on flat hypersurfaces. \\
Next we can construct the conservation equation for the curvature perturbation by starting with the perturbed conservation equation, \eq{PertEconspert} and evaluating for constant density hypersurfaces,
\begin{equation}
\label{PertEconsdens1}
\left[{\dot{C}}^{i}_{i} + \nabla^2 v a^{-1} \right] \left(\bar{\rho} + \bar{P} \right) \bigg|_{\delta \tilde{\rho}=0} + 3H \delta P \bigg|_{\delta \tilde{\rho}=0} = 0 .
\end{equation}
From the definition of $C_{i j}$ we find,
\begin{equation}
\label{CtoEandpsi}
\dot{C}^i_i \bigg|_{\delta \tilde{\rho}=0} = \dot{E},^i_i \bigg|_{\delta \tilde{\rho}=0} - 3 \dot{\psi} \bigg|_{\delta \tilde{\rho}=0} ,
\end{equation}
such that, coupled with the definition of $\zeta$, \eq{PertEconsdens1} when rearranged gives us the form of the evolution equation for the curvature perturbation, in the uniform density gauge,
\begin{equation}
\label{CurvPertEvolUdens1}
\dot{\zeta} = -\frac{H \delta P}{\left(\bar{\rho} + \bar{P} \right)} \bigg|_{\delta \tilde{\rho}=0}  -\frac{1}{3a} \nabla^2 \left( v + a \dot{E} \right)  \bigg|_{\delta \tilde{\rho}=0} ,
\end{equation}
which, if we take the large scale limit where the spatial gradient terms become negligible we find $\zeta$ is conserved for adiabatic fluids.\\
If we return to the perturbed conservation equation with the gauge unspecified, and separate the gradient and non-gradient terms we obtain,
\begin{equation}
\label{PertEconspertCurv1}
\left( \nabla^2 \left[\dot{E} + \frac{v}{a} \right] \right) \left(\bar{\rho} + \bar{P} \right) - 3 \dot{\psi} \left(\bar{\rho} + \bar{P} \right) + \delta \dot{\rho} + 3 H \left(\delta \rho + \delta P \right) = 0 .
\end{equation}
Again taking the large scale limit where the spatial gradients vanish, for simplicity and clarity in the derivations, we obtain,
\begin{equation}
\label{PertEconspertCurvLSL1}
- 3 \dot{\psi} \left(\bar{\rho} + \bar{P} \right) + \delta \dot{\rho} + 3 H \left(\delta \rho + \delta P \right) = 0 .
\end{equation}
Finally we show the invariance of this equation. \eq{PertEconspertCurvLSL1} in the uniform density gauge, expressed in terms of quantities in the perturbed manifold gives,
\begin{equation}
\label{PertEconspertCurvLSL2}
- 3 \dot{\tilde{\psi}} \left(\bar{\rho} + \bar{P} \right) \bigg|_{\delta \tilde{\rho}=0} + 3 H \delta \tilde{P} \bigg|_{\delta \tilde{\rho}=0} = 0 .
\end{equation}
If we substitute for the variables expressed in terms of the unperturbed manifold we recover the original gauge unspecified form of the perturbed conservation equation in the large scale limit; \eq{PertEconspertCurvLSL1}.
In the above work we set degrees of freedom, such as the density perturbation, to zero to define a hypersurface. This is called making a gauge selection. One or more degrees of freedom may be fixed in this way leading to a wide variety of gauges. Some common gauges are listed in Figure~\ref{gaugefig}. Note: Synchronous, Co-moving and Uniform Density are incomplete gauges and require additional gauge fixing conditions in order to remove all gauge artefacts e.g. Synchronous - and - comoving completely fixes the gauge.

\begin{table}
\centering
\begin{tabular}{|l|r|}
\hline 
\multicolumn{2}{|c|}{Common Gauges} \\ 
\hline {\bf Gauge Name} & {\bf Gauge Conditions}  \\
\hline 
Flat & $\psi = E = 0$ \\
\hline 
Longitudinal (Newtonian) & $ B = E = 0$ \\
\hline 
Synchronous & $\Phi = B = 0$ \\
\hline 
Co-moving & $v_i=0$ \\
\hline
Uniform Density& $ \delta \rho = 0 $ \\
\hline
\end{tabular}
\caption{Selected examples of commonly used gauges (see e.g.~Ref.\cite{MW2008}).}
\label{gaugefig}
\vspace*{20pt}
\end{table}

\chapter{Conserved Quantities in Lema{\^\i}tre-Tolman-Bondi Cosmology}
\label{ch:3}

In this chapter we study linear perturbations to a \emph{Lema{\^\i}tre-Tolman-Bondi} (LTB)
background spacetime following similar procedures as in Chapter 2 i.e. we study the transformation behaviour of the
perturbations under gauge transformations and construct gauge
invariant quantities. We show, using the perturbed energy conservation
equation, that there are conserved quantities in LTB, in particular a spatial metric trace perturbation, $\SMTP$, which is conserved on all scales. We then briefly
extend our discussion to the Lema{\^\i}tre spacetime, and construct
gauge-invariant perturbations in this extension of LTB spacetime, which unlike LTB allows for a background pressure.

\section{ Lema{\^\i}tre-Tolman-Bondi spacetime}
\label{ltb_sect}

In this section we first briefly review standard LTB cosmology at the background level. We
then extend the standard results by adding perturbations to the LTB
background. In order to remove any unwanted gauge modes, we study the
transformation behaviour of the perturbations, which then allows us to
construct gauge-invariant quantities, in particular the equivalent to
the curvature perturbation. We show under which conditions this
curvature perturbation is conserved.

Throughout this section we assume zero pressure in the background (see Section \ref{lemaitre} for the addition of non-zero background pressure) i.e. the matter content is pressureless dust. We do this since LTB gives an exact solution to the Einstein field equations in the absence of background pressure. We do however allow for a
pressure perturbation in the later subsections.

\subsection{Background}
\label{ltb_back}

The LTB metric can be written in various forms \cite{Bondi,Bonnor,Tim1}. Here we shall
use the following form of the metric \cite{Bondi,Ellis},
\begin{equation}
\label{LTBInterval}
ds^2 = -dt^2 + \sfx^2 (r,t) dr^2 
+ \sfy^2 (r,t) \left( d \theta^2 + \sin^2 \theta d \phi^2 \right) ,
\end{equation}
where $\sfx$ and $\sfy$ are scale factors dependent upon both the
radial spatial and time co-ordinates. The scale
factors are not independent and are related by,
\begin{equation}
\label{ScaleFactorsRelation}
\sfx = \frac{1}{W(r)} \frac{\partial \sfy}{\partial r} ,
\end{equation}
where $W(r)$ is an arbitrary function of $r$, following Bondi
\cite{Bondi}, arising from the Einstein field equations.\\

The 4-velocity in the background is given from its definition,
\eq{4vel}, as
\begin{equation}
\label{4velunpertLTB}
u^{\mu} = [1,0,0,0] \,,
\end{equation}
where the indices $0, 1, 2, 3$ are $t, r, \theta, \phi$ respectively, and since we assume we are comoving with respect to the background
coordinates $dr=d\theta=d\phi=0$, and therefore $d{\tau}^2
= dt^2$ (that is in the local rest frame).

From the definition of the energy-momentum tensor, \eq{SET}, we
immediately find that in the absence of pressure the only non-zero
component is, $T^{0 0} = \rho$. For later convenience we define Hubble
parameter equivalents for the two scale factors such that,
\begin{equation}
H_{\sfx} = \frac{\dot{\sfx}}{\sfx} \,, \qquad
H_{\sfy} = \frac{\dot{\sfy}}{\sfy} \,.
\end{equation}
where the ``dot'' denotes the derivative with respect to coordinate time $t$.

The Einstein equations are, from \eq{Einstein2}, for the $0-0$
component,
\begin{equation}
\label{HEinstein00}
\frac{1}{\sfy^2} + {H_{\sfy}}^2 + 2\frac{\sfx'\sfy'}{\sfx^3\sfy} + 2H_{\sfx} H_{\sfy} - \left(\frac{\sfy'}{\sfx\sfy}\right)^2 - 2\frac{\sfy''}{\sfx^2\sfy} = 8 \pi G \rho  \,,
\end{equation}
where a prime denotes a derivative with respect to the radial coordinate $r$. 
For the $0-r$ component we find,
\begin{equation}
\label{HEinstein01}
\frac{2}{\sfy} \left(\sfy'H_{\sfx} - \dot{\sfy}' \right) = 0  \,,
\end{equation}
for the $r-r$ component,
\begin{equation}
\label{HEinstein11}
\left(\frac{\sfy'}{\sfx\sfy}\right)^2 - \frac{1}{\sfy^2} - H_{\sfy}^2 - 2\frac{\ddot{\sfy}}{\sfy}  = 0 \,,
\end{equation}
and for ${\theta-\theta}$ and ${\phi-\phi}$ components we get,
\begin{equation}
\label{HEinstein22}
\frac{\sfy''}{\sfx^2\sfy} - \frac{\sfx'\sfy'}{\sfy\sfx^2} 
- \frac{\ddot{\sfy}}{\sfy} - H_{\sfx}H_{\sfy} 
- \frac{\ddot{\sfx}}{\sfx} = 0  \,.
\end{equation}
The other components are identically zero. The energy conservation
equation, obtained from \eq{Bianchi}, is
\begin{equation}
\label{HLTBEngCons}
\dot{\rho} + \rho(H_{\sfx} + 2 H_{\sfy}) = 0\,.
\end{equation}

\subsection{Perturbations}
\label{pert_ltb}

In this section we add perturbations to the LTB background. Unlike
recent works studying perturbed LTB models, e.g.~Refs.~\cite{Tim1}, we
do not decompose the perturbations into polar and axial scalars and vectors, and
multi-poles, which considerably simplifies our governing
equations.

We split quantities into a $t$ and $r$ dependent background part, and
a perturbation depending on all four coordinates. Compare this with FRW, as in Chapter 2, (see e.g.~\eq{decomp1}), where due to the Cosmological Principle, the background is only time dependent, while the perturbation depends upon all coordinates. For example, in LTB we
decompose the energy density $\rho$ as follows,
\be
\label{split_rho}
\rho=\bar\rho(t,r)+\delta\rho(x^\mu)\,,
\ee
where here and in the following a ``bar'' denotes a background
quantity, if there is a possibility for confusion.

We perturb the metric in a similar way as in the flat FRW case, the LTB
metric being very similar to flat FRW in spherical polar
coordinates, save for the two scale factors and the factor of $r$
being absorbed into $\sfy$. 

Hence we split the metric tensor as 
\begin{equation}
\label{metricsplit2}
g_{\mu \nu} = {\bar{g}}_{\mu \nu} + \delta g_{\mu \nu} , 
\end{equation}
where ${\bar{g}}_{\mu \nu}$ is given by \eq{LTBInterval}. For the
perturbed part of the metric, $\delta g_{\mu \nu}$, we make the ansatz,
\begin{equation}
\label{LTBMetricperturbations}
\delta g_{\mu \nu}=\begin{pmatrix}
  -2\Phi & \sfx B_r & \sfy B_\theta & \sfy \sin \theta B_\phi \\
  \sfx B_r & 2\sfx^2 C_{rr} & \sfx \sfy C_{r\theta} & \sfx \sfy \sin \theta C_{r\phi} \\
  \sfy B_\theta & \sfx \sfy C_{r\theta} & 2\sfy^2 C_{\theta\theta} & \sfy^2 \sin \theta C_{\theta\phi} \\
  \sfy \sin \theta B_\phi & \sfx \sfy \sin \theta C_{r\phi} & \sfy^2 \sin \theta C_{\theta\phi} & 2\sfy^2 \sin^2 \theta C_{\phi\phi}
\end{pmatrix} \,.
\end{equation}
Here $\Phi$ is the lapse function, and $B_n$, where $n=r,\theta,\phi$,
are the shift functions for each spatial coordinate. Similarly,
$C_{nm}$, where $n,m=r,\theta,\phi$, are the spatial metric
perturbations. Compare this with the perturbed metric in FRW, \eq{PertonlygFRWdd}, which is much more concise. As already pointed out, we do not decompose $B_n$ and
$C_{nm}$ further into scalar and vector perturbations (see however
Ref.~\cite{Tim1}).\\

Using the perturbed metric we can construct the perturbed
4-velocities using the definition, \eq{4vel}.
Proper time is to linear order in the perturbations given by,
\begin{equation}
\label{PropCoordPert}
d{\tau} = (1 +  \Phi) dt \,,
\end{equation}
and defining the 3-velocity as,
\be
\label{3veldef}
v^i = \frac{d x^i}{d t} , 
\ee
from \eq{4vel} we get the contravariant 4-velocity vector,
\begin{equation}
\label{4velpertLTB}
u^{\mu} = [(1 - \Phi),v^r,v^\theta,v^\phi] .
\end{equation}
By lowering the index using the perturbed metric we obtain the
covariant form,
\begin{equation}
\label{4velpertLTBDown}
u_{\mu} 
= [-(1 + \Phi),
\sfx\left(B_r + {\sfx} v^r\right),\,
\sfy\left( B_\theta + {\sfy} v^\theta \right),\,
\sfy\sin(\theta) \left(B_\phi + {\sfy}\sin(\theta) v^\phi\right)] \,.\\
\end{equation}
Conservation of the energy-momentum tensor, \eq{Bianchi}, allows us
together with its definition, \eq{SET}, to derive the perturbed
energy conservation equation, 

\bea
\label{PertEconsSphP}
&\delta \dot{\rho}& + \left(\delta \rho + \delta P\right)
\left(H_X+ 2 H_Y \right) + {\bar{\rho}} ' v^r + {\bar{\rho}}\Bigg(\dot{C}_{rr} + \dot{C}_{\theta\theta} + \dot{C}_{\phi\phi} \\ \nonumber &+& {v^r} ' + \partial_\theta v^\theta + \partial_\phi v^\phi + \left[\frac{\sfx '}{\sfx} + 2 \frac{\sfy '}{\sfy} \right] v^r + \cot \theta v^\theta\Bigg) =0 ,
\eea
where we used \eq{split_rho}, and the LTB background requires $\bar P=0$.
The perturbed momentum conservation equations are 
\bea
\label{PertMom1consSphP}
&\dot{\bar{\rho}} v^r&  + \bar{\rho} ({\dot{v}}^r  + \frac{{\dot{B}}_r}{\sfx} + \frac{B_r}{\sfx} H_{\sfx} + (3 H_{\sfx} + 2 H_{\sfy}) v^r) + \frac{1}{\sfx^2} \delta P ' = 0\,,\\
\label{PertMom2consSphP}
&\dot{\bar{\rho}} v^{\theta}& + \bar{\rho} ({\dot{v}}^{\theta}  + \frac{{\dot{B}}_{\theta}}{\sfy} + \frac{B_{\theta}}{\sfy} H_\sfy + (H_{\sfx} + 4 H_{\sfy}) v^\theta) + \frac{1}{\sfy^2} \partial_{\theta} \delta P = 0 \,,\\
\label{PertMom3consSphP}
&\dot{\bar{\rho}} v^{\phi}& + \bar{\rho} \left({\dot{v}}^{\phi}   + \frac{{\dot{B}}_{\phi}}{\sfy \sin \theta} + \frac{{B}_{\phi} H_\sfy}{\sfy \sin \theta} + (H_\sfx + 4 H_\sfy)v^{\phi} \right) \\ \nonumber  &+& \frac{1}{\sfy^2 \sin^2 \theta} \partial_{\phi} \delta P = 0 \,,
\eea
which we do not use in this work.

\subsection{Gauge Transformation}
\label{GaugeTransBeh}

In order to construct gauge-invariant perturbations, we have to study
the transformation behaviour of our matter and metric variables, as we saw in Chapter 2, Subsection~\ref{GaugeTrans}. Using the active point of view, linear order perturbations of a tensorial quantity $\mathbf{T}$ transform as \eq{gauge_gen}, in Chapter 2 Section~\ref{GaugeTrans} using the Lie derivative. The old and the new coordinate systems are related by \eq{gaugetrans1} where $\delta x^{\mu}=[\delta t, \delta x^i]$ is the gauge generator. The Lie derivative is denoted by ${\pounds}_{\delta x^{\mu}}$, defined in terms of the metric as in \eq{Lie1}.

\subsubsection{Metric and Matter Quantities}
\label{MetandMat}

From \eq{gauge_gen} and \eq{split_rho} we find that the density
perturbation transforms simply as,
\begin{equation}
\label{denspertgtransLTB}
\delta \tilde{\rho} = \delta \rho +  \dot{{\bar{\rho}}} \delta t
+  {\bar{\rho}}'  \delta r\,,
\end{equation}
since the background energy density depends on $t$ and $r$. c.f. \eq{actdenspert} for FRW which does not contain the ${\bar{\rho}}'  \delta r$ term. 
The perturbed spatial part of the 4-velocities, defined in \eq{4velpertLTB}
transform as, 
%
\begin{equation}
\label{vuiTrans}
{\tilde{v}}^i = v^i - \dot{\delta x^i} \,,
\end{equation}
where $i = r, \theta, \phi$. c.f. \eq{3velFRW} for FRW, which is similar but for factors of $\frac{1}{a}$ arising from the slightly different definition of the 4-velocity we use in LTB in \eq{4velpertLTB}.\\
The perturbed metric transforms, using \eq{gauge_gen}, as
\begin{equation}
\label{metricgtransgen2}
{\delta \tilde{g}}_{\mu \nu} = \delta g_{\mu \nu} + \delta x^{\gamma} \partial_{\gamma}  {\bar{g}}_{\mu \nu} +  {\bar{g}}_{\gamma \nu} \partial_{\mu} \delta x^{\gamma} +  {\bar{g}}_{\mu \gamma} \partial_{\nu} \delta x^{\gamma} .
\end{equation}
From the $0-0$ component of \eq{metricgtransgen2} we find that the
lapse function transforms as
\begin{equation}
\label{Phitrans}
\tilde{\Phi} = \Phi - \delta \dot{t} \,.
\end{equation}
For the perturbations on the spatial trace part of the metric we find 
for the $r$ coordinate from \eq{metricgtransgen2},
\begin{equation}
\label{CrrTrans}
{\tilde{C}}_{r r} = C_{r r} + \delta t \frac{\dot{\sfx}}{\sfx} + \delta r \frac{\sfx '}{\sfx} +  \delta r ' \,,
\end{equation}
for the $\theta$ coordinate,
\begin{equation}
\label{CthetheTrans}
{\tilde{C}}_{\theta \theta} = C_{\theta \theta} + \delta t \frac{\dot{\sfy}}{\sfy} + \delta r \frac{\sfy '}{\sfy} + \partial_{\theta} \delta \theta \,,
\end{equation}
and for the $\phi$ coordinate,
\begin{equation}
\label{CphiphiTrans}
{\tilde{C}}_{\phi \phi} = C_{\phi \phi} + \delta t \frac{\dot{\sfy}}{\sfy} + \delta r \frac{\sfy '}{\sfy} + \delta \theta \cot \theta + \partial_{\phi} \delta \phi \,.
\end{equation}
For later convenience
we define a spatial metric perturbation, $\psi_{\LT}$, as,
\begin{equation}
\label{CurvatureLTB1}
3 \psi_{\LT} = \delta g^k_{k} = C_{r r} + C_{\theta \theta} + C_{\phi \phi}\,,
\end{equation}
that is the trace of the spatial metric, in analogy with the curvature
perturbation $\psi_\frw$ in perturbed FRW spacetimes (see Section
\ref{FRWST} below). The relation between $\psi_{\LT}$ here and the curvature
perturbation in perturbed FRW can be most easily seen from the
perturbed expansion scalar, given in \eq{ExpFacSph2} below, which is
very similar to its FRW counterpart (see e.g.~Ref.~\cite{MW2008},
Eq.~(3.19)). The relation is not obvious from calculating the spatial
Ricci scalar for the perturbed LTB spacetime, as can be seen from
\eq{LTBR3EqPert}, given in the appendix.
From the above  $\psi_{\LT}$ transforms as 
\begin{equation}
\label{psiTrans2}
3\tilde{\psi}_{\LT} = 3\psi_{\LT} + \left[\frac{\dot{\sfx}}{\sfx} + 2 \frac{\dot{\sfy}}{\sfy} \right] \delta t + \left[\frac{\sfx '}{\sfx} + 2 \frac{\sfy '}{\sfy} \right] \delta r + \partial_i \delta x^i + \delta \theta \cot \theta \,,
\end{equation}
where $i = r, \theta, \phi$. c.f. \eq{GGTpsi} in FRW which is much simpler with only time derivatives and time coordinate artefacts.
In addition, from \eq{metricgtransgen2} the off diagonal spatial metric perturbations transform
as,
\bea
\label{CrtheTrans}
{\tilde{C}}_{r \theta} &=& C_{r \theta} + \frac{\sfy}{\sfx} \delta \theta ' + \frac{\sfx}{\sfy} \partial_{\theta} \delta r \,,\\
\label{CrphiTrans}
{\tilde{C}}_{r \phi} &=& C_{r \phi} + \frac{\sfy \sin \theta}{\sfx} \delta \phi ' + \frac{\sfx}{\sfy \sin \theta} \partial_{\phi} \delta r \,,\\
\label{CthephiTrans}
{\tilde{C}}_{\theta \phi} &=& C_{\theta \phi} + \frac{\sin \theta}{\sfx} \partial_{\theta} \delta \phi + \frac{1}{\sin \theta} \partial_{\phi} \delta \theta \,.
\eea
The mixed temporal-spatial perturbations of the metric, that is the
shift vector, from \eq{metricgtransgen2} transform as
\bea
\label{BrTrans}
{\tilde{B}}_r &=& B_r + \sfx \dot{\delta r} - \frac{\delta t '}{\sfx}\,,\\
\label{BTheTrans}
{\tilde{B}}_{\theta} &=& B_{\theta} + \sfy \dot{\delta \theta} - \frac{\partial_{\theta} \delta t}{\sfy}\,,\\
\label{BPhiTrans}
{\tilde{B}}_{\phi} &=& B_{\phi} + \sfy (\sin \theta) \dot{\delta \phi} - \frac{\partial_{\phi} \delta t}{\sfy (\sin \theta)}\,.
\eea

\subsubsection{Geometric Quantities}
\label{Geom}

The expansion scalar, as defined in \eq{ExpFacugen} with $n^\mu$ in place of $u^\mu$, calculated using the
4-velocity, given in \eq{4velpertLTB}, is,
\begin{equation}
\label{ExpFacSph2}
\Theta = \left( H_\sfx + 2 H_\sfy \right) + 3\dot{\psi}_{\LT} + \partial_i v^i  - \left( H_\sfx + 2 H_\sfy \right)\Phi + \left(\frac{\sfx '}{\sfx} + 2 \frac{\sfy '}{\sfy} \right) v^r + \left(\cot \theta \right) v^{\theta} \,,
\end{equation}
where $i = r, \theta, \phi$. Alternatively, the expansion factor
defined with respect to the unit normal vector field defined in
\eq{ExpFacugen}, is given by,
\begin{equation}
\label{ExpFacSphNorm}
\Theta_n = \left( H_\sfx + 2 H_\sfy \right) + 3\dot{\psi}_{\LT} - \left( H_\sfx + 2 H_\sfy \right)\Phi - \frac{B_r '}{\sfx} - \frac{\partial_\theta B_\theta}{\sfy} - \frac{\partial_\phi B_\phi}{\sfy \sin \theta} - \frac{2 B_r \sfy '}{\sfx \sfy} - \frac{B_\theta \cot \theta}{\sfy} \,.
\end{equation}
This is more complicated than the equivalent in FRW, \eq{expfFRW}, due to the additional scale factors and their additional radial spatial coordinate dependence.
In order to have the possibility to define later hypersurfaces of
uniform expansion, on which the perturbed expansion is zero, we have
to find the transformation behaviour of the expansion scalar. We find,
that e.g.~$\Theta_n$ transforms as,
\begin{eqnarray}
\label{ExpFacSphNormTrans2}
\nonumber {\tilde{\Theta}}_n &=& \Theta_n + \left[ {\dot{H}}_\sfx + 2 {\dot{H}}_\sfy \right] \delta t +  \left[ H_\sfx + 2 H_\sfy \right] \dot{\delta t} + \left(\frac{\dot{\sfx} '}{\sfx} - \frac{\dot{\sfx} \sfx '}{{\sfx}^2} + 2 \frac{\dot{\sfy} '}{\sfy} - 2 \frac{\dot{\sfy} \sfy '}{{\sfy}^2} \right) \delta r \\ &+& \left[ \frac{1}{\sfx^2} \partial_{r r} + \frac{1}{\sfy^2} \partial_{\theta \theta} + \frac{1}{\sfy^2 \sin^2 \theta} \partial_{\phi \phi} \right] \delta t + \frac{2 \sfy '}{\sfy \sfx^2} \delta t ' + \frac{\cot \theta}{\sfy^2} \partial_\theta \delta t \,.
\end{eqnarray}
We immediately see that the transformation behaviour of $\Theta_n$ is
rather complicated, and we therefore do not use it to specify a gauge.

\subsection{Gauge invariant quantities}
\label{TheCurvePert}

We can now use the results from the previous section, to construct
gauge-invariant quantities. Luckily, we can use the results derived
for the FRW background spacetime, as above and in Chapter 2, as guidance to get the evolution equations. We showed that the evolution equation for the curvature perturbation on uniform density hypersurfaces, $\zeta$, as seen in \eq{CurvPertEvolUdens1}, can be derived solely from the energy conservation equations (on large scales).\\

\subsubsection{FRW spacetime}
\label{FRWST}

We will first consider the construction of gauge-invariant quantities in perturbed FRW spacetime, which is the homogeneous limit of LTB.
As per Chapter 2, the perturbed FRW metric is,
\begin{equation*} 
\label{ds2}
ds^2=-(1+2\phi)dt^2+2aB_{,i}dt dx^i
+a^2\left[(1-2\psi_\frw)\delta_{ij}+2E_{,ij}\right]dx^idx^j \,, 
\end{equation*}
where we have performed a scalar-vector-tensor decomposition, and kept only the scalar part.\ 
\eq{gauge_gen} and \eq{gaugetrans1} then give \cite{MW2008}
\bea
\label{psi_frw}
\wt{\psi_\frw}&=&{\psi_\frw}+\frac{\dot a}{a}\delta t\,, \\
\wt{\delta\rho_\frw}&=&\delta\rho_\frw+ {\dot {\bar \rho}} \delta t\,, \\
\tilde{E} &=& E + \delta x \,.
\eea
where as before $a=a(t)$ is the scale factor (as compared with, $\sfx (r,t)$ and $\sfy (r,t)$, the two time and radial spatial coordinate scale factors in LTB) and $\bar\rho=\bar\rho(t)$ is the
background energy density.
We can now choose a gauge condition, to get rid of the gauge
artefacts, here $\delta t$. To this end, the uniform density gauge can then be specified by the choice $\wt{\delta\rho_\frw}\equiv 0$, which implies
\be
\label{delta_t_frw}
\delta t=-\frac{\delta\rho_\frw}{\dot{\bar{\rho}}}\,.
\ee
Combining \eq{psi_frw} and \eq{delta_t_frw}, we are then led to define
\be
-\zeta \equiv \psi_\frw+\frac{\dot a/a}{\dot{\bar{\rho}}}\delta\rho_\frw \,,
\ee
as before in \eq{unidenshyp3}, which is gauge-invariant under \eq{gauge_gen}, as can be seen by
direct calculation.

\subsubsection{LTB spacetime}
\label{subsec:LTBst}

We can now proceed to construct gauge-invariant quantities in the
perturbed LTB model, taking the FRW case as guidance.  From the
transformation equation of the perturbed spatial metric trace, $\psi_{\LT}$, \eq{psiTrans2}, we see
that here we have to substitute for $\delta t$ and $\delta x^i$, that
is we have to choose temporal \emph{and} spatial hypersurfaces.

From the density perturbation transformation, \eq{denspertgtransLTB}, choosing
uniform density hypersurfaces, $\delta \tilde{\rho}=0$, to
fix the temporal gauge, we get
\be
\label{temporal_ltb}
\delta t\Big|_{\delta \tilde{\rho}=0}= -\frac{1}{\dot{{\bar{\rho}}}}\left[\delta \rho
+  {\bar{\rho}}'  \delta r\right]\,.
\ee
Substituting this into \eq{psiTrans2}, the transformation of the
metric trace, we get
\begin{equation}
\label{zetafull1}
- \SMTP = \psi_{\LT} 
- \frac{1}{3}\left[\frac{\dot{\sfx}}{\sfx} + 2 \frac{\dot{\sfy}}{\sfy} \right] \left(\frac{\delta \rho + \bar{\rho} ' \delta r}{\dot{\bar{\rho}}} \right) 
+ \frac{1}{3}\left\{\left[\frac{\sfx '}{\sfx} + 2 \frac{\sfy '}{\sfy} \right] \delta r + \partial_i \delta x^i + \delta \theta \cot \theta\right\} \,,
\end{equation}
where $\SMTP$ is the Spatial Metric Trace Perturbation and we chose the sign convention and notation to coincide with the
FRW case.
We can now choose comoving hypersurfaces to fix the remaining spatial
gauge freedom. This gives for the spatial gauge generators from the
transformation of the 3-velocity perturbation, \eq{vuiTrans},
\be
\label{spatial_ltb}
\delta x^i=\int v^i dt\,.
\ee
Substituting the above equations into \eq{zetafull1} we finally get
the gauge-invariant spatial metric trace perturbation on comoving, uniform
density hypersurfaces,
\bea
\label{zetafull4}
\nonumber - \SMTP &=& \psi_{\LT} + \frac{\delta \rho}{3\bar{\rho}} 
+ \frac{1}{3}\Bigg\{
\left(\frac{\sfx '}{\sfx} + 2 \frac{\sfy '}{\sfy}
+ \frac{\bar{\rho} '}{\bar{\rho}} \right) \int v^r dt + \partial_r \int v^r dt + \partial_{\theta} \int v^{\theta} dt \\ &+& \partial_{\phi} \int v^{\phi} dt + \cot \theta \int v^{\theta} dt\Bigg\} \,,
\eea
i.e. $\SMTP = -\frac{1}{3} \wt{\delta g^k_{k}} \big|_{\wt{\delta \rho} = 0, v = 0}$. We can check by direct calculation, i.e.~by substituting
\eq{psiTrans2}, \eq{denspertgtransLTB}, and \eq{vuiTrans} into
\eq{zetafull4}, that $\SMTP$ is gauge invariant.\\

Instead of using $\delta\rho$ to specify our temporal gauge, we can
just as easily use the spatial metric trace perturbation, that is
define hypersurfaces where $\wt\psi_{\LT}\equiv0$. This gives for $\delta t$
\be
\label{deltatfix}
\delta t= - \frac{1}{H_\sfx + 2 H_\sfy} \left[\psi_{\LT} + \left(\frac{\sfx '}{\sfx} + 2 \frac{\sfy '}{\sfy} \right) \delta r + \partial_i \delta x^i + \delta \theta \cot \theta \right].
\ee
This allows us to construct another gauge invariant quantity, the
density perturbation on uniform spatial metric trace perturbation
hypersurfaces, using \eq{denspertgtransLTB}, as
\bea
\label{GIdenspertZeta}
\nonumber \delta \tilde{\rho} \Big|_{\psi_{\LT}=0} &=& \delta \rho + \bar{\rho} \Bigg\{3 \psi_{\LT} +
\left(\frac{\sfx '}{\sfx} + 2 \frac{\sfy '}{\sfy}
+ \frac{\bar{\rho} '}{\bar{\rho}} \right) \int v^r dt + \partial_r \int v^r dt + \partial_{\theta} \int v^{\theta} dt \\ &+& \partial_{\phi} \int v^{\phi} dt + \cot \theta \int v^{\theta} dt\Bigg\} \,,
\eea
where the spatial gauge generators were eliminated by selecting the
comoving gauge \eq{spatial_ltb} again.
The density perturbation defined in \eq{GIdenspertZeta} can be
written in terms of $\SMTP$, defined in \eq{zetafull4}, simply as
\begin{equation}
\label{GIdenspertZeta2}
\delta \tilde{\rho} \Big|_{\psi_{\LT}=0} = - 3 \bar{\rho} \SMTP\, .
\end{equation}
This expression allows us to relate the density perturbation at different times to the
spatial metric trace perturbation, which, as we shall see in Section
\ref{zeta_evol}, is conserved or constant in time on all scales for barotropic fluids.\\

Alternatively, in both cases above, \eq{zetafull4} and
\eq{GIdenspertZeta}, we could have used the shift functions instead of
the 3-velocities to define the spatial gauge, in analogy with the
Newtonian or longitudinal gauge condition in perturbed FRW. In this
case the spatial gauge generators are
\bea
\label{fixedrshift}
\delta r &=& - \int dt \left[ \frac{\partial_r}{\sfx^2} \left(\frac{\delta \rho}{\dot{\bar{\rho}}} + \frac{B_r}{\sfx} \right) \right] - \int dt \left[ \frac{\partial_r}{\sfx^2} \left(\frac{\delta r \bar{\rho} '}{\dot{{\bar{\rho}}}} \right) \right] \,,\\
%
%
\label{fixedtheshift}
\delta \theta &=& - \int dt \left[ \frac{\partial_\theta}{\sfy^2} \left(\frac{\delta \rho}{\dot{\bar{\rho}}} + \frac{B_\theta}{\sfy} \right) \right] - \int dt \left[ \frac{\partial_\theta}{\sfy^2} \left(\frac{\delta r \bar{\rho} '}{\dot{{\bar{\rho}}}} \right) \right] \,,\\
\label{fixedphishift}
\delta \phi &=& - \int dt \left[ \frac{\partial_\phi}{\sfy^2 \sin^2 \theta} \left(\frac{\delta \rho}{\dot{\bar{\rho}}} + \frac{B_\phi}{\sfy \sin \theta} \right) \right] - \int dt \left[ \frac{\partial_\phi}{\sfy^2 \sin^2 \theta} \left(\frac{\delta r \bar{\rho} '}{\dot{{\bar{\rho}}}} \right) \right] \,.
\eea
Since the expressions are considerably longer than \eq{spatial_ltb} above, we
did not pursue this choice of spatial gauge any further.

Another alternative would be to choose a more geometric definition of
the longitudinal or Newtonian gauge, namely use a zero shear condition
to fix temporal and spatial gauge, again in analogy with FRW, i.e.,
\be
\wt{\delta\sigma_{ij}}=0\,.
\ee
However, again we find that this leads to much more complicated gauge
conditions (since we do not decompose into axial and polar scalar and vector parts), and we
here do not pursue this further. See however appendix \ref{Shear} for
the components of the shear tensor.

\subsection{Evolution of $\SMTP$}
\label{zeta_evol}

Before we derive the evolution equation for spatial metric trace perturbation
$\SMTP$, we briefly discuss the decomposition of the pressure
perturbation in the LTB setting. We assume that the pressure
$P=P(\rho,S)$, where $\rho$ is the density and $S$ the entropy of the
system. We can then expand the pressure as
\be
\label{deltaPlong}
\delta P= \frac{\p P}{\p\rho}\bigg|_{S={\rm{const}}} \delta\rho+\frac{\p P}{\p S}\bigg|_{\rho={\rm{const}}} \delta S\,, 
\ee 
or,
\be
\label{deltaP}
\delta P= \cs2\delta\rho+\dpn\,, 
\ee 
where 
\be
\label{dpnaddefn}
\dpn = \frac{\p P}{\p S}\bigg|_{\rho={\rm{const}}} \delta S ,
\ee
is the entropy or non-adiabatic pressure perturbation, and
the adiabatic sound speed is defined as
\be
\label{cs2def}
\cs2\equiv \left.\frac{\p P}{\p\rho}\right|_S\,,
\ee
for a pedagogical introduction to this topic see
e.g.~Ref.~\cite{Christopherson:2008ry}.
Since in LTB background quantities are $t$ and $r$ dependent, therefore 
allowing for now $P \equiv P(t,r)$, we find that
\be 
\label{cs2_tandr}
\cs2 = \frac{\dot{\bar{P}} + \bar{P}' v^r}{\dot{\bar{\rho}} +
  \bar{\rho}' v^r} \,. 
\ee

However, since in LTB $\bar P=0$, we have that on uniform density
hypersurfaces $\delta P=\dpn$.

The evolution equation for spatial metric trace perturbation on uniform
density and comoving hypersurfaces, $\SMTP$, using the time derivative
of \eq{zetafull4}, \eq{PertEconsSphP} and background conservation
equation, \eq{HLTBEngCons}, is
\begin{equation}
\label{zetaevol5}
 \DOTSMTP =  \frac{H_\sfx + 2 H_\sfy}{3\bar{\rho}} \dpn\, .
\end{equation}
This result is valid on all scales.
We see that $\SMTP$ is conserved for $\dpn=0$, e.g.~for barotropic
fluids. While this result is similar to the FRW case \cite{WMLL},
 we do not have to assume the large scale limit here, which is a striking contrast to be discussed in Section \ref{LTB Conclusion}.

\subsection{Spatial Metric Trace Perturbation in FRW}
\label{CompareFRWSMTP}
 
In this subsection we will now compare the behaviour of the $\SMTP$ variable that we defined in LTB with the spatial metric trace perturbation on comoving constant density hypersurfaces in FRW spacetime, including background pressure. From \eq{ds2}, the trace of the perturbed part of the spatial metric can be seen to be given in FRW by
\begin{equation}
\label{SMTPtoPsiFRW}
 \delta {g^k_{k}}_\frw =  - 3 \psi_\frw + \nabla^2 E \,.
\end{equation}
This quantity can be seen to transform under \eq{gauge_gen} as
\begin{equation}
\label{SMTPPsiFRWTrans}
\wt{\delta g^k_{k}} = \delta {g^k_{k}} -3 H \delta t + \nabla^2 \delta x  \,.
\end{equation}
The 3-velocity transformation has the same form as in LTB, and is given by \eq{vuiTrans}.
Additionally, the density perturbation evolves as
\be
\label{FRWPertE}
\delta \dot{\rho} + 3 H \left(\delta \rho + \delta P\right) - 3 \left(\bar{\rho} + \bar{P}\right) {\dot{\psi}}_\frw + \left(\bar{\rho} + \bar{P}\right) \frac{\nabla^2}{a^2} \left(a v + a^2 \dot{E} \right)
 =0 \,.
\ee
Taking the time derivative of \eq{SMTPtoPsiFRW} and substituting into \eq{FRWPertE} we then find that the spatial metric trace perturbation on comoving constant density hypersurfaces evolves as
\be
\label{FRWSMTP}
-\frac{1}{3} \dot{\wt{\delta g^k_{k}}} \big|_{\wt{\delta \rho} = 0, v = 0} =  \frac{H}{(\bar{\rho} + \bar{P})} \dpn\, .
\ee
This equation is again valid on all scales, and can again be seen to demonstrate that the spatial metric trace perturbation on comoving constant density hypersurfaces\footnote{$-\frac{1}{3} \dot{\wt{\delta g^k_{k}}} \big|_{\wt{\delta \rho} = 0, v = 0} \equiv \DOTSMTP$} is conserved for barotropic fluids. It should be noted that in order to relate this spatial metric trace perturbation on comoving constant density hypersurfaces in FRW to observables such as the density perturbation both the density perturbation and 3-velocity need to be specified on flat hypersurfaces. It should also be noted that this quantity is not the same as the curvature perturbation, $\zeta$, from the standard FRW literature. Both \eq{FRWSMTP} and \eq{zetaevol5} differ from the result for the Lema{\^\i}tre spacetime, as shall be seen in Section \ref{lemaitre} below.

\section{The Lema{\^\i}tre spacetime}
\label{lemaitre}

Although the main focus of this chapter is on LTB cosmology, we here
briefly also discuss perturbations around a Lema{\^\i}tre background
spacetime. The Lema{\^\i}tre spacetime is a generalisation of LTB,
allowing for non-zero pressure in the background
\cite{Hellaby}. Although no exact solutions are known in this case, we
nevertheless think it is interesting to extend the discussion of the
previous sections to this spacetime.

The Lema{\^\i}tre background metric is given by
\begin{equation}
\label{LInterval}
ds^2 = -f^2 dt^2 + \sfx^2 (r,t) dr^2 + \sfy^2 (r,t) \left( d \theta^2 + \sin^2 \theta d \phi^2 \right) \,,
\end{equation}
where $f$ is an additional factor, $f \equiv f(t,r)$. The background
four velocity, from \eq{4vel}, is,
\begin{equation}
\label{4velunpertL}
u^{\mu} = \left[\frac{1}{f},0,0,0\right] ,
\end{equation}

and energy-momentum tensor, from \eq{SET}, becomes,
\begin{equation}
\label{LStress}
T^{\mu \nu}=\begin{pmatrix}
  \frac{\rho}{f^2} & 0 & 0 & 0 \\
  0 & \frac{P}{\sfx^2} & 0 & 0 \\
  0 & 0 & \frac{P}{\sfy^2} & 0 \\
  0 & 0 & 0 & \frac{P}{\sfy^2 \sin^2 \theta}
 \end{pmatrix} \,.
\end{equation}
Energy conservation is similar to LTB but with an additional pressure term,
\begin{equation}
\label{LEngCons}
\dot{\rho} + (\rho + P)(H_{\sfx} + 2 H_{\sfy}) = 0 \,.
\end{equation}
If we now perturb the metric in a similar way to LTB, \eq{LTBMetricperturbations}, we get,
\begin{equation}
\label{LMetricperturbations}
\delta g_{\mu \nu}=\begin{pmatrix}
  -2 f^2 \Phi & f \sfx B_r & f \sfy B_\theta & f \sfy \sin \theta B_\phi \\
 f \sfx B_r & 2\sfx^2 C_{rr} & \sfx \sfy C_{r\theta} & \sfx \sfy \sin \theta C_{r\phi} \\
 f \sfy B_\theta & \sfx \sfy C_{r\theta} & 2\sfy^2 C_{\theta\theta} & \sfy^2 \sin \theta C_{\theta\phi} \\
 f \sfy \sin \theta B_\phi & \sfx \sfy \sin \theta C_{r\phi} & \sfy^2 \sin \theta C_{\theta\phi} & 2\sfy^2 \sin^2 \theta C_{\phi\phi}
\end{pmatrix} \,.
\end{equation}
The perturbed 4-velocity, from \eq{4vel}, is,
\begin{equation}
\label{4velpertL}
u^{\mu} = \frac{1}{f} \left[(1 - \Phi),v^r,v^\theta,v^\phi \right] \,,
\end{equation}

As in the LTB case, we can now study how the perturbations in this
case change under the transformation \eq{gaugetrans1}. The perturbed
energy density $\delta\rho$, and the 3-velocities, $v^i$, transform as
in the LTB background \eq{denspertgtransLTB} and \eq{vuiTrans}.
The perturbed metric components transform as
\be
\label{LPhitrans}
\tilde{\Phi} = \Phi - \frac{\dot{f}}{f} \delta t  - \frac{f '}{f} \delta r  + \dot{\delta t} \,,
\ee
and
\bea
\label{LBrTrans}
{\tilde{B}}_r &=& B_r + \frac{\sfx}{f} \dot{\delta r} - \frac{f \delta t '}{\sfx}\,,\\
\label{LBTheTrans}
{\tilde{B}}_{\theta} &=& B_{\theta} + \frac{\sfy}{f} \dot{\delta \theta} - \frac{f \partial_{\theta} \delta t}{\sfy}\,,\\
\label{LBPhiTrans}
{\tilde{B}}_{\phi} &=& B_{\phi} + \frac{\sfy (\sin \theta)}{f} \dot{\delta \phi} - \frac{f \partial_{\phi} \delta t}{\sfy (\sin \theta)}\,.
\eea
The transformation behaviour of the perturbed metric components
$C_{ij}$, and hence $\psi$,  are unchanged from
the LTB case, see \eq{CrrTrans} - \eq{CphiphiTrans} and \eq{psiTrans2} - \eq{CthephiTrans}.\\

The perturbed energy conservation equation is,
\begin{eqnarray}
\label{LPertEconsSphP}
\delta \dot{\rho} &+& \left(\delta \rho + \delta P\right)\left(\frac{\dot{\sfx}}{\sfx} + 2 \frac{\dot{\sfy}}{\sfy}\right) + \left({\bar{\rho}} ' + {\bar{P}} '\right)v^r + \frac{f B_r}{\sfx} {\bar{P}}'+ \left(\partial_\theta \frac{B_\theta}{\sfy} + \partial_\phi \frac{B_\phi}{\sfy \sin \theta}\right) f {\bar{P}} \nonumber \\
&+& \left({\bar{\rho}} + {\bar{P}}\right)\left(\dot{\psi} + {v^r} ' + \partial_\theta v^\theta + \partial_\phi v^\phi + \left[\frac{f '}{f} + \frac{\sfx '}{\sfx} + 2 \frac{\sfy '}{\sfy} \right] v^r + \frac{B_r f '}{\sfx} + \cot \theta v^\theta\right) \nonumber \\
&=& 0 \,.
\end{eqnarray}

As in the previous section, we can now construct gauge-invariant
quantities.  We choose hypersurfaces of vanishing perturbed energy
density to define the temporal gauge, that is,
\begin{equation}
\label{Consdenshyp}
\delta t = \frac{\delta \rho}{\dot{\bar{\rho}}}
+ \frac{\bar{\rho}'}{\dot{\bar{\rho}}} \delta r \,,
\end{equation}
and choose again co-moving gauge, where $v^i = 0$, to get for
the spatial coordinate shifts
\begin{equation}
\label{LvuiTransComoving}
\delta x^i = \int v^i dt \,.
\end{equation}
Then using the transformation for perturbed metric trace $\psi$, given
above in \eq{CurvatureLTB1}, we can construct the gauge-invariant
spatial metric trace perturbation on uniform density and comoving hypersurfaces,
\bea
\label{LPresszetafull4}
\nonumber - \SMTP &=& \psi + \frac{\delta \rho}{3(\bar{\rho} + \bar{P})} + \frac{1}{3}\Bigg\{\left(\frac{\sfx '}{\sfx} + 2 \frac{\sfy '}{\sfy}
+ \frac{\bar{\rho} '}{\bar{\rho} + \bar{P}} \right) \int v^r dt + \partial_r \int v^r dt + \partial_{\theta} \int v^{\theta} dt \\ &+& \partial_{\phi} \int v^{\phi} dt + \cot \theta \int v^{\theta} dt \Bigg\}\,.
\eea
The evolution equation for $\SMTP$ is then found from
\eq{LPertEconsSphP}, using the decomposition of the pressure
perturbation, \eq{deltaP}, and the definition of the adiabatic sound
speed, \eq{cs2_tandr}, as
%
%
\begin{eqnarray}
\label{AdPresszetaevol1b}
\nonumber - \DOTSMTP &=& 
 \frac{ \dot{\bar{\rho}}}{\left({\bar{\rho}} + {\bar{P}}\right)^2}\dpn - \frac{\bar{P} '}{\left({\bar{\rho}} + {\bar{P}}\right)}v^r 
- \frac{f B_r}{\sfx \left({\bar{\rho}} + {\bar{P}}\right)} {\bar{P}} ' 
- \left(\partial_\theta \frac{B_\theta}{\sfy} + \partial_\phi \frac{B_\phi}{\sfy \sin \theta}\right) \frac{f \bar{P}}{\left({\bar{\rho}} + {\bar{P}}\right)} \\ &+& \left[\partial_t \left(\frac{\sfx '}{\sfx} + \frac{\sfy '}{\sfy} + \frac{\bar{\rho} '}{\bar{\rho} + \bar{P}} \right) \right] \int v^r dt - \frac{f '}{f} v^r  + \frac{B_r f '}{\sfx} \,.
\end{eqnarray}

By transforming the coordinates to Cartesian using the chain rule and taking the spatial derivatives to be negligible on large scales, \eq{AdPresszetaevol1b}, reduces to,
\begin{eqnarray}
\label{AdPresszetaevol3}
\DOTSMTP &=&  \frac{H_\sfx + 2 H_\sfy}{3({\bar{\rho}} + {\bar{P}})}
\, \dpn \,.
\end{eqnarray}
This can be seen to be similar to that for LTB, \eq{zetaevol5}, but as with the standard $\zeta$ in FRW, and unlike $\SMTP$ in both LTB and FRW, is only valid at large scales.

\section{Discussion on Gauge-invariants in Inhomogeneous Spacetimes}
\label{LTB Conclusion}

In this chapter we have constructed gauge-invariant quantities in
perturbed LTB spacetime.  In particular we have constructed the
gauge-invariant spatial metric trace perturbation on comoving, uniform density
hypersurfaces, $\SMTP$. We derived the evolution equation for $\SMTP$
and found that it is conserved on all scales for barotropic fluids
(when $\dpn=0$). We found this result for the evolution equation for $\SMTP$ also holds for FRW. This is in contrast to the standard FRW result,
where an equivalent gauge-invariant quantity, $\zeta$, is only conserved on
large scales. It was also found that the evolution equation for $\SMTP$, in Lema{\^\i}tre spacetime which would be conserved in the case of barotropic fluids is only found in the large scale limit, as with the result for the standard $\zeta$ in FRW.\\

Deriving these results in LTB is more involved than in the FRW case, because
the background is $t$ and $r$ dependent, whereas the FRW background is
homogeneous and isotropic, and hence only $t$ dependent.
Additional complications often arise in LTB because it suggests a 1+1+2
decomposition, and not ``simply'' a 1+3 one, as in FRW. The 1+3 decomposition makes a multi-pole decomposition much more complicated, and hence we did not
use such a multi-pole decomposition here to construct conserved quantities.\\

The difference in the behaviour of the LTB $\SMTP$ found here, to the
curvature perturbation in FRW may prove useful in studying the
differences in structure formation in the two models.

As pointed out in Ref.~\cite{Meyer:2014qla} the gauge-invariant quantity we have constructed would be particularly useful in numerical simulations of structure formation in regions of the universe best modelled using LTB e.g. large voids or overdensities. This is because in numerical simulations initial conditions, for -for example- densities and velocities are set and therefore known. These can then be compared with their values at the end of the simulation, as opposed to the limited information available through actual observations at different times.

In addition, further
extensions of this research into similar and related spacetimes, such
as Lema{\^\i}tre, may provide similar tools for comparing
inhomogeneous spacetimes with the standard FRW model, as was highlighted with reference to our research in Ref.~\cite{Faraoni:2015uma}. This is of
particular interest to research trying to explain the effects of
Dark Energy using inhomogeneous spacetimes. For example, LTB is difficult to fit to all observations simultaneously e.g. Baryon Accoustic Oscillations (BAOs) and supernovae data (see e.g.~Refs.~\cite{Clarkson:2012bg,Vargas:2015ctw}). Specifically, to explain the observed dimming of distant supernovae it is possible to use a spherically symmetric inhomogeneous model such as LTB with a local underdensity. However density profiles for such a void which allow BAO observations to match observations at all times differ from those needed to fit the supernovae data (the former requiring a greater void density than the latter). In fact density profiles which work well with the supernovae data stretch the BAO scale at lower redshifts i.e. near the centre of the void.  However other inhomogeneous cosmologies, such as Lema{\^\i}tre might still prove a better fit to observations while providing an alternative explanation for accelerated expansion observations but without DE.



\chapter{The Growth of Structure in Assisted Coupled Quintessence Cosmologies}
\label{ch:4}


\section{Introduction to Assisted Coupled Quintessence}
\label{ACQIntroduction}
In this chapter we investigate \emph{assisted coupled quintessence} (ACQ) models of DE. These models contain multiple CDM fluids and DE scalar fields coupled with each other. We study the behaviour of linear perturbations to these models in order to compare the growth of structure in those models, and one other recently researched DE model, \emph{multi-coupled dark energy} (McDE)~\cite{Piloyan:2014gta}, against $\Lambda$CDM. This chapter is set out as follows. Section \ref{ACQModel} describes the ACQ model used. Subsection \ref{IntDEBack} describes those aspects of the background equations specific to the models studied. Subsection \ref{GenNoGauge} contains the general gauge unspecified perturbed equations. Subsection \ref{Fixing} describes fixing the gauge in order that the equations can be solved numerically. Section \ref{PythonBack1} then describes the resulting \PY~code. Section \ref{Obs} reviews the observational quantities against which our results can be compared. Finally, section \ref{PythonPert1} details our numerical investigation of specific ACQ and related models.  We conclude in Section \ref{Conclusion1}.

\section{The model}
\label{ACQModel}

In the ACQ models, the dark sector of the universe is modelled by $J$ different 
dark matter  fluids, with arbitrary equation of state, and $K$ different 
scalar fields. 
We also include two further fluids which model baryonic matter, and radiation. The general energy-momentum tensor for any perfect fluid, taken from~\eq{SET} but here with mixed indices, is given by
\be
\label{MatSEM}
{T^{\mu}_{\nu}}^{(M_\alpha)} = (\rho_\alpha + P_\alpha) u^\mu _{(\alpha)} u_{\nu (\alpha)} + \delta^\mu_\nu P_{\alpha} \,,
\ee
where the subscript $\alpha$  labels the $J+2$ fluids, $\rho_\alpha$ is the density of any given fluid and $P_\alpha$  the corresponding pressure, and $u^\mu _{(\alpha)}$ is the four velocity for a given fluid.
The equation of state for a given fluid is defined as in \eq{eostake2} such that,
\be
\label{EOS}
w_\alpha=\frac{P_\alpha}{\rho_\alpha}\,. 
\ee
Here and throughout Greek indices $\mu$ and $\nu$ label coordinates running over time and relative dimensions in space, and we use lower case Latin indices to label only spatial dimensions. The energy-momentum tensor for the scalar fields is given by
\be
\label{FieSEM}
{T^{\mu}_{\nu}}^{(\ph)} = g^{\lambda \mu} \sum_I \partial_\lambda \ph_I \partial_\nu\ph_I - \delta^\mu_\nu \left (\frac{1}{2} \sum_I g^{\rho \sigma} \partial_\rho \ph_I \partial_\sigma \ph_I + V(\ph_1,\dots,\ph_M) \right) \,,
\ee
where $V$ is the potential energy, $\ph_I$ the ``$I^{th}$" scalar field, and upper case Roman indices label the $K$ fields. In addition,
\be
\label{TotSEM}
{T^{\mu}_{\nu}} = {T^{\mu}_{\nu}}^{(M_\alpha)} + {T^{\mu}_{\nu}}^{(\ph)} ,
\ee
where ${T^{\mu}_{\nu}}$ is the total energy-momentum tensor. In order to model the interaction of the matter fluids with the scalar fields, we assume \cite{Amendola:1999dr,Amendola:2014kwa}

\begin{equation}
\label{econsint1}
\nabla_\mu {T^{\mu}_{\nu}}^{(\ph)} = \kappa \sum\limits_{\alpha, I} \C_{I \alpha} T_{(M_\alpha)} \nabla_\nu \ph_I \quad , \quad \nabla_\mu {T^{\mu}_{\nu}}^{(M_\alpha)} = - \kappa \sum\limits_{I} \C_{I \alpha} T_{(M_\alpha)} \nabla_\nu \ph_I \, ,
\end{equation}
where $\kappa = (8 \pi G)^{\frac{1}{2}}$ and $\C_{I \alpha}$ are coupling constants. Here $T_{(M_\alpha)}$ is the trace of energy-momentum tensor,
\be
\label{trace}
T_{(M_\alpha)} = T^\mu_{\mu(M_\alpha)} \,, 
\ee 
for a given fluid. Equations \eq{econsint1} respect energy-momentum conservation of the total matter content. In what follows we will set the relevant components of the $C$ matrix such that there is no interaction between baryons or radiation and the scalar fields.

\subsection{Background cosmology}
\label{IntDEBack}

We take a flat FRW spacetime as our 
background with the metric \eq{FRWspacetime}. We assume the fluids to be comoving with the expansion of the universe such that
\begin{equation}
\label{4velunpertDE}
\bar{u}_{0 (\alpha)} = -1 \,, \bar{u}_{i (\alpha)} = 0 \,.
\end{equation}
Here we use ``bars" to denote background quantities. The background stress energy tensor for the fluids then becomes
\bea
\label{BackSEMDE}
\nonumber {\bar{T}}_{0 0} &=& \sum\limits_{\alpha} \bar{\rho}_{\alpha} + \sum\limits_{I} \frac{\dot{\bar{\ph}}_{I}^2}{2} + V \,, {\bar{T}}_{0 j} = 0 \,, {\bar{T}}_{i j} = \delta_{i j} a^2 \left(\sum\limits_{\alpha} \bar{P}_{\alpha} + \sum\limits_{I}\frac{\dot{\bar{\ph}}_{I}^2}{2} - V\right)\,,\\
\eea
where an overdot indicates a derivative with respect to cosmic time. \eq{econsint1} leads to the evolution equation for each fluid
\begin{equation}
\label{IntDEMatBack2}
\dot{\bar{\rho}}_{\alpha} + 3 H (\bar{\rho}_{\alpha} + \bar{P}_{\alpha}) = - \kappa \sum\limits_{I} \C_{I \alpha} (\bar{\rho}_{\alpha} - 3 \bar{P}_{\alpha}) \dot{\bar{\ph}}_{I} ,
\end{equation}
where $H$ is the Hubble parameter, and to the Klein-Gordon equation for each field
\begin{equation}
\label{IntDESFBack2}
\ddot{\bar{\ph}}_{I} + 3 H \dot{\bar{\ph}}_{I} + {V}_{, \ph_I} = \kappa \sum\limits_{\alpha} \C_{I \alpha} (\bar{\rho}_{\alpha} - 3 \bar{P}_{\alpha}) \,.
\end{equation}
The background Friedmann equation is
\be
\label{FEIntDEBack}
H^2 = \frac{\kappa^2}{3} \left[\sum\limits_{\alpha} \bar{\rho}_{\alpha} + \sum\limits_{I} \frac{\dot{\bar{\ph}}_{I}^2}{2} + V \right] \,.
\ee
Finally, we define the density parameter for a given fluid as per \eq{DensityParameter}, such that,
\be
\label{densparam}
\Omega_\alpha = \frac{\bar{\rho}_{\alpha}}{\rho_{\rm c}} \,,
\ee
where $\rho_{\rm c}$ is the critical density defined as in \eq{CriticalDensity}.

\subsection{Linear perturbations}
\label{IntDEperts}
\subsubsection{General Perturbed Equations Gauge Unspecified}
\label{GenNoGauge}
The line element for  perturbations about a  flat FRW spacetime with the gauge unspecified is given by \eq{PertlineFRW}. The perturbed 4-velocity is derived from \eq{4vel} such that,
\be
\label{pertvelgen}
u_{0(\alpha)} = -(1+\Phi) \qquad , \qquad u_{i(\alpha)} = a(v+B),_i \,,
\ee
and the total perturbed energy-momentum tensor for our model is given by
\bea
\label{PertSEMDE}
\delta T_{0 0} &=&
  \sum\limits_{\alpha} \delta \rho_{\alpha} + \sum\limits_{I} (- \Phi {\dot{\bar{\ph}}_{I}}^2 + \delta \ph_{I} \dot{\bar{\ph}}_{I} + V,_{\ph_I} \delta \ph_I) ,\\ \nonumber
 \delta T_{0 j} &=& a\left[\sum\limits_{I}\dot{\bar{\ph}}_{I}\left(\dot{\bar{\ph}}_{I} B_{,i} + \frac{1}{a} \delta \ph_{I,i}\right) - \sum\limits_{\alpha} (\bar{\rho}_{\alpha} + \bar{P}_{\alpha} )v_{(\alpha),i} \right] ,\\ \nonumber
 \delta T_{i j} &=& \delta_{i j} a^2 \left(\sum\limits_{\alpha} \delta P_{\alpha} - \sum\limits_{I}( \Phi {\dot{\bar{\ph}}_{I}}^2 - \dot{\delta \ph}_{I}\dot{\bar{\ph}}_{I} + V,_{\ph_I} \delta \ph_I)  \right)\,.
\eea
We now move to Fourier space, where tensor perturbations may be decomposed into eigenmodes of the spatial Laplacian such that,
\be
\label{fourierdecomp}
\nabla^2 = - \frac{k^2}{a^2} ,
\ee
where $k$ is the wavenumber. The evolution equations for density fluctuations are then given by 
\bea
\label{IntDEPertEcons3}
\nonumber &\dot{\delta \rho}_{\alpha}& - \left(\frac{k^2 v_{\alpha}}{a} +  k^2{\dot{E}} +3 \dot{\psi} \right) (\bar{\rho}_{\alpha} + \bar{P}_{\alpha} ) + 3 H (\delta \rho_{\alpha} + \delta P_{\alpha}) = - \kappa \sum\limits_{I} \C_{I \alpha} (\bar{\rho}_{\alpha} - 3 \bar{P}_{\alpha}) \dot{\delta \ph}_{I} \\ &-& \kappa \sum\limits_{I} \C_{I \alpha} (\delta \rho_{\alpha} - 3 \delta P_{\alpha}) \dot{\bar{\ph}}_{I} \,,
\eea
momentum conservation gives the constraint
\be
\label{IntDEPertMomcons}
\dot{v}_{\alpha} =  \kappa \sum\limits_{I} \C_{I \alpha} (\bar{\rho}_{\alpha} - 3 \bar{P}_{\alpha}) \frac{\delta \ph_{I}}{a} + 3H \frac{\dot{\bar{P}}_{\alpha}}{\dot{\bar{\rho}}_{\alpha}} (v_{\alpha} + B) - H(v_{\alpha} + B) - \frac{\Phi}{a} - \frac{\delta P_{\alpha}}{a({\bar{\rho}_{\alpha}} + {\bar{P}_{\alpha}})} - \dot{B} \,,
\ee
and the evolution of scalar field perturbations is given by
\bea
\label{IntDEPertEconsSF}
\nonumber &\ddot{\delta \ph}_{I}&  + 3 H \dot{\delta \ph}_{I} + \sum\limits_{J} V,_{\ph_I \ph_J} \delta \ph_J - (k^2 {\dot{E}} + 3 \dot{\psi}) \dot{\bar{\ph}}_{I} + \frac{k^2}{a^2} \delta \ph_{I} + \frac{\dot{\bar{\ph}}_{I}}{a} k^2 B - \dot{\bar{\ph}}_{I} \dot{\Phi} + 2 V,_{\ph_I} \Phi\\ &-& 2 \kappa \sum\limits_{\alpha} \C_{I \alpha} (\bar{\rho}_{\alpha} - 3 \bar{P}_{\alpha}) \Phi - \kappa \sum\limits_{\alpha} \C_{I \alpha} (\delta \rho_{\alpha} - 3 \delta P_{\alpha}) = 0 \,.
\eea
The Einstein Field Equations are as follows. From the $0-0$ component we get
\be
\label{G00DEEFE}
3 H (\dot{\psi} + H \Phi) + \frac{k^2}{a^2}(\psi + H[a^2\dot{E} - aB]) = - \frac{\kappa^2}{2} \left[\sum\limits_{\alpha}\delta \rho_{\alpha} + \sum\limits_{I}(- \Phi \dot{\bar{\ph}}^2_{I} + \dot{\delta \ph}_{I} \dot{\bar{\ph}}_{I} + V,_{\ph_I} \delta \ph_I)\right] \,,
\ee
from the $0-i$ component 
\be
\label{Gi0DEEFE}
\dot{\psi} + H \Phi = - \frac{\kappa^2}{2} \left[\sum\limits_{\alpha}a(v_{\alpha} + B)(\bar{\rho}_{\alpha} + \bar{P}_{\alpha}) - \sum\limits_{I} \dot{\bar{\ph}}_{I} \delta \ph_{I} \right] \,,
\ee
from the trace of the $i-j$ component 
\be
\label{TraceGijDEEFE}
\ddot{\psi} + 3 H \dot{\psi} + H \dot{\Phi} + (3 H^2 + 2 \dot{H}) \Phi = \frac{\kappa^2}{2} \left[\sum\limits_{\alpha}\delta P_{\alpha} + \sum\limits_{I} (- \Phi \dot{\bar{\ph}}^2_{I} + \dot{\delta \ph}_{I} \dot{\bar{\ph}}_{I} - V,_{\ph_I} \delta \ph_I)\right] \,,
\ee
and from the trace-free part of the $i-j$ component 
\be
\label{TraceFreeGijDEEFE}
{\dot{\sigma}}_s + H \sigma_s - \Phi + \psi = 0 \,,
\ee
where $\sigma_s$ is the scalar shear and $\sigma_s = a^2 \dot{E} - a B$.

\subsubsection{Governing equations in flat gauge}
\label{Fixing}

As we saw at the end of Chapter 2, it is possible to define hypersurfaces on which given quantities are zero and thereby ``fix'' the gauge. This may be done by fixing one or more degrees of freedom leading to many different possible choices of gauge. Previously in the literature (see e.g.~Refs.\cite{Amendola:2014kwa,Raccanelli:2015qqa}) a common choice of gauge for studies of coupled quintessence models has been the longitudinal gauge ($\tilde{B} = \tilde{E} = 0$), and we present the equations of motion for perturbations in this gauge in Appendix~\ref{Long2f2dm}. However, we found that this gauge is not a good choice for the numerical integration of the full equations of motion.  This is due to the prefactor term in \eq{Phiconstraint}. The magnitude of the second term in this prefactor is orders of magnitude smaller than the first, except when the first touches zero, which can occur as the fields oscillate. This leads to a loss of accuracy at these times and to a numerical instability.  For our numerical integration we therefore use the flat gauge which does not suffer from this problem. The \PY~code is covered in more detail in Chapter 5.\\
The flat gauge is defined by the conditions $\tilde{\psi}=0$ and $\tilde{E}=0$. Defining the new quantity
\be
\label{FlatNewv}
\hat{v}_\alpha = v_\alpha + B\,,
\ee
in this gauge 
\eq{IntDEPertEcons3} reduces to
\bea
\label{FlatIntDEPertEcons4}
\nonumber &\dot{\delta \rho_\alpha}& + 3 H (\delta \rho_\alpha + \delta P_\alpha) - \frac{k^2 (\hat{v}_\alpha - B)}{a}({\bar{\rho}}_\alpha + {\bar{P}}_\alpha) = - \sum\limits_{I} \kappa \C_{I \alpha} ({\bar{\rho}}_\alpha - 3 {\bar{P}}_\alpha) {\dot{\delta \ph}}_I \\ &-& \sum\limits_{I} \kappa \C_{I \alpha} (\delta \rho_\alpha - 3 \delta P_\alpha) {\dot{\bar{\ph}}}_I \,.
\eea
and \eq{IntDEPertMomcons} to
\be
\label{FlatIntDEPertMomconsLong}
\dot{\hat{v}}_\alpha =  \kappa \sum\limits_{I} \C_{I \alpha} (\bar{\rho}_\alpha - 3 \bar{P}_\alpha) \frac{\delta \ph_I}{a} + 3H \frac{\dot{\bar{P}}_\alpha}{\dot{\bar{\rho}}_\alpha} \hat{v}_\alpha - H\hat{v}_\alpha - \frac{\Phi}{a} - \frac{\delta P_\alpha}{a({\bar{\rho}_\alpha} + \bar{P}_\alpha)}\,.
\ee
The evolution equation for the fields, \eq{IntDEPertEconsSF}, becomes
\bea
\label{FlatDEPertEconsSF2}
\nonumber &{\ddot{\delta \ph}}_I& + 3 H {\dot{\delta \ph}}_I + \sum\limits_{J} V,_{\ph_I \ph_J} \delta \ph_J - {\Bigg[} \frac{\kappa^2}{2H} \left( \sum\limits_{\alpha} \delta P_\alpha - \sum\limits_{I} (\Phi \dot{\bar{\ph}}^2_I - {\dot{\delta \ph}}_I {\dot{\bar{\ph}}}_I + V,_{\ph_I} \delta \ph_I ) \right) \\ \nonumber &-& \frac{(3 H^2 + 2 \dot{H})}{H} \Phi {\Bigg]} {\dot{\bar{\ph}}}_I + \frac{k^2}{a^2} \delta \ph_I + \frac{k^2 B}{a} {\dot{\bar{\ph}}}_I + 2 V,_{\ph_I} \Phi - 2 \sum\limits_{\alpha} \kappa \C_{I \alpha} ({\bar{\rho}}_\alpha - 3 {\bar{P}}_\alpha) \Phi \\ &-& \sum\limits_{\alpha} \kappa \C_{I \alpha} (\delta \rho_\alpha - 3 \delta P_\alpha) = 0 \,. 
\eea
From \eq{G00DEEFE}, we get
\be
\label{FlatG00DEEFE2}
3 H^2 {\Phi} - \frac{k^2 B}{a} H = - \frac{\kappa^2}{2} \left[\sum\limits_{\alpha} \delta \rho_\alpha + \sum\limits_{I}(-\Phi \dot{\bar{\ph}}^2_I + {\dot{\delta \ph}}_I {\dot{\bar{\ph}}}_I + V,_{\ph_I} \delta \ph_I)\right] \,,
\ee
and from \eq{Gi0DEEFE} 
\be
\label{FlatGi0DEEFE2}
\Phi = - \frac{\kappa^2}{2H} \left[\sum\limits_{\alpha} a \hat{v}_\alpha (\bar{\rho}_\alpha + \bar{P}_\alpha) - \sum\limits_{I} {\dot{\bar{\ph}}}_I \delta \ph_I \right] \,,
\ee
which allows us to replace  $\Phi$ in terms of field and fluid perturbations. For completeness we note that \eq{TraceGijDEEFE} gives
\be
\label{FlatTraceGijDEEFE2}
H \dot{\Phi} + (3 H^2 + 2 \dot{H}) \Phi = \frac{\kappa^2}{2} \left[\sum\limits_{\alpha} \delta P_\alpha - \sum\limits_{I} \left(\Phi \dot{\bar{\ph}}^2_I - {\dot{\delta \ph}}_I {\dot{\bar{\ph}}}_I + V,_{\ph_I} \delta \ph_I\right) \right] 
\ee
and from \eq{TraceFreeGijDEEFE} we have
\be
\label{FlatTraceFreeGijDEEFE2}
\dot{B} + 2HB  = - \frac{\Phi}{a} \,.
\ee
Combining \eq{FlatG00DEEFE2} and \eq{FlatGi0DEEFE2} we find 
\bea
\label{FlattildeBNewVar}
\nonumber &B& = \frac{3 \kappa^2 a}{2k^2} \Bigg[ \frac{1}{3H} \left( \sum\limits_{\alpha} \delta \rho_\alpha - \sum\limits_{I} (\Phi \dot{\bar{\ph}}^2_I - {\dot{\delta \ph}}_I {\dot{\bar{\ph}}}_I - V,_{\ph_I} \delta \ph_I ) \right) + \sum\limits_{I} {\dot{\bar{\ph}}}_I \delta \ph_I \\ &-& \sum\limits_{\alpha} a \hat{v}_\alpha (\bar{\rho}_\alpha + \bar{P}_\alpha) \Bigg] \,.
\eea
which allows us to replace $B$ is terms of field and fluid perturbations.

\section{Numerical solutions}
\label{PythonBack1}

We can now solve the closed system of equations derived in the previous section numerically. The system of background equations for the scalar fields and the energy densities of the fluids, \eq{IntDEMatBack2} and \eq{IntDESFBack2}, together with the Friedmann constraint \eq{FEIntDEBack}, is solved simultaneously with the evolution equations for the perturbations $\delta\rho_\alpha$, $\hat v_\alpha$ and $\delta \varphi_I$, \eq{FlatIntDEPertEcons4} to \eq{FlatDEPertEconsSF2}, together with the constraint equations for $\Phi$ and $B$, \eq{FlatGi0DEEFE2} and \eq{FlattildeBNewVar}. The numerical code, named \PY, is written in Python and publicly available on Bitbucket~\cite{Bitbucket} and on the \PY~website~\cite{Pyweb} under an open source modified BSD license, with documentation available in Ref.~\cite{PYDOCREF}.

\subsection{Initial Conditions}
\label{CQIC}

\subsubsection{Background Initial Conditions}

We set the initial conditions for the background energy densities of the fluids and the background field amplitudes such that the background evolution follows closely that of the $\Lambda$CDM model. Due to the potentials used in the models tested we have analytical solutions for the background evolution equations, which enables us to set the background initial conditions in terms of their values today. We are free to choose an initial time, and select $N=-14$, where $N$ is the number of e-folds from today ($N=0$), which fixes the initial value for the scale factor $a$ and coordinate time, $t$. This also ensures we are well into the radiation dominated epoch. In particular, we demand that the model satisfies constraints on present day energy densities from Planck data \cite{Ade:2015xua}. These are $\Omega_{\Lambda} = 0.6911 \pm 0.0062$ for the cosmological constant, $\Omega_{r} = 9.117 \times 10^{-5}$ for radiation, $\Omega_{b} = 0.0486 \pm 0.0003$ for baryons and $\Omega_{CDM} = 1 - \Omega_{DE} - \Omega_{r} - \Omega_{b}$ for cold dark matter. To do so, we assume that the scalar fields will collectively replace $\Lambda$, and the dark matter fluids collectively replace the single cold dark matter species of the $\Lambda$CDM model. Initially we take the fields' velocity to be zero, $\dot \varphi_I=0$. Of course we need to check on a case by case basis whether the fields really do generate acceleration in a way that accounts for observations, and that dark matter components behave in a viable way, such that the background evolution is compatible with current limits.

\subsubsection{Perturbed Initial Conditions}
\label{perticflat}

We start our simulations at sufficiently early times to ensure radiation domination and that all the $k$ modes studied lie outside the horizon at that time. For simplicity, we choose the initial conditions for the field velocity and  field perturbations to be zero
\be
{\dot{\delta \ph}}_I =  \delta \ph_I = 0 \,,
\ee
though we find the evolution is insensitive to this choice. The initial conditions for all other perturbations can be given in terms of observational constraints on the power spectrum of the gauge invariant curvature perturbation $\zeta$, as defined earlier in \eq{unidenshyp3}, 

\be
\label{zetadef}
\left\langle\zeta^2\right\rangle = \delta^3 ({\bf{k-k}}') \frac{2 \pi^2}{k^3} {\cal{P}}_\zeta (k) \,.
\ee
On superhorizon scales the power spectrum can be parametrised as 
\be
\label{Ps}
{\cal{P}}_\zeta (k) = A_s \left( \frac{k}{k_*} \right)^{n_s - 1} \,,
\ee
where \cite{Planck:2013jfk} $ A_s = 2.142 \times 10^{-9}$ is the scalar amplitude at the Planck pivot scale
$k_* = 0.05$ Mpc$^{-1}$,
and $n_s=0.9667$ is the spectral index~\cite{Ade:2015xua}. \\
From \eq{unidenshyp3} we then get a relation between the curvature perturbation and the total energy density perturbation in flat gauge, such that,
\be
\delta\rho_{\rm flat}=-\frac{\dot{\bar\rho}}{H}\zeta\,.
\ee
This allows us to set the initial condition for the individual fluids. In addition we assume that the initial conditions are adiabatic, which gives a relation between the fluid density perturbations initially. The gauge-invariant relative entropy perturbation between two non-interacting fluids~\cite{Malik:2004tf} is given by
\be
\label{entpert}
{\cal{S}}_{\alpha \beta} = 
- 3 H \left( \frac{\delta \rho_\alpha}{\dot{\bar{\rho}}_\alpha} - \frac{\delta \rho_\beta}{\dot{\bar{\rho}}_\beta} \right) \,.
\ee
Adiabatic initial conditions require that ${\cal{S}}_{\alpha \beta} =
0$. Combining \eq{entpert} with \eq{IntDEMatBack2} for radiation and
baryons, which for these models, as specified in
Section~\ref{ACQIntroduction} have couplings of zero, we find \be
\label{deltarels}
\delta_{b} = \frac{3}{4} \delta_r \,,
\ee
where we introduced the density contrast for a given fluid species, $\alpha$, as
\be
\label{denscontdefn}
\delta_\alpha 
\equiv \frac{\delta \rho_\alpha}{\bar{\rho}_\alpha}  \,.
\ee
Finally we can set the initial conditions for the 3-velocities, $\hat{v}_\alpha$.  We checked numerically that the late time evolution of the system is not very sensitive to the actual value for the 3-velocities, and we therefore set $\hat{v}_\alpha=0$ initially. While studying the initial conditions we found that aside from the initial radiation density perturbation the results are fairly insensitive to small changes in the initial conditions, due to the integration starting well inside radiation domination. Small variations in the initial conditions for the other constituents, for a given $k$ mode, soon converged to a common trajectory within approximately one e-fold from the start of the simulations. This meant there was negligible difference in the observable growth of the density perturbations.

\subsubsection{Relating Longitudinal Gauge to Flat Gauge}
\label{pertininlong}

In the previous sections we have presented the system of governing equations and the initial conditions for the code in flat gauge. However, in order to connect to previous studies in the literature we present our results in terms of the density contrast in longitudinal gauge.\\
Using the background and perturbed densities as defined in \eq{BackSEMDE} and \eq{PertSEMDE}, the total density contrast is defined as,
\be
\label{totdenscontdefn}
\delta = \frac{\sum\limits_{\alpha} \delta \rho_\alpha + \sum\limits_{I} \delta \rho_{\ph_I}}{\sum\limits_{\alpha} \bar{\rho}_\alpha + \sum\limits_{I} \bar{\rho}_{\ph_I}}  \,.
\ee
Using the transformations for the metric and matter variables given in appendix~\ref{Flat to Long Gauge Relations}, and the constraint Eqns.~(\ref{FlattildeBNewVar}), we find
\bea
\label{KAMdeltalinkbulk}
\nonumber \delta_{\rm long}&=&\delta_{\rm flat} + \frac{{\dot{\bar{\rho}}}^2}{\bar\rho} a \Bigg( \frac{3 \kappa^2 a}{2k^2} \Big[ \frac{1}{3H} \left( \sum\limits_{\alpha} \delta \rho_\alpha - \sum\limits_{I} (\Phi \dot{\bar{\ph}}^2_I - {\dot{\delta \ph}}_I {\dot{\bar{\ph}}}_I - V,_{\ph_I} \delta \ph_I ) \right) \\ &+& \sum\limits_{I} {\dot{\bar{\ph}}}_I \delta \ph_I - \sum\limits_{\alpha} a \hat{v}_\alpha (\bar{\rho}_\alpha + \bar{P}_\alpha) \Big] \Bigg)  \,,
\eea
which reduces initially to
\be
\label{KAMdeltalinkIC}
\delta_{\rm long}=\delta_{\rm flat}
+\left(\frac{k}{a}\right)^{-2}\left[4\pi\, G\delta_{\rm flat}
-\frac{{\dot{\bar{\rho}}}^2}{3H\bar\rho}
a\sum_\alpha (\bar{\rho}_\alpha + \bar{P}_\alpha)   \hat v_\alpha\right]   
\,.
\ee

\section{Observations}
\label{Obs}

Two key parameters which are constrained by observational data are the growth factor and growth function. We therefore apply our code to calculate these quantities. The growth factor is defined as 
\be
\label{growthdef}
g = \frac{\delta}{\delta_0} \,,
\ee
where $\delta$ is the total density contrast defined in the longitudinal gauge \cite{Raccanelli:2015qqa}, 
and $\delta_0$ is the total density contrast today. The growth function, $f$, is defined as
\be
\label{fdefn}
f = \frac{\delta'}{\delta} \,,
\ee
where the prime in this case denotes a derivative with respect to the number of e-folds \cite{Raccanelli:2015qqa}. 
Typically observational results are presented as constraints on the combinations $fg$ and $f\sigma_8$, since, for example, these quantities can be extracted directly from redshift space distortions  (see e.g.~Ref.~\cite{Bull:2015lja}). $\sigma_8$ is the amplitude of the matter power spectrum at a scale of $8h^{-1}$Mpc \cite{Raccanelli:2015qqa,Macaulay:2013swa}. The experimental uncertainty of $\sigma_8$, taken from \emph{Dark Energy Survey} (DES), which overlaps two other data sets which are in some tension \emph{Canada-France-Hawaii Telescope Lensing Survey} (CFHTLenS) and Planck), is $0.81^{+0.16}_{-0.26}$\cite{Abbott:2015swa}. In Subsection~\ref{Lspaceexplore} we use $\sigma_8 = 0.81$~\cite{Ade:2015xua} since this is consistent with the other Planck based parameter values we have used. Future surveys hope to have the sensitivity to pick up $k$ dependence in the growth of structure. \emph{Square Kilometer Array} (SKA)~\cite{Bull:2015lja,Raccanelli:2015qqa}, for example, should be sensitive to measurements of growth at approximately the percent level (or better) for $42H_0 < k < 420H_0$ at a redshift $z \approx 1$~\cite{Bull:2015lja}. For $k > 42H_0$ this sensitivity falls to $\approx 30\%$, for example, being at this level around $k = 21H_0$. According to the author~\cite{Bull:2015lja} this combined four survey approach (SKA1-MID Band 1 and Band 2 IM (intensity mapping) surveys, H$\alpha$ and SKA2) should therefore have sufficient accuracy to distinguish between GR (General Relativity) + $\Lambda$CDM and alternative models, such as coupled quintessence. This accuracy is potentially increased still further through multiple tracer analysis, cross-correlating with other surveys such as Euclid. The combined redshift range for SKA and Euclid is $0.5 \gtrsim z \gtrsim 2$.

Current surveys offer far looser constraints on the growth of structure. Below we use observational data from \emph{6dF Galaxy Survey} (6dFGS), \emph{Luminous Red Galaxy} (LRG)$_{200}$, LRG$_{60}$ (where $200$ and $60$ refer to the sample size of luminous red galaxies selected), \emph{Baryon Oscillation Spectroscopic Survey} (BOSS), WiggleZ and \emph{VIMOS Public Extragalactic Redshift Survey} (VIPERS) with associated errors~\cite{Macaulay:2013swa} in our plots for $fg$. $fg$ itself was extracted from the values for $f \sigma_8$ from these surveys, and applying the value of $\sigma_8 = 0.81$ as detailed above. The values for $f \sigma_8$ themselves are obtain by assuming a weak or negligible $k$ dependence in the growth of structure and a linear dependence in the same on $\sigma_8$. These values are obtained by averaging over a range of scales, for example with the LRG results specific scales were selected in the range $30h^{-1}Mpc$ to $200h^{-1}Mpc$~\cite{Samushia:2011cs} and averaged over. This approach would therefore hide any k dependence in the growth. These current surveys also have a shorter redshift range than that predicted for future surveys ($z \lesssim 0.8$) and constrain growth at only $\approx 10 - 20 \%$ level. In single field coupled quintessence there is an observational constraint on the magnitude of the coupling between DE and CDM as $ \C < 0.1 \sqrt{\frac{2}{3}}$~\cite{Amendola:1999er}. For this class of models couplings greater than this give unrealistic background cosmologies, through deviations in the sound horizon at decoupling from that obtained in $\Lambda$CDM (see e.g.~Ref.~\cite{Amendola:1999er}). The McDE models first described in Section~\ref{McDE} (1 scalar field and 2 CDM species) give viable background cosmologies through the effect of the opposite charges and symmetric magnitudes of the CDM species~\cite{Piloyan:2014gta}. We restrict our background analysis to ensure that the relative background densities match today's values, and that the evolution moves from radiation domination, through a period of CDM domination to a final epoch of DE domination.

Subsequent to the initial submission of this thesis and the submission of \cite{ACQ} to Physical Review D we received the referees report for this paper raising questions over our treatment of the background cosmology. The point was raised, with some justification, that while we had ensured that for the models studied the various components density parameters had reached values in agreement with those today, those values themselves are derived assuming the 
$\Lambda$CDM model. This is problematic since interacting DE models such as those we investigated could lead to density parameters of the various components at the time of decoupling which vary greatly from that obtained assuming $\Lambda$CDM (see e.g.~Ref.~\cite{Valiviita:2015dfa}). This in turn then leads to different density parameters today than those derived assuming $\Lambda$CDM. See also Appendix A of~Ref.~\cite{Jennings:2009qg} for a succinct but detailed discussion of this topic. 
To ensure the validity of our results we propose confirming that the models studied are sufficiently close to $\Lambda$CDM at the time of decoupling such that their background evolutions would show negligible difference. Given the initial conditions were set such that the final density parameters corresponded to the Planck $\Lambda$CDM derived values today, this check should be sufficient. Since $\Omega_\Lambda$ is always orders of magnitude subdominant to $\Omega_M$ (the total matter density parameter) in those models we claim satisfy background constraints we shall simply compare the $\Omega_M$ obtained from these models at decoupling with that from $\Lambda$CDM to ensure no significant deviation. However, if there is a wide deviation in the energy densities at the time of decoupling from $\Lambda$CDM this would change our estimates of the current energy densities based on CMB observations interpreted using these interacting DE models, and as such would also change our values for the growth factors and $fg$. Such a possible degeneracy in the results could also remove the ability to distinguish the models studied from $\Lambda$CDM using the SKA and Euclid future survey data.

\section{Example models}
\label{PythonPert1}

\begin{figure}
\centerline{\includegraphics[angle=0,height=85mm]{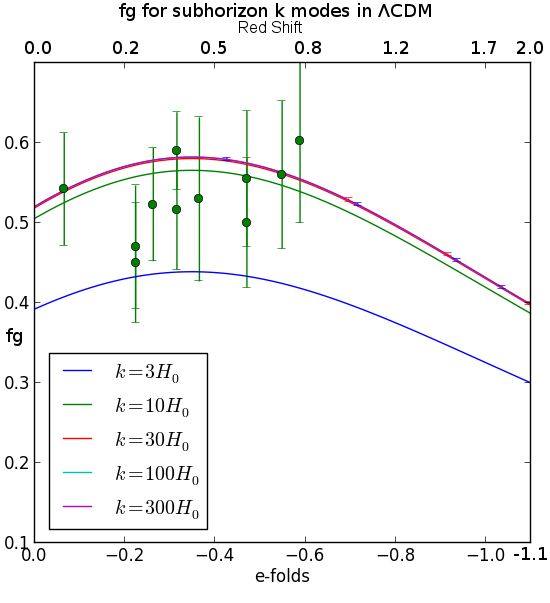}\includegraphics[angle=0,height=85mm]{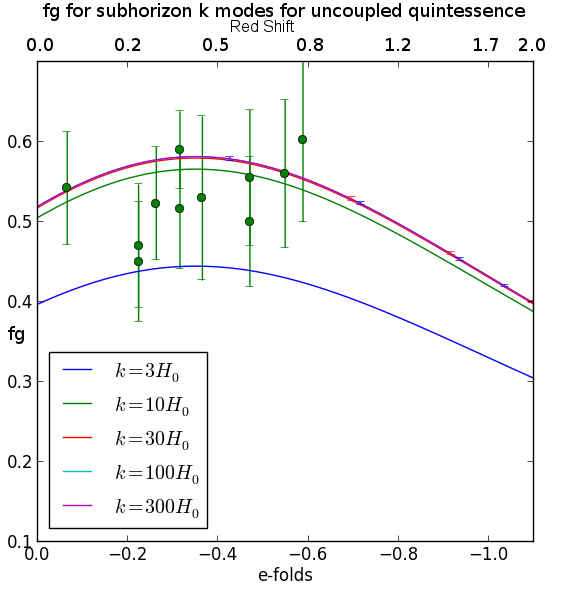}}
\centerline{\includegraphics[angle=0,height=85mm]{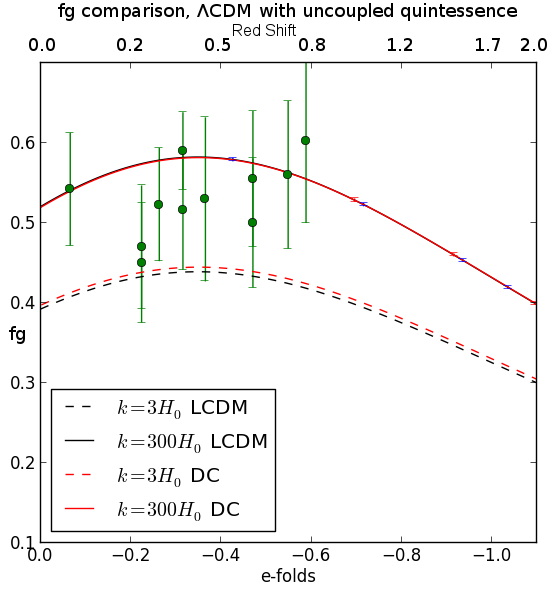}}
\caption{The top left plot shows the growth function, $fg$, on sub-horizon scales for $\Lambda$CDM, for the region of redshifts relevant for current and future surveys. The green points are observational data from 6dFG
S, LRG$_{200}$, LRG$_{60}$, BOSS, WiggleZ and VIPERS with associated errors~\cite{Macaulay:2013swa}. The red error bars are the Euclid forecasts and the blue the SKA forecasts~\cite{Raccanelli:2015qqa} applied to the $k=300H_0$ plot. The forecast error bars are approximately the line width. The top right plot shows the same for uncoupled two field two CDM species quintessence, $\lambda=0.1$. The bottom plot compares $fg$ for $\Lambda$CDM with uncoupled quintessence (DC) for $k=300H_0$ and $k=3H_0$.}
\label{fLCDMsubhorLONG}
\end{figure}

In order to compare models against the standard model, we first applied our code to produce results for the $\Lambda$CDM cosmology. Figure~\ref{fLCDMsubhorLONG} shows the results for the behaviour of $fg$ together with current observational constraints. We also applied our code to a uncoupled quintessence model with two scalar fields and two CDM species. In this case, and for all subsequent models including McDE, the potential for the scalar fields is taken to be a sum of exponentials,
\be
\label{sumofexppot}
V(\ph_1,\ldots,\ph_I) = M^4 \sum\limits_{I} e^{-\kappa \lambda_I \ph_I} \,,
\ee
where $\lambda_I$ is the slope of the potential for field $I$ and $M$ is the scale of the potential. The sum of exponentials potential was selected since it gives analytic solutions for the evolution of background quantities which in turn made setting the initial conditions more straightforward. In addition this potential also gives a simpler matrix of the derivatives of the potential in terms of the fields, which simplified the analysis. The results for uncoupled quintessence are also shown in Figure~\ref{fLCDMsubhorLONG}. We can see that for large $k$ there is a negligible difference from $\Lambda$CDM in the growth, and even for small $k$, the difference is still too small to be detectable by future surveys such as SKA and Euclid.

\subsection{Multi-coupled Dark Energy - McDE}
\label{McDE}
Next, we investigated the recently proposed subclass of coupled quintessence, McDE, as described in Refs.~\cite{Baldi:2012kt,Piloyan:2013mla,Piloyan:2014gta}. The McDE model has two CDM species coupled to one DE scalar field. The couplings of each DM species have the same magnitude but opposite signs. In order to compare directly with the results of Ref.~\cite{Piloyan:2014gta}, we set the baryon density to zero for this model. In previous work, perturbations in this model have been studied using an approximation to the full system of equations \cite{Piloyan:2014gta,Amendola:2014kwa,AmenTsuji,Amendola:1999er}. This simplification is valid for modes on subhorizon scales and allows scalar field fluctuations to be written in terms of density perturbations. The dimensionality of the system can therefore be reduced and an autonomous system of equations formed for the density perturbations alone. We use the system of ODEs, taken from Ref.~\cite{Piloyan:2014gta}, to evolve the density perturbations. We also use the same initial conditions to generate results using our implementation of the full equations. This provides a useful examination of the applicability of the subhorizon approximation. Finally, for comparison, we produce $\Lambda$CDM results with the assumption of zero baryonic content, using the McDE subhorizon approximations equations and our full system of equations.

We take the initial conditions used in Figure 7 of Ref.~\cite{Piloyan:2014gta}. The couplings are symmetric and set to $\beta$ = $\pm 0.03$ where $\beta$ $\equiv$ $\left(\sqrt{\frac{3}{2}}\right) \C$ and $\alpha=0.12$ where $\alpha \equiv \lambda$. The potential is as \eq{sumofexppot}, for $I=1$, $\alpha=2$. The initial conditions were set non-adiabatically with $A_{IC}=2$, where  
\be
\label{adiab}
A_{IC} = \frac{\Omega_- \delta_{-i}}{\Omega_+ \delta_{+i}}\,, 
\ee
and $A_{IC}$ is the measure of the deviation from adiabaticity, `$-$' denote the negatively charged CDM species and `$+$' the positively charged. One further parameter is the asymmetry between these two species, $\mu$, and is defined
\be
\label{mu}
\mu=\frac{\Omega_+ - \Omega_-}{\Omega_+ + \Omega_-}\,. 
\ee
Initially $\mu=0.5$, however we found the final results to be insensitive to this initial condition. Once again we generated plots using the reduced system and the full equations for a range of $k$s. For quantities which were absent in Ref.~\cite{Piloyan:2014gta}; radiation perturbations, perturbations to the scalar field, these were initially set to zero.

\begin{figure}
\centerline{\includegraphics[angle=0,height=85mm]{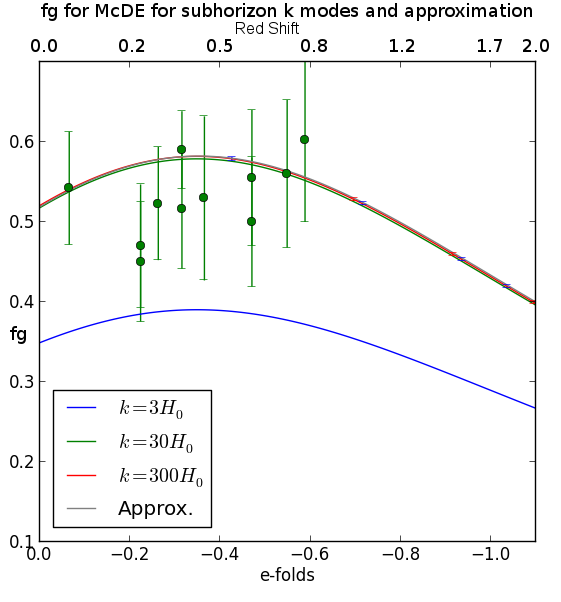}\includegraphics[angle=0,height=85mm]{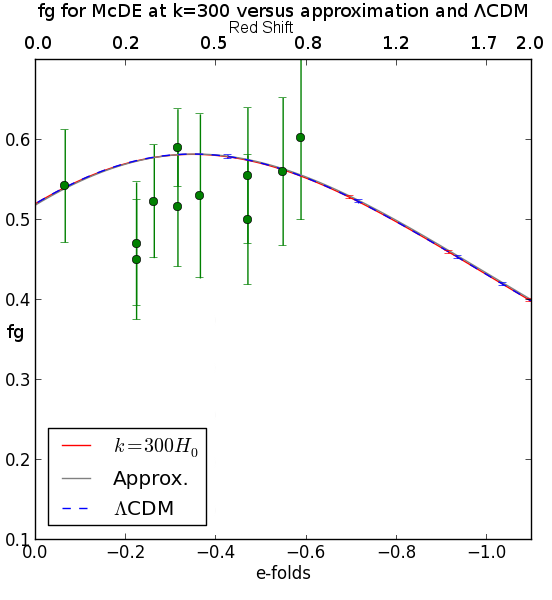}}
\centerline{\includegraphics[angle=0,height=85mm]{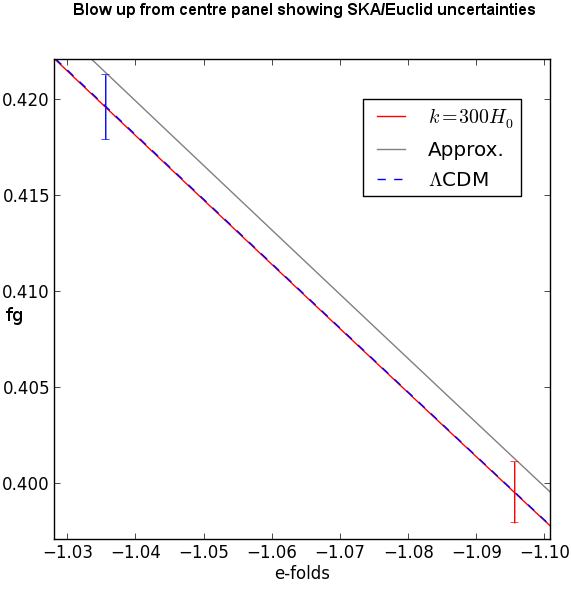}}
\caption{The top left hand panel shows $fg=\frac{\delta'}{\delta_0}$ for McDE with $\Omega_{\Lambda}=0.692$, no baryons, one CDM species and unperturbed radiation, $\lambda=012$, $\C=\pm0.03\sqrt{\frac{2}{3}}$. A range of subhorizon $k$ modes are shown with convergence towards a $k$ independent evolution of growth with larger $k$s. The result for the subhorizon approximation from Ref.~\cite{Piloyan:2014gta} is shown in grey. The top right panel shows $fg$ for McDE for $k=300H_0$ for the full equations,  the subhorizon approximation from Ref.~\cite{Piloyan:2014gta} and $\Lambda$CDM for $k=300H_0$. In each panel, the green points are observational data from 6dFGS, LRG$_{200}$, LRG$
_{60}$, BOSS, WiggleZ and VIPERS with associated errors~\cite{Macaulay:2013swa}. The red error bars are the Euclid forecasts and the blue the SKA forecasts~\cite{Raccanelli:2015qqa} applied to the $k=300H_0$ plot. The bottom panel reproduces a magnified area of the lower panel, showing that the approximation results differ from the full equations by more than the uncertainties.}
\label{fgLCDMBALDI}
\end{figure}

The results are presented in terms of the evolution of $fg$ and are shown in Figure~\ref{fgLCDMBALDI}. For the simplified $\Lambda$CDM model, with the baryon content set  to zero,  and the radiation unperturbed (initially for our full code, while radiation perturbation equations are not included in the subhorizon approximation) the results are shown in Figure~\ref{fgLCDMBALDI} together with present and future constraints.
Examining this figure, we see that for the largest $k$ modes the results converge with the result generated using the subhorizon approximation. It should be noted however that there is a noticeable difference in the evolution of growth between the different $k$ modes down to the scale of $k=300H_0$, and as such the subhorizon approximation is masking this $k$ dependence over this range of $k$s.

As in Ref.~\cite{Piloyan:2014gta} we found that the evolution provided by the subhorizon approximation gives an evolution for $fg$ close to $\Lambda$CDM but with a deficit at lower red shifts. The larger $k$ modes have mostly converged with the approximation, however, there is a small deviation such that at late times $fg$ is closer to $\Lambda$CDM than the approximations. As with all full equation results produced, the growth results are converging with increasing $k$, as expected. However, even at scales of $k=300H_0$ the small scale approximation appears insufficient for this model, even for the conservative predicted precision for SKA and Euclid measurements. We can see in the right hand plot of Figure~\ref{fgLCDMBALDI} that the approximation deviates from the full equations results by more than the predicted observational precision at these higher redshifts. Additionally, for the full equations at $k=300H_0$ the evolution of $fg$ for McDE and $\Lambda$CDM models can not be distinguished from the predicted observational precision.

\subsection{Assisted Coupled Quintessence}
\label{ACQ}
\subsubsection{Transient Matter Domination}
\label{CQSS}

\begin{figure}
\centerline{\includegraphics[angle=0,width=85mm]{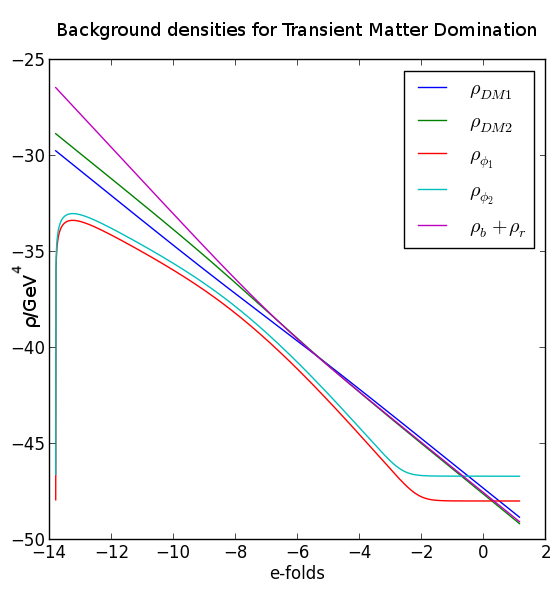}\includegraphics[angle=0,width=85mm]{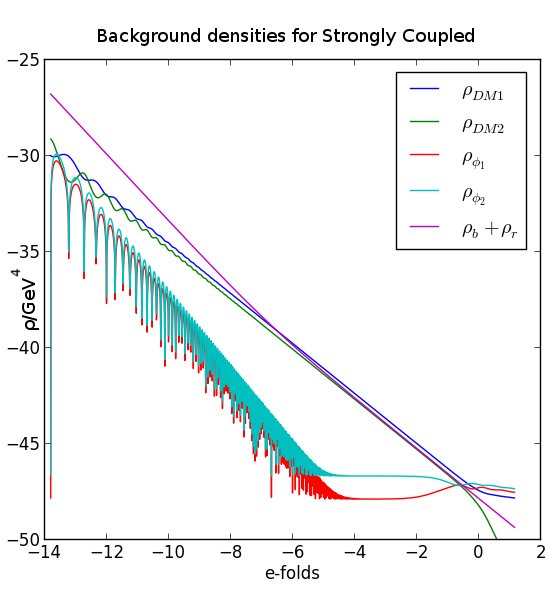}} 
\caption{The left hand plot shows the evolution of the background densities of components for the transient matter domination ACQ model. The scale is logarithmic. Subscript $b$ denotes baryons, subscript $r$ denotes radiation. Couplings, $\C_{11}=-0.2$, $\C_{12}=0.4$, $\C_{21}=-0.3$, $\C_{22}=0.6$. Slopes for the potentials, $\lambda_1 = \lambda_2 = 0.1$. The right hand plot shows the evolution of the background densities of components for the strongly coupled matter dominated coupled quintessence model. Subscript $b$ denotes baryons, subscript $r$ denotes radiation. Couplings, $\C_{11}=-20$, $\C_{12}=40$, $\C_{21}=-30$ and $\C_{22}=60$. Slopes for the potentials, $\lambda_1 = \lambda_2 = 10$}
\label{backmatdom}
\end{figure}

Next we considered the ACQ model introduced in Ref.~\cite{Amendola:2014kwa}. The model contains two pressureless dark matter fluids coupled to two scalar fields. Initially we choose small couplings ($\C_{11}=-0.2$, $\C_{12}=0.4$, $\C_{21}=-0.3$ and $\C_{22}=0.6$) and small slopes for the potentials, $\lambda_I$, ($\lambda_1 = \lambda_2 = 0.1$). The evolution of the background densities for this model is shown in the left hand panel of Figure~\ref{backmatdom}. These small couplings give rise to a tracking behaviour, by which we mean that the scalar fields densities between $e$-folds of within the interval  $-13$ and $-3$ approximately follow the evolution of the energy densities of the other components. While this may not alleviate the coincidence problem and instead restate the problem in terms of the value of the potential/effective potential minimum, some may find this aesthetically more acceptable given the additional dynamism in the field and a more ``natural" interacting scalar fields driven explanation.
This model also still gave a transition to a near constant energy density for the scalar fields at late times and domination of the scalar field energy densities at late times, as required to produce similar background behaviour to $\Lambda$CDM.

\begin{figure}
\centerline{\includegraphics[angle=0,height=85mm]{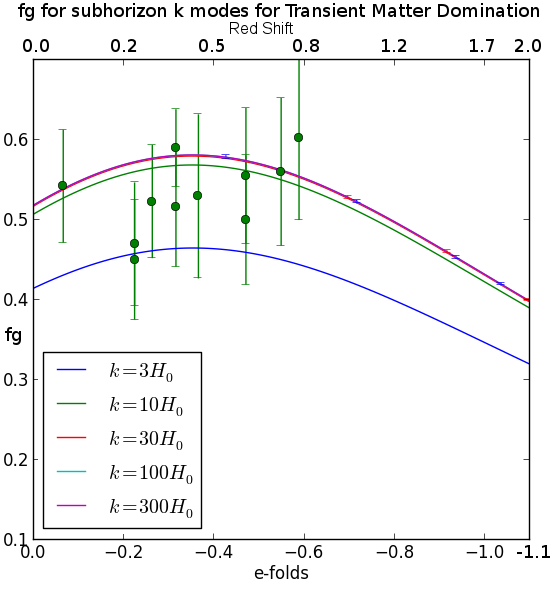}\includegraphics[angle=0,height=85mm]{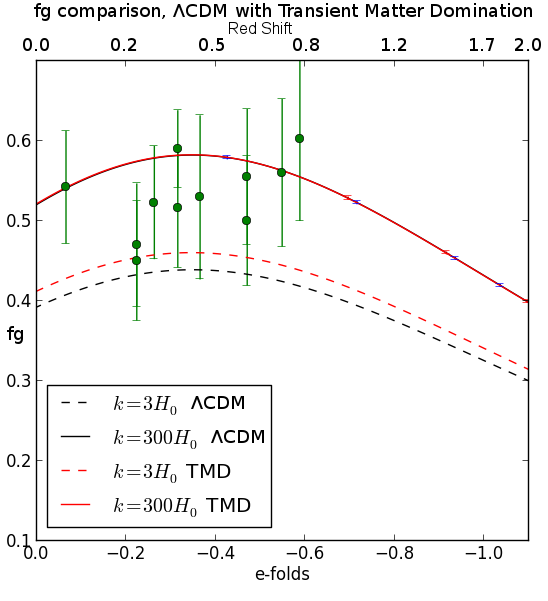}}
\centerline{\includegraphics[angle=0,height=85mm]{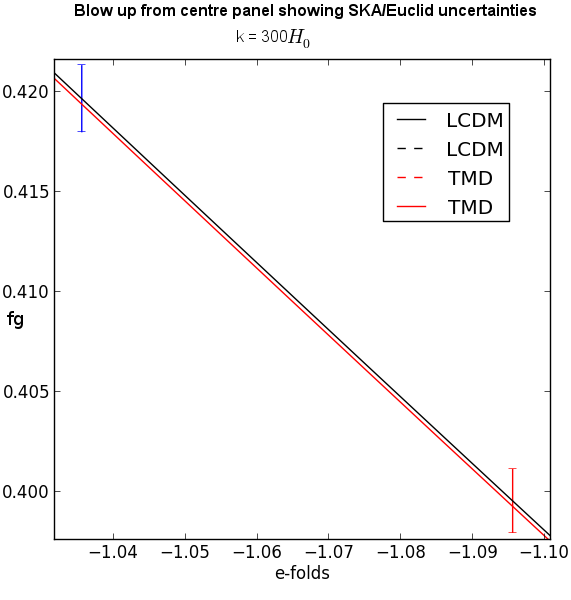}}
\caption{The top left plot shows the growth function, $fg$, sub-horizon scales, for the transient matter domination ACQ model, for the region of redshifts relevant for current and predicted future surveys. Couplings, $\C_{11}=-0.2$, $\C_{12}=0.4$, $\C_{21}=-0.3$, $\C_{22}=0.6$. Slopes for the potentials, $\lambda_1 = \lambda_2 = 0.1$.  The green points are observational data from 6dFGS, LRG$_{200}$, LRG$_{60}$, BOSS, WiggleZ and VIPERS with associated errors~\cite{Macaulay:2013swa}. The red error bars are the Euclid forecasts and the blue the SKA forecasts~\cite{Raccanelli:2015qqa} applied to the $k=300H_0$ plot. The top right plot compares the $fg$ between $\Lambda$CDM and transient matter dominated model (TMD) for $k=300H_0$ and $k=3H_0$. The bottom panel zooms in on the top right panel to show the results versus the SKA/Euclid uncertainties for $k=300H_0$.}
\label{fCQMDsubhorLONG}
\end{figure}

The right hand panel of Figure~\ref{fCQMDsubhorLONG} is the evolution of $fg$ for $k=300H_0$, and shows the conservative predicted observational precision would not be enough to distinguish between this model and $\Lambda$CDM. However, if optimal performance were achieved leading to an order of magnitude improvement in the observational uncertainties this could be sufficient to distinguish the two models.

\subsubsection{Strongly Coupled Matter Domination}
\label{CQNO}
Taking again the same setup, next we choose the couplings  $\C_{11}=-20$, $\C_{12}=40$, $\C_{21}=-30$ and $\C_{22}=60$ and the slopes for the potentials  $\lambda_1 = \lambda_2 = 10$. The background evolution of this system was also studied in Ref.~\cite{Amendola:2014kwa} and can be seen in the right hand plot in Figure~\ref{backmatdom}. The initial oscillations in the scalar fields are caused by the initial conditions for the fields, which are set above the minimum of the effective potential and subsequently oscillate around this minimum. The average behaviour of the scalar fields' energy densities is similar to the transient matter domination model. Initially there is a nearly tracking period at early times, followed by transition to nearly constant energy densities for the fields. Unlike the transient matter domination model, one of the CDM fluids then scales with the scalar fields' energy densities as shown in the right panel of Figure~\ref{backmatdom}. Although there is oscillatory behaviour at early times in the growth factor it does not exceed unity, and the average behaviour is very similar to that of the weaker coupled transient matter dominated model. As such the model  is consistent with present observations.

\begin{figure}
\centerline{\includegraphics[angle=0,height=85mm]{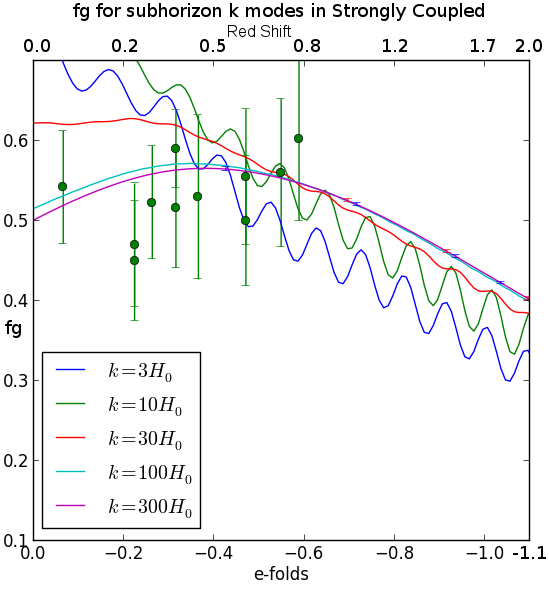}\includegraphics[angle=0,height=85mm]{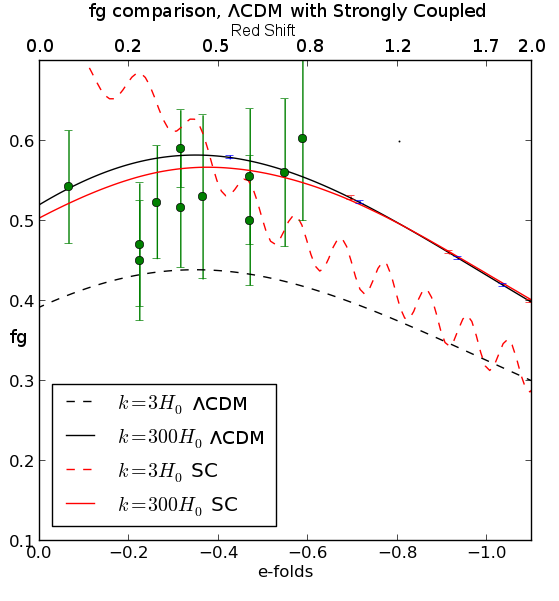}} 
\caption{The left plot shows the growth function, $fg$, sub-horizon scales, for strongly coupled matter dominated ACQ model, for the region of redshifts relevant for current and future surveys. Couplings $\C_{11}=-0.2$, $\C_{12}=0.4$, $\C_{21}=-0.3$ and $\C_{22}=0.6$. Slopes for the potentials, $\lambda_1 = \lambda_2 = 0.1$. The right hand plot compares the $fg$ between $\Lambda$CDM and the strongly coupled model (SC) for $k=300H_0$ and $k=3H_0$.}
\label{NOfsubhorlong}
\end{figure}
\subsubsection{Scaling Solution}
\label{CQScaling}

\begin{figure}
\centerline{\includegraphics[angle=0,height=85mm]{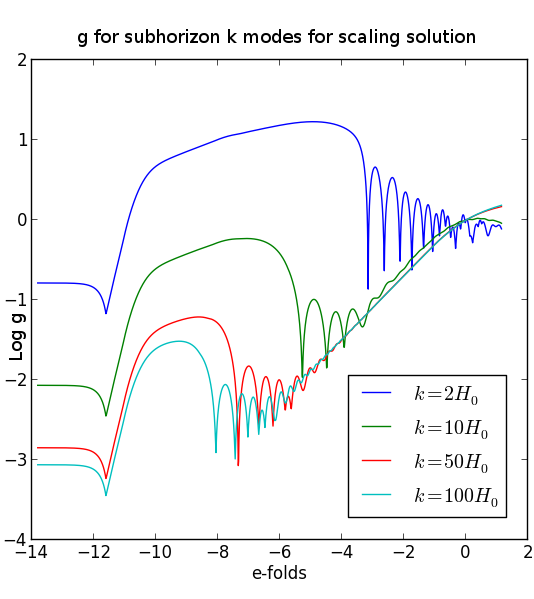}} 
\caption{The plot shows the log of growth factor, g, for scaling solution ACQ model, for subhorizon $k$ modes.}
\label{Scalingg}
\end{figure}

As a second example we followed Ref.~\cite{Amendola:2014kwa}, and considered the same setup and potential, but chose couplings which give rise to a scaling behaviour. The resultant system is, however,  not consistent with observations. It even lacks dark matter domination at earlier epochs. In this example $\C_{11}=90$, $\C_{12}=-8$, $\C_{21}=-63$ and $\C_{22}=-10$ and  the slopes of the potentials were taken to be $\lambda_1 = 10 , \lambda_2 = 5.4$. For this example we calculated the growth factor, $g$, shown in Figure \ref{Scalingg}. It can clearly be seen that it becomes greater than unity on subhorizon scales, although less pronounced with increasing $k$, showing this model to be unrealistic at both the background and perturbed level.

\subsubsection{Exploration of Potential Slope Space for Strongly Coupled Matter Domination}
\label{Lspaceexplore}
\begin{figure}
\centerline{\includegraphics[angle=0,height=85mm]{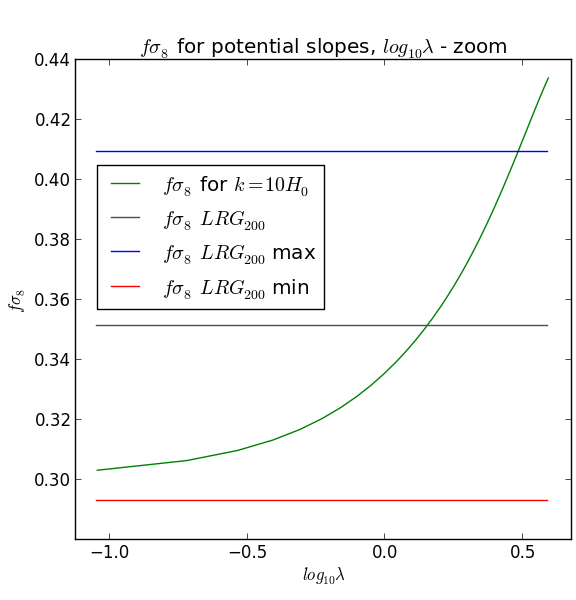}} 
\caption{$f\sigma_8$ for the matter dominated model with varying slopes for the potentials, $\lambda$. The wavenumber was set to $k=42H_0$ for these runs. Couplings, $\C_{11}=-20$, $\C_{12}=40$, $\C_{21}=-30$, $\C_{22}=60$. The observational values with uncertainties used for comparison were those from LRG$_{200}$, for $z=0.25$. The plot is a subsection from a region of $\lambda$ parameter space from $\lambda=10$ down to $\lambda=0.01$ where the results are consistent with observations.}
\label{NOslopesfs8COARSE}
\end{figure}

We now explore how changes in the slopes of the potentials (the $\lambda_I$ terms in \eq{sumofexppot}) in the matter dominated model affects the cosmology. Since, for the couplings in the strongly coupled model, the original large value of the slopes, $\lambda_1 = 10$, $\lambda_2 = 10$ produced excessive growth, we investigated the slope parameter space. This was done from $\lambda_I=10$ down to $\lambda_I=0.01$. This region including observationally consistent models is shown in Figure~\ref{NOslopesfs8COARSE}.
In producing this figure, the wavenumber of $k=42H_0$ was selected since it is the smallest $k$ mode for which SKA is predicted to still attain its highest precision~\cite{Bull:2015lja}. The LRG$_{200}$ data set was selected simply to serve as an example for comparison (see Section~\ref{Obs} for more details on  observations used for comparison). Different data sets would move the value of $f\sigma_8$ slightly, and alter the range of the error bars. There is a range of slopes for which these models not only gave a realistic background cosmology but also gave growth consistent with observations. In this region the parameter values are at least an order of magnitude smaller than the original values used. The background cosmologies for this region are very close in behaviour to Figure~\ref{backmatdom}. For slopes much smaller than $\lambda=0.01$ the potential is becoming increasingly flat and the results become noise dominated. As such they were excluded from our analysis.

\subsubsection{Exploration of Couplings Space for Strongly Coupled Matter Domination}
\label{Cspaceexplore}

For completeness a coarse exploration of the full parameter space of couplings was conducted and the growth function calculated. The range of couplings investigated was from $-50 \leq \C \leq 50$ with a stepping of $10$. The slopes for the potentials and initial conditions were left as before i.e. $\lambda_1 = \lambda_2 = 10$. For the portions of coupling space where the couplings satisfied the background constraints for these models, all exhibited excessive growth in the perturbations.\\

\begin{figure}
\centerline{\includegraphics[angle=0,height=85mm]{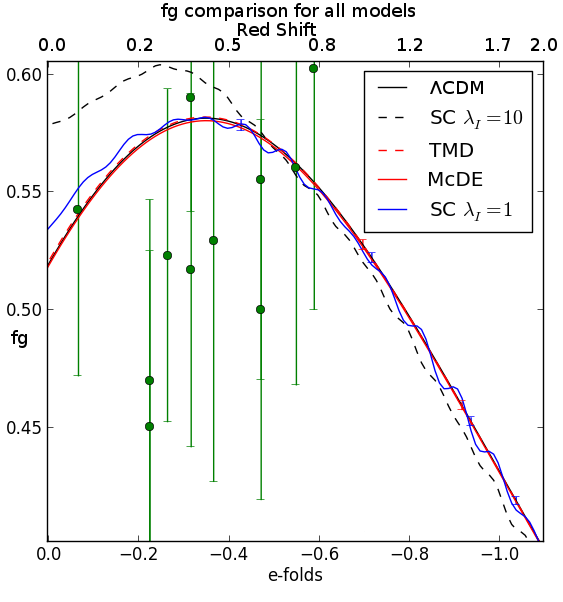}\includegraphics[angle=0,height=85mm]{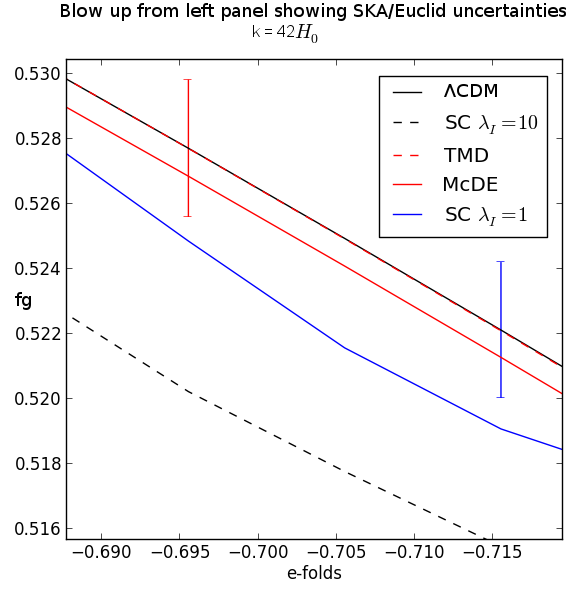}} 
\caption{This plot compares the $fg$ between $\Lambda$CDM and the strongly coupled model (SC) for both $\lambda_I=10$ and $\lambda_I=1$, the transient matter dominated model (TMD) and McDE model. All models are evaluated at $k=42H_0$. The insert zooms in on an example region in redshift space where future surveys should measure $fg$ sufficiently accurately to compare different model predictions.}
\label{allfg}
\end{figure}

Finally in Figure~\ref{allfg} we show $fg$ for a sample of the models studied against $\Lambda$CDM compared with the SKA and Euclid predicted precisions. This was carried out for mode $k=42H_0$ as it corresponds to the largest scale for which the highest predicted precision should be achieved for SKA~\cite{Bull:2015lja}. We can see that unless the best possible predicted precision is achieved it may still be hard to distinguish models with small couplings and slopes from $\Lambda$CDM. However, models with larger couplings should be easily identified. The strongly coupled model with $\lambda_I = 1$ was chosen since it lay within one of the viable regions discovered in Subsection~\ref{Lspaceexplore}. For this model it is clear that this would be distinguishable from $\Lambda$CDM given even the conservative predicted precision for SKA and Euclid. Therefore, there is a region of parameter space between the transient matter domination parameters and the strongly coupled parameters we initially tested in which subregions satisfy both background constraints and give growth results distinguishable from $\Lambda$CDM by future surveys, as the strongly coupled model does.

\section{Discussion of ACQ Results}
\label{Conclusion1}

In this chapter we have presented the full equations for perturbations in ACQ models, produced a numerical package to evolve these perturbations, \PY, and used this package to compare a set of example models with observations. We found that the longitudinal gauge, often employed in previous studies of less general systems, is not ideal for the numerical evolution of the full system, and we therefore used the flat gauge.

We found that there are examples of ACQ models which lie within current observational bounds, however,  distinguishable from $\Lambda$CDM models with future surveys such as Euclid and SKA, as they will attain a precision in $fg$ at the percent level or better \cite{Raccanelli:2015qqa}. On the other hand, we also found examples such as the strongly coupled model defined in Ref.~\cite{Amendola:2014kwa}, were $fg$ is incompatible with current observations, ruling out the model. This confirms  the conclusion in Ref.~\cite{Amendola:2014kwa}, that while ``large'' couplings might give a realistic background model, the perturbations experience excessively strong growth (or damping) and are, therefore, unrealistic. However, we found that it did not require both the couplings and the slopes to be reduced simultaneously in order for a region of viable background and perturbed cosmologies to be recovered, as discussed in Subsection~\ref{Lspaceexplore}, since when $\lambda \lesssim 2$ this leads to a viable parameter space region.

We have found for the McDE model, and the transient matter dominated case for the ACQ models studied, that they give realistic background cosmologies while apparently exceeding the allowed coupling strength for single field ACQ, $\C \lesssim 0.1 \sqrt{\frac{2}{3}}$ (see e.g.~Ref.~\cite{Amendola:1999er}). This difference in behaviour between single field (and single CDM species) and multiple CDM species models results from the relative signs of the couplings. In Ref.~\cite{Piloyan:2014gta}, the McDE model with couplings significantly greater than $0.1 \sqrt{\frac{2}{3}}$ gave rise to viable background and perturbed cosmologies. This is attributed to the unique way in which the CDM species are oppositely charged with respect to the DE scalar field (couplings are also of the same magnitude). In our ACQ models each CDM species has an opposite charge relative to each scalar field i.e.~CDM species 1 has a negative coupling to scalar field 1 while CDM species 2 has a positive coupling, and similarly for scalar field 2. Although the couplings are no longer symmetric in magnitude, this partial balance of charge still has a similar effect as in McDE, both in giving viable background cosmologies and in controlling the growth of structure. However, of the models studied only the transient matter dominated model satisfied both the background evolution and the evolution of growth through $fg$ for low redshift.

Finally, we have also addressed the question of the applicability of the large $k$ approximation, and investigated at which scales it may be considered a good approximation. The deviation of the full equation results for large $k$ modes from the approximation is frequently greater than the experimental uncertainty in future surveys. In Section~\ref{McDE} we showed that using a subhorizon approximation gave a difference in results for growth from the full equations which would be larger than the predicted observational precision for SKA and Euclid. The approximation already deviates from the full equations by more than the predicted precision of SKA~\cite{Bull:2015lja} at $k=300H_0$ and becomes progressively worse towards $k=42H_0$, the boundary for which SKA is predicted to have the highest precision. Hence results from the full equations should be used for comparison with future observations instead of those obtained using the approximation. This is therefore an important aspect to take into account in the analysis of large scale structure from near future experiments.


\chapter{\PY~- Assisted Coupled Quintessence Linear Perturbation Python Code}
\label{ch:5}


\section{Introduction}
\label{Intro}

In this chapter we discuss the construction and operation of the \PY~ Python code used to obtain the results discussed in Chapter 4. \PY~is designed to evolve linearly perturbations to coupled quintessence models with multiple CDM fluid species and multiple DE scalar fields.

The code allows two main approaches to investigating the viability of ACQ and related models. Firstly, the ``stability" of these perturbations may be investigated. Here we use the word ``stability" rather loosely to mean the perturbations might experience runaway growth or ``explode"; models in which the perturbations have runaway growth may be excluded. In addition models where the growth factor, $g$, exceeds unity for subhorizon $k$ modes may also be excluded (see e.g. Subsection~\ref{CQScaling}). Secondly, the power spectra or growth functions may be calculated to compare with observations e.g. $k$ dependent $fg$ as in Section~\ref{PythonPert1}. They may either prove to be outside current observational bounds (see e.g. Ref.~\cite{Macaulay:2013swa}) such as the strongly coupled ACQ model in Subsection~\ref{CQNO}, or provide deviations from the standard $\Lambda$CDM model of cosmology which would be detectable in future surveys e.g. SKA~\cite{Raccanelli:2015qqa} or Euclid~\cite{Kitching:2015fra} as seen in Section~\ref{Conclusion1}.\\
The code is designed to be flexible, with the form of the potential and other model specific parameters set in the MODEL.py module (while the equations within the code allow for more exotic forms of dark matter which might have non-zero equations of state, as in \emph{warm dark matter} (WDM), see e.g. Ref.~\cite{Pagels:1981ke}). For any given model the code either produces directly, or allows the calculation of, quantities such as such as $g$, the evolution of the density contrast normalised by today's value i.e. $\frac{\delta}{\delta_0}$, or $f$, the e-fold derivative\footnote{As before in Subsection~\ref{subsec:InfSRA}, e-fold is the logarithmic measure of time in terms of the expansion of the universe, as in \eq{Nloga} such that $N = \ln (\frac{a}{a_0})$, where $N$ is the number of e-folds, $a$ is the scale factor at a given time and $a_0$ is the scale factor today.} of the density contrast scaled to the density contrast i.e. $\frac{\delta'}{\delta}$ may be calculated from the data output and compared with, for example, $f\sigma_8$ measurements (see e.g. Ref.~\cite{Raccanelli:2015qqa}) or $fg$ (see e.g. Refs.~\cite{Piloyan:2014gta,ACQ}) as seen in Section~\ref{Obs}. The evolution of the density perturbations for the CDM species is produced directly by the code, which allows the power spectrum for the density perturbations to be generated by running the code for a range of wavenumbers, $k$. The code gives results in flat gauge but these may be converted into whichever gauge is required for a given task, or for comparison with existing literature e.g. the frequently used longitudinal or Newtonian gauge~\cite{Amendola:2014kwa,Raccanelli:2015qqa,Padmanabhan:2006kz}.\\
An advantage of this code is that it is relatively small and therefore fast. It generates observables which allow the ruling out of regions of parameter space (or potentially a given model entirely) as with, for example, the slopes and couplings parameter space investigation in Subsection~\ref{Lspaceexplore} and Subsection~\ref{Cspaceexplore} respectively, before embarking on more detailed analysis using larger codes with broader functionality e.g. CLASS~\cite{Blas:2011rf} or CAMB~\cite{Lewis:1999bs}.

The rest of this chapter is set out as follows; Section \ref{Req} outlines the system requirements for the {\sc{Pyessence}} package. Section \ref{Modules} details the variable names and other code specific features, as well as listing each of the modules; CONSTANTS.py, BACKGROUND.py, PERTURBED.py and MODEL.py. In the PERTURBED.py Subsection we also discuss problems encountered while constructing the code and also some of the reliability testing. Finally, Section \ref{Examples} details some further example applications of the {\sc{Pyessence}} code complementing those from Chapter 4, which were used while testing the code in development. 

\section{Requirements}
\label{Req}

{\sc{Pyessence}} was written and tested using Python 2.7.3 and should therefore work on higher versions of Python. It may work with earlier versions but this has not been tested.

The core modules use Numpy, and these were developed and tested using v1.6.2, and also use Scipy, using v0.10.1.

The various {\sc{Pyessence}} application examples e.g. EXAMPLE1.py, use Matplotlib to demonstrate plotting of results but this is not required for the core modules.

\section{Modules}
\label{Modules}

The variable and function labels are listed in a table in the README.txt file. The variables are stored in an array labeled $\rm{In[x]}$, where $\rm{x}$ runs from zero to $\rm{5+3A+4I}$ where $\rm{I}$ is the number of scalar fields and $\rm{A}$ is the number of CDM fluids as defined by the dimensions of the couplings matrix.

The MODEL.py is model specific and would therefore need configuring for each model studied. It contains the matrix for the couplings as an array labelled $\rm{C}$ in the code, corresponding to $\C$ in the equations below and in Chapter 4. The value of $k$ is also set here, however, this may be overridden by the python module constructed to call the {\sc{Pyessence}} modules if, for example, stepping through $k$ space is required e.g. constructing power spectra. To explore the range of viable couplings for a given model, or the potential slopes give in the $\rm{L}$ matrix in the code, the $\rm{C}$ or $\rm{L}$ matrix values may be overridden in a similar way, as in Subsections~\ref{Cspaceexplore} and~\ref{Lspaceexplore}. We shall return to these settings in more detail in Section \ref{MODEL}.

\subsection{CONSTANTS.py}
\label{CONSTANTS}

This module is the smallest and simply contains the constants used within the {\sc{Pyessence}} package. Any additional constants required if the code is modified should be put here. It contains the gravitational constant, $G$, 
\be
\label{kappa}
\kappa=(8 \pi G)^{\frac{1}{2}} , 
\ee
the Hubble parameter, $h$, and the critical density today,

\be
\label{critdens}
\rho_{c(0)} (=\frac{3 H_0^2}{\kappa^2}). 
\ee

\subsection{BACKGROUND.py}
\label{BACKGROUND}

This module contains the background equations. In this subsection we list the equations coded in the module in the order in which they appear. When  listed in previous chapters we refer back. The only equations seen in this subsection are those included in the code in a specific form, or which were included in the code but were not part of the system integrated. These non-integrated equations will be those useful for calculating or plotting other quantities. A bar is used to denote background quantities. Please note, all the pressure terms in the equations throughout are replaced with appropriate equations of state terms within the code. The non-integrated equations are below. The first is \eq{FEIntDEBack}, the Friedmann equation, for the fluids and fields. The coordinate time derivative is denoted by ``dot". Derivatives with respect to fields are denoted by a ``comma". The time derivative of the Hubble parameter is given by,
\be
\label{FEdotIntDEBack}
\dot{H} = \frac{\kappa^2}{6 H} \left[\sum\limits_{\alpha} \dot{\bar{\rho}}_{\alpha} + \sum\limits_{I} (\dot{\bar{\ph}}_{I} {\ddot{\bar{\ph}}_{I}} + \dot{\bar{\ph}}_{I} V,_{\ph_{I}}) \right] .
\ee
The energy density of the scalar fields are given by,
\be
\label{FieldDens}
\sum\limits_{I} \bar{\rho}_{\ph_{I}} = \sum\limits_{I} \frac{\dot{\bar{\ph}}_{I}^2}{2} + V  .
\ee
The integrated equations are as follows. The evolution equation for the radiation energy density can be taken from \eq{Conservation} with $w=\frac{1}{3}$, and is given by,

\be
\label{dotrhorad}
\dot{\bar{\rho}}_r = - 4 H \bar{\rho}_r . 
\ee
The evolution equation for the baryon energy density can be taken from \eq{Conservation2}, and is given by,
\be
\label{dotrhobar}
\dot{\bar{\rho}}_b = - 3 H \bar{\rho}_b .
\ee
Next in the module is the evolution equation for the CDM fluids energy densities; \eq{IntDEMatBack2}.

The subsequent equations simply equate the functions for the time derivatives of the scalar fields, labelled $\rm{dx}$ in the code, with the variables for the same, labelled $\rm{y}$, within the code.
Finally, the second time derivatives of the scalar fields are given by \eq{IntDESFBack2}.

\subsection{PERTURBED.py}
\label{PERTURBED}

This module contains the perturbed equations. In constructing the \PY~code we initially selected longitudinal gauge (See Appendix~\ref{Long2f2dm}) for our perturbed equations. However we discovered a numerical instability, first noticed through all available integration methods (dopri5, LSODA, vode, zvode and dop853) failing to converge using longitudinal gauge, due to the problems caused by the first term in \eq{Phiconstraint} outlined in Subsection~\ref{Fixing}. To avoid this we switched to flat gauge where the constraint equations did not suffer from this numerical instability. By incrementally increasing the relative and absolute tolerances (the $\rm{rtol}$ and $\rm{atol}$ settings for the ODE package respectively) we were able to ensure convergence occurred well before the tolerances eventually used for all the runs reproduced here and in Chapter 4; specifically, $\rm{rtol=10^{-14} , atol=10^{-14}}$.

In this subsection we list the equations coded in the module in the order in which they appear. When listed in previous chapters we refer back. The only equations seen in this subsection are those included in the code in a specific form, or which were included in the code but were not part of the system integrated. These non-integrated equations will be those useful for calculating or plotting other quantities.
The non-integrated equations are below.
The first is the constraint for the metric potential, $\Phi$, \eq{FlatGi0DEEFE2}.
Next is the constraint equation for $B$ \eq{FlattildeBNewVar}.
The perturbed energy density for the scalar fields, taken from \eq{PertSEMDE}, is given by,

\be
\label{pertphdens}
\delta {\rho}_{\ph_{I}} = - \Phi \dot{\bar{\ph}}^2_I + \dot{\bar{\ph}}_I {\dot{\delta \ph}}_I + V, _{\ph_I} \delta {{\ph}}_I .
\ee
The equation for the gauge invariant curvature perturbation, $\zeta$ (see e.g. Ref.~\cite{Carrilho:2015cma}), taken from \eq{unidenshyp3} and expressed in terms of the sums of the components is,
\be
\label{zeta}
\zeta = - \psi - H \left(\frac{\sum\limits_{\alpha} \delta \rho_\alpha}{\sum\limits_{\alpha} \dot{\rho}_\alpha} \right) .
\ee
The integrated equations are as follows. The evolution equation for the radiation perturbed energy density is taken from \eq{FlatIntDEPertEcons4} with zero couplings and $w=\frac{1}{3}$, and is given by,

\be
\label{dotpertrhorad}
\dot{\delta \rho_r} = - 4 H \delta \rho_r + \frac{4 k^2}{3 a}(\hat{v}_r - B){\bar{\rho}}_r .
\ee
The evolution equation for the baryon perturbed energy density is taken from \eq{FlatIntDEPertEcons4} with zero couplings and $w=0$, and is given by,
\be
\label{dotpertrhobar}
\dot{\delta \rho_b} = - 3 H \delta \rho_b + \frac{k^2}{a}(\hat{v}_b - B){\bar{\rho}}_b .
\ee
The evolution equation for the CDM fluids perturbed energy densities is  \eq{FlatIntDEPertEcons4}.
The evolution equation for the 3-velocity for the radiation fluid is taken from \eq{IntDEPertMomcons} with zero couplings and $w=\frac{1}{3}$, and is given by,
\be
\label{dotvrad}
\dot{\hat{v}}_r = - \frac{\Phi}{a} - \frac{\delta \rho_r}{4 a\bar{\rho}_r} .
\ee
The evolution equation for the 3-velocity for the baryon fluid is taken from \eq{IntDEPertMomcons} with zero couplings and $w=0$, and is given by,
\be
\label{dotvbar}
\dot{\hat{v}}_b = - H \hat{v}_b - \frac{\Phi}{a} .
\ee
The evolution equation for the 3-velocity for the CDM fluids is \eq{IntDEPertMomcons}.\\
Similarly to the BACKGROUND.py module, the next equation simply equates the function for the time derivative of the perturbed scalar fields, labelled $\rm{dpx}$ within the code, with the variable for the same, labelled $\rm{py}$, within the code.
Next the second time derivatives of the perturbed scalar fields is given by \eq{FlatDEPertEconsSF2}. Finally, the last function included at the end of the module, $\rm{df}$, is the array of all functions passed to the integrator, both background and perturbed.

\subsection{MODEL.py}
\label{MODEL}

This module defines the model being studied. The wavenumber, $k$, is also set here, for convenience in the code structure. Additional modules calling \PY~may override this value locally, for example if looping through $k$ values, or plotting functions from multiple saved data sets for different $k$s. Many of the parameters in the MODEL.py file included are specific to the sum of exponentials potential, \eq{sumofexppot}, used to test the \PY~code and give the first scientific results. This is simply one example of a possible potentials, and the example MODEL.py contains parameters set specifically for this example potential. These would need to be altered to configure for a different potential e.g. the derivatives of the potential. Below only general quantities will be discussed.\\
In this subsection we list the functions and variables coded in the module. $\rm{C}$ is the array of the couplings. This has been entered directly in the included MODEL.py but may be loaded from a Numpy save file created separately. For larger models with many CDM fluids and many scalar fields this would be a more practical method. The module uses the dimensions of this array to determine the number of CDM fluids, assigned to variable $\rm{A}$, and number of scalar fields, assigned to variable $\rm{I}$. It also uses these to initialise the array of all integrated variables, $\rm{In}$, the initial condition array, $\rm{f\_0}$, and the CDM fluids equations of state array, $\rm{w}$.\\
Function $\rm{V}$ is the potential for a given model. Function $\rm{VP}$ is the array of  derivatives of the potential with respect to the scalar fields. Function $\rm{VPP}$ is the array of second derivatives of the potential with respect to the scalar fields. Those included in the MODEL.py file are for the sum of exponentials potential and have been entered manually as for a two CDM fluid, two scalar field model. More generally, for many CDM fluids and many scalar fields loading $\rm{VP}$ and $\rm{VPP}$ functions from saved arrays would be more practical, as per the $\rm{C}$ array.

\section{Examples}
\label{Examples}

The example Python files included with the package distribution were created during the testing and initial use of the \PY~code. They are included to give some guidance as to how the code may be used, but are not intended to be prescriptive.

\subsection{Example 1 - Matter and Radiation only Universe}
\label{Ex 1}

\begin{figure}[H]
\centerline{\includegraphics[angle=0,width=80mm]{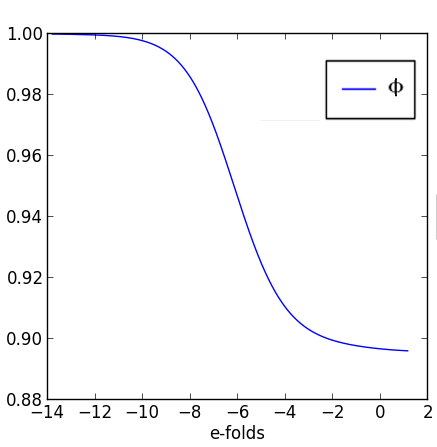}} 
\caption{Evolution of metric potential, $\Phi$, for $k=0.01k_{eq}$, where $k_{eq}$ is the wavenumber for the horizon size at the time of matter-radiation equality. $N= -6$ e-folds corresponds to the time of horizon crossing and these numerical results matched those expected c.f Ref.~\cite{Padmanabhan:2006kz}.}
\label{padphi}
\end{figure}

This file was designed to evolve perturbations to just matter and radiation, with no CDM fluids or dark energy. This was compared to the same results in Ref.~\cite{Padmanabhan:2006kz}. The file is included as EXAMPLEPAD.py in the official release on Refs.\cite{Bitbucket,Pyweb}. The corresponding model file is also included as MODELPAD.py.\\
In this subsection we list the settings and functions coded in the module in the order in which they appear. The imports section heads the file. Next, $\rm{t\_i}$ is the initial time, $\rm{t\_f}$ the final time and the $\rm{step}$ is the step size, in e-folds. After these are the initial conditions. In this example some of the perturbed initial conditions depended upon functions of the background, hence the split in the setting of the initial conditions seen in the module. For convenience an array of all times is created, $\rm{t\_out}$.\\
Next the integrator is set up. The $\rm{dopri5}$ integration method is being used in this example, as was the case for all the scientific results shown in Chapter 4, but other possible integration methods are shown ``hashed out" within the code.\\
After the integrator has finished the results are saved ($\rm{fulloutput}$), along with the time array ($\rm{t\_out}$).\\
The next section contains plotting routines. The first two sections plot the background and perturbed energy density for matter and radiation. The next two sections plot the density contrasts in the flat gauge in which the code is written and then plots the density contrasts converted to longitudinal gauge as in \cite{Padmanabhan:2006kz}. The next section plots the shear, $\sigma$ (where $\sigma = \frac{3}{4} \delta_r - \delta_m$). This is followed by the comoving curvature perturbation, $\zeta$, and then the metric potential, $\Phi$, in flat gauge, the shift, $B$, and then $\Phi$ in longitudinal gauge shown in Figure~\ref{padphi} for comparison with Ref.\cite{Padmanabhan:2006kz}.\\
The final section plots the 3-velocities for matter and radiation.

\subsection{Example 2 - $\Lambda$CDM}
\label{Ex 2}

\begin{figure}[H]
\centerline{\includegraphics[angle=0,width=100mm]{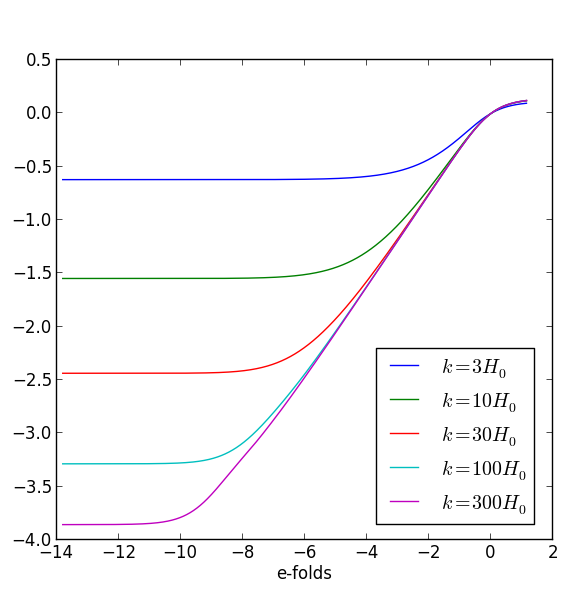}} 
\caption{Log of growth factor, g, $\frac{\delta}{\delta_0}$, subhorizon scales, for $\Lambda$CDM. $H_0$ is the Hubble constant.}
\label{gLCDMlong}
\end{figure}

The file for this example is included in the official release on Refs.\cite{Bitbucket,Pyweb} as LCDM.py. The model file corresponding to this is also included as MODELLCDM.py. The layout is much as for Section \ref{Ex 1} with the following exceptions. This code was adapted from a test for one scalar field interacting with one CDM fluid. Standard $\Lambda$CDM behaviour was then achieved by flattening the potential and setting it to the same energy density as a cosmological constant today. In this subsection we list the settings and functions coded in the module in the order in which they appear. The initial conditions for the scalar field and the scalar field velocity are then set to zero as are the field perturbation and field perturbation velocity. A small additional Python code called gANDfPLOTTER(long)LCDM.py, held in the Data folder, was used to produce plots of the growth functions in longitudinal gauge over a range of $k$s. Figure~\ref{gLCDMlong} is included as an example output for log of the growth factor, $g$ ($=\frac{\delta}{\delta_0}$).
This shows the expected behaviour for $\Lambda$CDM, specifically near constant growths over a range of e-folds prior to horizon crossing, followed by a linear increase in growths over a range of e-folds once the mode has re-entered the horizon, before the slopes begin to decrease as we enter $\Lambda$ domination.

\subsection{Example 3 - Assisted Coupled Quintessence - Transient Matter Domination}
\label{Ex 3}

\begin{figure}[H]
\centerline{\includegraphics[angle=0,width=100mm]{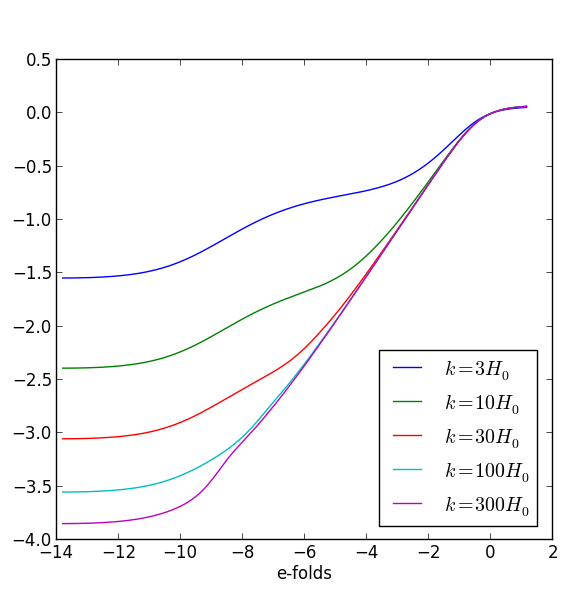}} 
\caption{Log of growth factor, g, subhorizon scales, for ACQ with couplings, $\C_{11}=-0.2$, $\C_{12}=0.4$, $\C_{21}=-0.3$, $\C_{22}=0.6$. Slopes for the potentials, $\lambda_1 = \lambda_2 = 0.1$. $H_0$ is the Hubble constant.}
\label{gCQMDlong}
\end{figure}

The file for this example is included as EXAMPLE1.py in the official release on Refs.\cite{Bitbucket,Pyweb} as MODELLCDM.py. The model file corresponding to this is also included as MODEL.py. The layout is much as for Section \ref{Ex 1} with the following exceptions. This code is for two scalar fields interacting with two CDM fluids. After initial radiation domination, an epoch of matter domination is entered which finally transitions to one of dark energy domination. Again, gANDfPLOTTER(long).py was used to produce plots of the growth functions in longitudinal gauge over a range of $k$s. Figure~\ref{gCQMDlong} is included as an example output for log of the growth factor, $g$ ($\frac{\delta}{\delta_0}$).
Compared to $\Lambda$CDM (Figure~\ref{gLCDMlong}), there are deviations from these results with additional fluctuations in the growth during matter domination. These become more pronounced with decreasing wavenumber.

\section{Concluding Remarks}
\label{Conclusion5}

As detailed above, \PY~is designed to be a fast code for quickly performing initial testing of coupled quintessence models by evolving perturbations, such as CDM density perturbations, which may be compared with observations. Models may be constrained through the regions of parameter space matching observations, or possible eliminated altogether. For example, while conducting the slope parameter space exploration (see Subsection~\ref{Lspaceexplore}), $\approx100$ runs were conducted \emph{sequentially} to produce the results shown in $\approx20$ hours. This was performed on a quad core desktop PC with Intel Core i5-2400 3.10GHz CPUs, 7.7 Gb of RAM running Debian release 7.11. The code can be optimised for parallel runs, which should significantly increase its efficiency. \PY~is released under an open source BSD license which can be found in the LICENSE.txt file included with this distribution on Refs.\cite{Bitbucket,Pyweb}.


\chapter{Conclusions and Further Work}
\label{ch:Conclusions}

In this chapter we summarise the work contained in this thesis, highlighting key methods and results. We will also look at possible directions for future work from a theoretical perspective. Finally we shall give an outlook to further work in the field in the light of future observations.

\section{Summary}
\label{sec:finalsummary}

The research in this thesis, conducted using CPT at linear order and shown in Chapters 3, 4 and 5, highlights the importance of CPT as a tool for studying the universe. In Chapter 3 we applied techniques previously used in standard flat FRW models - to construct gauge invariant and conserved quantities - to inhomogeneous cosmologies, specifically LTB and Lema{\^\i}tre. We constructed a gauge invariant curvature perturbation, $\SMTP$ (or Spatial Metric Trace Perturbation), in LTB and Lema{\^\i}tre cosmologies and also examined its behaviour in standard flat FRW cosmology. Specifically we found it was conserved on all scales in LTB and FRW, while only being conserved on large scales in Lema{\^\i}tre. With two gauge fixing conditions, specifically constant density and comoving hypersurfaces, $\SMTP$ is best suited to numerical simulations of the evolution of density perturbations i.e. structure in the universe, where it can provide an analytical check against these numerical results. This is because both conditions must be known at one time such that they can then be analysed in combination at another time. This is would not be possible observationally since even if density and velocity perturbations were known at the current time, $\SMTP$ would only reveal their combined value at early times, which could not be separated observational. However in numerical simulations both conditions are known at the start and end of the simulations and as such $\SMTP$ provides a consistency check in such simulations.

In Chapters 4 and 5 we derived the full system of evolution equations for linear perturbations in ACQ cosmologies before fixing the gauge (longitudinal, synchronous and flat). We used equations in flat gauge for a Python package,~\PY, designed to quickly investigate the evolution of these perturbed quantities. We used the code to evolve the CDM density perturbations in order to then calculate the growth of density perturbations, through $f$, $g$, $fg$ and $f\sigma_8$, which we could then compare with observations. We also compared the ACQ models and McDE against standard $\Lambda$CDM. We found which models were inconsistent with current observations, specifically the ACQ scaling (Subsection~\ref{CQScaling}) and strongly coupled (Subsection~\ref{CQNO}) solutions. We also found models which would be indistinguishable from $\Lambda$CDM even with future observations, namely McDE (Subsection~\ref{McDE}) and the transient matter domination ACQ solution (Subsection~\ref{CQSS}). A region of potential slope and coupling parameter space in which ACQ models would match background and current perturbed observations, but which also should have growth distinguishable from $\Lambda$CDM by future surveys was identified in Subsection~\ref{Cspaceexplore}. It was shown in Subsection~\ref{McDE} that the small scale approximation would be inadequate to describe growth when measured in future surveys, given the results they produce differ from the full equations by more than the predicted observational precision.

\section{Future Work}
\label{sec:Future work}

LTB as a global cosmology, describing the universe on the largest scales, has problems matching all observations simultaneously i.e. e.g. Baryon Accoustic Oscillations (BAOs) and supernovae data (see e.g.~Refs.~\cite{Clarkson:2012bg,Vargas:2015ctw}). Although there is still some recent ongoing work in the field (see e.g.~Ref.~\cite{I.:2016gam}) LTB is most useful as a toy model. However, though LTB is not a viable cosmology on the largest scales, it can be used when modelling large voids (see e.g.~Refs.~\cite{Meyer:2014qla,Sussman:2015bea}). Recent research,~Refs.~\cite{Das:2015vda,Aragon-Calvo:2016vye}, also considers structure growth and galaxy formation in large voids from an observational perspective. If a void or overdensity on a sufficiently large scale were discovered, such that linear order CPT would be applicable, understanding the behaviour of perturbations in these models, and therefore applying $\SMTP$ to numerical simulations in this area could be an avenue for future research.\\
Other inhomogeneous spacetimes e.g. Lema{\^\i}tre, while perhaps unlikely to provide a better observational match than $\Lambda$CDM in flat FRW, are still interesting to construct gauge invariant and conserved quantities in. These may in turn provide useful checks for numerical work conducted in this area. We have already constructed $\SMTP$ for Lema{\^\i}tre and the future applications described above for LTB are equally applicable to Lema{\^\i}tre. Besides Lema{\^\i}tre, other inhomogeneous models are actively being researched, for example Szekeres (see e.g.~Refs.~\cite{Sussman:2015wna,Musoke:2015kql}). For completeness, an extension to our research would be to perturb these cosmologies and construct a $\SMTP$, or similar gauge invariant conserved quantity, which could be related to the perturbed matter content. It may also be possible to extend the work to anisotropic cosmologies such as Bianchi (for recent research see e.g.~Refs.~\cite{Reddy:2016wlp,Camci:2016yed}) or Kantowski-Sachs (for recent research see e.g.~Refs.~\cite{Camci:2016yed,Keresztes:2015pxa}.\\

The first extension to our work in ACQ should be a finer grained exploration of both the slopes and couplings parameter space to further constrain and identify the regions of observationally consistent cosmologies. Within this, the regions of parameter space where such cosmologies would be distinguishable from $\Lambda$CDM should also be determined. In addition, extending the work to include more fluids and fields would enable us to see whether this allows for larger couplings and/or slopes and yet gives growth consistent with observations i.e. whether the effect of oppositely charged CDM species increases with the numbers of species - namely by increasing the suppression of excessive growth.

We have only used one potential, the sum of exponentials potential, \eq{sumofexppot}, largely because this allows initial conditions to be set relatively simply and also simplified the matrix of derivatives of the potential with respect to the field. Other potentials should be explored, but given the almost limitless choice of potentials making a selection could be problematic. One route would be to look at assisted inflation (see e.g. Ref.~\cite{Kanti:1999vt}) and choose potentials employed there e.g. quadratic, quartic, generalised monomial potential. We might prefer potentials which give a minimum, since the field should eventually settle there giving cosmological constant-like behaviour, driving late time accelerated expansion. However this requirement may not even be necessary, since with multiple fields even potentials without a minimum, such as the exponential potential, in combination can produce an effective minimum, which has the same effect - the field will come to rest there and drive late time accelerated expansion. The space of possible potentials is therefore a promising area for future exploration.

Finally, other models of interacting dark energy should be explored beyond ACQ. Our ACQ model assumed constant couplings but we could extend this to couplings as functions. Beyond ACQ itself models include Chameleon Dark Energy (for recent research see e.g. Refs.~\cite{Elder:2016yxm,Tamanini:2016klr} and k-essence (for recent research see e.g. Refs.~\cite{Bouhmadi-Lopez:2016cja,Guendelman:2015jii}). With Chameleon Dark Energy the 
Chameleon scalar fields allow large couplings between the fields and ordinary matter as well as CDM. Screening mechanisms are then used which introduce additional terms into the Lagrangian density to mask fifth force effects - i.e. the effect of the couplings - at small scale e.g. solar system scales. With k-essence the Lagrangian density is entirely kinetic i.e. no potential, and the k-essence field or fields can take on the combined role of CDM and DE. Chameleon DE and k-essence would both require modifications to the governing equations derived for ACQ, and consequently to the \PY~code to reflect these differences, but this would then allow further research into these classes of models.

\section{Observational Outlook}
\label{sec:ObsOut}

Future galaxy surveys and observations in the context of large voids or overdensities will be most relevant to the application of LTB or other inhomogeneous cosmologies on these scales, where large cosmological scale voids or overdensities might be discovered. These surveys often overlap with those probing the nature DE since they are measuring both the structure of the universe through markers of the matter distribution such as luminous matter or gravitational lensing effects, and its growth e.g. DES~\cite{Bonnett:2015pww}, SKA~\cite{Raccanelli:2015qqa} or Euclid~\cite{Kitching:2015fra}.\\
Future surveys relevant to the nature of DE will therefore be relevant both to studying the viability of inhomogeneous cosmologies, such as those studied in Chapter 3, or the examples given in Subsection~\ref{sec:Future work} for future work, and interacting or dynamical DE models such as ACQ studied in Chapters 4 and 5. For SKA~\cite{Bull:2015lja} we saw in Chapter 4, Subsection~\ref{Obs}, that it has a predicted observational precision for growth in the range $42H_0<k<420H_0$ at $z=1$ is $1-2\%$. This seems to be true out to $z=2$, although there is some uncertainty due to increasing survey area versus less efficient foreground removal. In addition, for smaller $k$ ($k<42H_0$), where differences in growth are often more pronounced, the predicted observational precision drops to $\approx30\%$. These predicted precisions would be sufficient to distinguish some ACQ models from $\Lambda$CDM as already seen in Chapter 4. However, we can see in Ref.~\cite{Bull:2015lja} that there is already some uncertainty in these predictions. More efficient foreground removal or longer survey time could improve upon these, thereby allowing more models to be ruled out through SKA observations. We can see that the Euclid predicted observational precision given in Ref.~\cite{Raccanelli:2015qqa} is similar to SKA at most redshift where they overlap. However, it does offer higher precision in the range $1.5<z<2.0$ and extends to a higher redshift than SKA. The predicted precision of Euclid might well also be improved with increased survey time, and foreground cleaning. As such, Euclid might give us the first observations of DE beyond the $\Lambda$CDM model. Failing that, the next generation of surveys, for example the Maunakea Spectroscopic Explorer (see e.g. Ref.~\cite{Maunakea}) or \emph{Large Synoptic Survey Telescope} (LSST) (see e.g. Ref.~\cite{Bacon:2015dqe}) will hopefully achieve an order of magnitude improvement on SKA and Euclid and give us the required evidence.

\appendix

\chapter{Additional material for LTB}

In this section of the appendix we present some material that is not
essential to follow the main body the LTB research in Chapter 3. However, since it might be useful and save time in reproducing or extending some or all of the calculations, we reproduce it here.

\section{Contravariant LTB Metric Perturbations}
\label{Contmetpert}

Using the constraint \eq{MetricConstraint}, acting on the covariant perturbed metric, \eq{LTBMetricperturbations}, we get the contravariant perturbed metric components,
\begin{equation}
\label{LTBMetricperturbationsuppy}
\delta g^{\mu \nu}=\begin{pmatrix}
  2\Phi & \frac{B_r}{ \sfx } & \frac{B_\theta}{ \sfy } & \frac{B_\phi}{ \sfy \sin \theta} \\
  \frac{B_r}{ \sfx } & - \frac{2C_{rr}}{\sfx^2} & - \frac{C_{r\theta}}{ \sfx \sfy } & - \frac{C_{r\phi}}{ \sfx \sfy \sin \theta}  \\
  \frac{B_\theta}{ \sfy } & - \frac{C_{r\theta}}{ \sfx \sfy } & - \frac{2C_{\theta\theta}}{\sfy^2}  & - \frac{C_{\theta\phi}}{ \sfy^2 \sin \theta}  \\
 \frac{B_\phi}{ \sfy \sin \theta} & - \frac{C_{r\phi}}{ \sfx \sfy \sin \theta} & - \frac{C_{\theta\phi}}{ \sfy^2 \sin \theta} & - \frac{2 C_{\phi\phi}}{\sfy^2 \sin^2 \theta }
\end{pmatrix} .
\end{equation}

\section{LTB Shear}
\label{Shear}
The shear, as discussed in Subsection~\ref{subsec:LTBst}, is given by,
\be
\label{genshear}
\sigma_{\mu \nu} = \frac{1}{2} {\cal{P}}_\mu^\alpha {\cal{P}}_\nu^\beta (\nabla_\beta n_\alpha + \nabla_\alpha n_\beta) - \frac{1}{3} \Theta {\cal{P}}_{\mu \nu} .
\ee
The $t-t$ component of the shear is zero. To linear order we find that the $r-r$ component is,

\begin{eqnarray}
\label{shearrr}
\nonumber \sigma_{r r} &=& -\frac{1}{3} \sfx^2 \bigg( \dot{\psi} - 2 (1 - \Phi)(H_\sfx - H_\sfy) - 4 C_{r r} (H_\sfx - H_\sfy) - 2 \frac{B_r}{\sfx \sfy} \sfy ' - \frac{B_\theta \cot \theta}{\sfy} \\ &+& \frac{2}{\sfx} B_r ' - \frac{1}{\sfy} \partial_\theta B_\theta - \frac{1}{\sfy \sin \theta} \partial_\phi B_\phi - 3 \dot{C_{r r}} \bigg) ,
\end{eqnarray}
the $\theta-\theta$ component,

\begin{eqnarray}
\label{shearthethe}
\nonumber \sigma_{\theta \theta} &=& -\frac{1}{3} \sfy^2 \bigg( \dot{\psi} + (1 - \Phi)(H_\sfx - H_\sfy) + 2 C_{\theta \theta} (H_\sfx - H_\sfy) + \frac{B_r \sfy '}{\sfx \sfy} - \frac{B_\theta}{\sfy} \cot \theta \\ &-& \frac{B_r '}{\sfx} + \frac{2}{\sfy} \partial_{\theta} B_{\theta} - \frac{\partial_\phi B_\phi}{\sfy \sin \theta} - 3 {\dot{C}}_{\theta \theta} \bigg) ,
\end{eqnarray}
and the $\phi-\phi$ component,

\begin{eqnarray}
\label{shearphiphi}
\nonumber \sigma_{\phi \phi} &=& -\frac{1}{3} \sfy^2 \sin^2 \theta \bigg( \dot{\psi} +  \left( 1 - \Phi \right) \left( H_\sfx - H_\sfy \right) + 2 C_{\phi \phi} \left( H_{\sfx} - H_{\sfy} \right) + \frac{B_r \sfy '}{\sfx \sfy} + 2 \frac{B_\theta}{\sfy} \cot \theta \\ &-& \frac{ B_r '}{\sfx} - \frac{\partial_\theta B_\theta}{\sfy} - \frac{\partial_\phi B_\phi}{\sfy \sin \theta} - 3 {\dot{C}}_{\phi \phi} \bigg) .
\end{eqnarray}
We also need the off-diagonal components. For the mixed temporal-spatial components we get,

\begin{equation}
\label{shearti}
\sigma_{t r} = \frac{2 B_r \sfx}{3} \left(H_{\sfx} - H_{\sfy} \right) ,
\quad
\sigma_{t \theta} = - \frac{B_\theta \sfy}{3} \left(H_{\sfx} - H_{\sfy} \right) ,
\quad
\sigma_{t \phi} = - \frac{B_\phi \sfy \sin \theta}{3} \left(H_{\sfx} - H_{\sfy} \right) .
\end{equation}

For the mixed spatial components we get,

\bea
\label{shearij}
\nonumber &\sigma_{r \theta}& = \frac{1}{3} C_{r \theta} \sfx \sfy \left( H_{\sfx} - H_{\sfy} \right) + \dot{C_{r \theta}} \sfx \sfy - \frac{1}{2} \sfy B_{\theta} ' + \frac{1}{2} B_{\theta} \sfy ' - \frac{1}{2} \sfx \partial_{\theta} B_r , \\ & & \\ \nonumber &\sigma_{\theta \phi}& = - \frac{1}{3} C_{\theta \phi} \sfy^2 \sin \theta \left( H_{\sfx} - H_{\sfy} \right) + \frac{1}{2} \dot{C_{\theta \phi}} \sfy^2 \sin \theta - \frac{1}{2} \sfy \sin \theta \partial_\theta B_{\phi} + \frac{1}{2} \sfy \cos \theta B_{\phi} , \\ & & \\ \nonumber &\sigma_{r \phi}& = \frac{1}{6} C_{r \phi} \sfx \sfy \sin \theta \left( H_{\sfx} - H_{\sfy} \right) + \frac{1}{2} \dot{C_{r \phi}} \sfx \sfy \sin \theta - \frac{1}{2} \sfy \sin \theta B_{\phi} ' + \frac{1}{2} \sin \theta B_{\phi} \sfy ' .\\ & & 
\eea

\section{The LTB Ricci 3-scalar}
\label{LTBR3}
The Ricci scalar on the spatial 3-hypersurfaces is given, in the background, as,
\begin{eqnarray}
\label{LTBR3EqBack}
{\bar{R}}^{(\mathrm{3})} &=& \frac{4 \sfx ' \sfy '}{\sfx^3 \sfy} - \frac{2  \sfy '^2}{\sfx^2 \sfy^2} - \frac{4 \sfy ''}{\sfx^2 \sfy} + \frac{2}{\sfy^2} ,
\end{eqnarray}
and the perturbed Ricci scalar is given by,
\begin{eqnarray}
\label{LTBR3EqPert}
\delta R^{(\mathrm{3})} &=& \frac{4 C_{r r} }{\sfx^2 \sfy} \left(\frac{\sfy '^2}{\sfy} + 2 \sfy '' - \frac{2 \sfx ' \sfy '}{\sfx} \right) - \frac{2}{\sfy^2} (2 C_{\theta \theta}) - \frac{2}{\sfx^2}(C_{\theta \theta} '' + C_{rr} '') \\ \nonumber &+& \frac{2 C_{r \theta} \cot \theta}{\sfx^2 \sfy} (\sfx ' - \sfy ') + \frac{2 \sfx ' (\partial_\theta C_{r \phi})}{\sfx^2 \sfy} + \frac{2 \sfx ' C_{\theta \theta} '}{\sfx^3} + \frac{4 \sfy '  C_{r r} '}{\sfx^2 \sfy} + \frac{2 \sfx ' C_{\phi \phi} '}{\sfx^3} \\ \nonumber &+& \frac{2 \cot \theta (\partial_\theta C_{\theta \theta})}{\sfy^2} - \frac{6 \sfy ' ( C_{\theta \theta} ' + C_{\phi \phi} ')}{\sfx^2 \sfy} - \frac{4 \cot \theta (\partial_\theta C_{\phi \phi})}{\sfy^2} - \frac{2 (\partial_{\theta \theta} C_{\phi \phi})}{\sfy^2} + \frac{2 (\partial_{\theta} C_{r \theta} ')}{\sfx \sfy} \\ \nonumber &+& \frac{2 (\partial_{\phi} C_{r \phi} ')}{\sfx \sfy \sin \theta} + \frac{2 (\partial_{\theta \phi} C_{\theta \phi})}{\sfy^2 \sin \theta} .
\end{eqnarray}

\chapter{The Spatial Metric Trace Perturbation in 2+2 Spherical Harmonic Formalism}
\label{ZetaGS}

\section{Background}
\label{GSBack}

The background LTB metric in the Clarkson, Clifton and February formalism is \cite{Tim1}
\be
\label{GSds2}
d s^2 = -d t^2 + \frac{a_{\parallel}^2(t,r)}{(1-\kappa r^2)} d r^2 + a_{\perp}^2(t,r)
r^2 d \Omega^2 .
\ee
This is the same metric in the same coordinates as that used in Chapter 3, \eq{LTBInterval}. This allows us to compare directly the perturbed metric components once the relations between the background functions are known. In the rest of this section, where a symbol is used in the Clarkson, Clifton and February formalism which has a different meaning to the same symbol in this thesis we have made it calligraphic, except for ``$v$'' which is made ``$\rm{v}$''. Also, a radial derivative is later defined which differs slightly from that used in earlier sections of this thesis. To distinguish this alternative radial derivative we use a dagger in place of the prime used in Ref.~\cite{Tim1}. From \eq{GSds2} and \eq{LTBInterval} we get,

\begin{equation}
\label{sfconvert}
\sfx = \frac{a_{\parallel}}{\sqrt{(1-\kappa r^2)}} \,, \qquad
\sfy = a_{\perp} r \,, \qquad
H_\parallel \equiv \frac{\dot{a_{\parallel}}}{a_{\parallel}} = H_\sfx \,, \qquad
H_\perp \equiv \frac{\dot{a_{\perp}}}{a_{\perp}} = H_\sfy \,,
\end{equation}
where $\kappa \equiv \kappa (r)$. The radial derivative defined in Ref.~\cite{Tim1} for an arbitrary function, $F$, is
\be
\label{RadDer}
F^\dagger = \frac{\sqrt{(1-\kappa r^2)}}{a_{\parallel}} F' = \frac{F '}{\sfx} \,, 
\ee
where the time derivative of the above radial derivative behaves as
\be
\label{RadDerDot}
(\dot{F})^\dagger - (F^\dagger \dot{)} = H_\parallel F^\dagger = H_\sfx \frac{F '}{\sfx} \,,
\ee

\section{Perturbations}
\label{GSPert}

The perturbed portion of the metric for axial perturbations~\cite{Tim1} i.e. perturbations which are odd modes of the spherical harmonic functions, ${\cal{Y}}^{(lm)}$, is given as
\begin{align}
\label{GSpertmetax}
\delta g_{\mu \nu} 
&\equiv
\left( \begin{array}{cc}
0 & h_A^{\text{axial}} \bar {\cal{Y}}_a\\ \nonumber
h_A^{\text{axial}} \bar {\cal{Y}}_a \; & h \; \bar {\cal{Y}}_{ab} \end{array} \right) ,\\ 
\end{align}
and for the polar perturbations~\cite{Tim1}, i.e. perturbations which are even modes of the spherical harmonic functions, ${\cal{Y}}^{(lm)}$, as
\begin{align}
\label{GSpertmetpol}
\delta g_{\mu \nu} 
&\equiv
\left( \begin{array}{cc}
h_{AB} {\cal{Y}} & h_A^{\text{polar}} {\cal{Y}}_{a}\\ \nonumber
h_A^{\text{polar}} {\cal{Y}}_{a} \; \; & a_{\perp}^2 r^2 (K {\cal{Y}} \gamma_{ab}+G {\cal{Y}}_{:ab}) \end{array} \right). \\ 
\end{align}
In the above equations ${\cal{Y}} \equiv {\cal{Y}}^{(lm)}$ and are the various spherical harmonic functions for scalar, vector and tensor equivalent perturbations (see Ref.~\cite{Tim1}). The bar indicates odd modes, no bar even. The index $A$ runs over $t$ and $r$, while $a$ runs over $\theta$ and $\phi$. The colon represents the covariant derivative with respect to the metric on the unit sphere. $h_A^{\text{axial}} , h , h_{AB} , h_A^{\text{polar}} , K , G $ are the perturbation variables and are functions of $x^A$.
By direct comparison between the perturbed metrics in both formalisms i.e. \eq{GSpertmetax} and \eq{GSpertmetpol} with \eq{LTBMetricperturbations} we find,
\be
\label{psiGS}
\psi = \frac{1}{3} (C_{rr} + C_{\theta \theta} + C_{\phi \phi}) = \frac{1}{6} \left(\frac{h_{r r} {\cal{Y}}}{\sfx^2} + \frac{h \; \bar {\cal{Y}}_{\theta \theta}}{\sfy^2}+ \frac{h \; \bar {\cal{Y}}_{\phi \phi}}{\sfy^2 \sin^2 \theta} + 2 K {\cal{Y}} + G {\cal{Y}}_{:\theta \theta} + \frac{G {\cal{Y}}_{:\phi \phi}}{\sin^2 \theta} \right)
 ,
\ee

where we have used Bondi's scale factors, $\sfx$ and $\sfy$ as in Subsection~\ref{ltb_back}, for brevity. The covariant form of the axial perturbed 4-velocities is
\be
\label{4vaxGMGcov}
\delta u_{\mu} = (0, {\bar{\rm{v}}} \; \bar {\cal{Y}}_a) ,
\ee
and the scalar perturbed 4-velocities are
\be
\label{4vscalGMGcov}
\delta u_{\mu} = \left[ \left(\tilde{w} {\hat{n}}_A+\frac{1}{2}h_{AB} {\hat{u}}^B
\right) {\cal{Y}}, \tilde{\rm{v}} \; {\cal{Y}}_a \right] ,
\ee
where ${\bar{\rm{v}}}, \tilde{w}, \tilde{\rm{v}}$ are all functions of $x^A$, and $\hat{n}_A$ is the unit spacelike radial vector and $\hat{u}^A$ is the unit timelike vector.
The contravariant form of the perturbed metric for axial perturbations is
\begin{align}
\label{GSpertmetaxuppy}
\delta g^{\mu \nu} 
&\equiv
\left( \begin{array}{cccc}
0 & 0 & \frac{1}{\sfy^2} h_t^{\text{axial}} \bar {\cal{Y}}_\theta & \frac{1}{\sfy^2 \sin^2 \theta} h_t^{\text{axial}} \bar {\cal{Y}}_\phi\\ \nonumber
0 & 0 & -\frac{1}{\sfx^2 \sfy^2} h_r^{\text{axial}} \bar {\cal{Y}}_\theta & -\frac{1}{\sfx^2 \sfy^2 \sin^2 \theta} h_r^{\text{axial}} \bar {\cal{Y}}_\phi\\ \nonumber
\frac{1}{\sfy^2} h_t^{\text{axial}} \bar {\cal{Y}}_\theta & -\frac{1}{\sfx^2 \sfy^2} h_r^{\text{axial}} \bar {\cal{Y}}_\theta & - \frac{1}{\sfy^4} h \; \bar {\cal{Y}}_{\theta \theta} \; & - \frac{1}{\sfy^4 \sin^2 \theta} h \; \bar {\cal{Y}}_{\theta \phi}\\ \nonumber
\frac{1}{\sfy^2 \sin^2 \theta} h_t^{\text{axial}} \bar {\cal{Y}}_\phi & -\frac{1}{\sfx^2 \sfy^2 \sin^2 \theta} h_r^{\text{axial}} \bar {\cal{Y}}_\phi & - \frac{1}{\sfy^4 \sin^2 \theta} h \; \bar {\cal{Y}}_{\theta \phi} & - \frac{1}{\sfy^4 \sin^4 \theta} h \; \bar {\cal{Y}}_{\phi \phi} \end{array} \right)\\
\end{align}
and for the polar perturbations is
{\begin{footnotesize}
\begin{align}
\label{GSpertmetpoluppy}
\delta g^{\mu \nu} 
&\equiv
\left( \begin{array}{cccc}
- h_{tt} {\cal{Y}} & \frac{1}{\sfx^2} h_{tr} {\cal{Y}} & \frac{1}{\sfy^2} h_t^{\text{polar}} {\cal{Y}}_{\theta} \; \; & \frac{1}{\sfy^2 \sin^2 \theta} h_t^{\text{polar}} {\cal{Y}}_{\phi}\\ \nonumber
\frac{1}{\sfx^2} h_{tr} {\cal{Y}} & - \frac{1}{\sfx^4} h_{rr} {\cal{Y}} & - \frac{1}{\sfx^2 \sfy^2} h_r^{\text{polar}} {\cal{Y}}_{\theta} \; \; & - \frac{1}{\sfx^2 \sfy^2 \sin^2 \theta} h_r^{\text{polar}} {\cal{Y}}_{\phi}\\ \nonumber
\frac{1}{\sfy^2} h_t^{\text{polar}} {\cal{Y}}_{\theta} \; \; & - \frac{1}{\sfx^2 \sfy^2} h_r^{\text{polar}} {\cal{Y}}_{\theta} \; \; & - \frac{1}{\sfy^2} (K {\cal{Y}} +G {\cal{Y}}_{:\theta \theta}) \; \; & - \frac{1}{\sfy^2 \sin^2 \theta} G {\cal{Y}}_{: \theta \phi}\\ 
\nonumber
\frac{1}{\sfy^2 \sin^2 \theta} h_t^{\text{polar}} {\cal{Y}}_{\phi} & - \frac{1}{\sfx^2 \sfy^2 \sin^2 \theta} h_r^{\text{polar}} {\cal{Y}}_{\phi} & - \frac{1}{\sfy^2 \sin^2 \theta} G {\cal{Y}}_{: \theta \phi} & - \frac{1}{\sfy^2 \sin^4 \theta} (K {\cal{Y}} \sin^2 \theta +G {\cal{Y}}_{: \phi \phi}) \end{array} \right)\\
\end{align}
\end{footnotesize}}
where we have once again used Bondi's scale factors, $\sfx$ and $\sfy$, for brevity.
The perturbed 4-velocity in contravariant form is
{\begin{footnotesize}
\bea
\label{4vGMGcont}
u^{\mu} &=& \Bigg[ \qquad 1 + \frac{1}{2} h_{t t} {\cal{Y}}, \qquad \qquad \qquad \qquad \qquad \qquad - \frac{{\cal{Y}}}{\sfx^2} \left(\frac{1}{2} h_{t r} + \sfx \tilde{w} \right) ,\\ \nonumber & & \frac{1}{\sfy^2} \left( {\bar{\rm{v}}} \; \bar {\cal{Y}}_\theta + \tilde{\rm{v}} \; {\cal{Y}}_\theta -  h_t^{\text{axial}} \bar {\cal{Y}}_\theta - h_t^{\text{polar}} {\cal{Y}}_{\theta}  \right), \qquad \frac{1}{\sfy^2 \sin^2 \theta} \left( {\bar{\rm{v}}} \; \bar {\cal{Y}}_\phi + \tilde{\rm{v}} \; {\cal{Y}}_\phi -  h_t^{\text{axial}} \bar {\cal{Y}}_\phi - h_t^{\text{polar}} {\cal{Y}}_{\phi}  \right)  \Bigg] ,
\eea
\end{footnotesize}}
where the last three terms correspond directly with $v^r, v^\theta, v^\phi$ respectively in the formalism of Chapter 3.
Substituting \eq{4vGMGcont}, \eq{psiGS}, \eq{RadDer} and \eq{sfconvert} into \eq{zetafull4} we get
\bea
\label{GSzetafull}
- \SMTP &=& \frac{1}{6} \Bigg( \frac{(1-\kappa r^2)}{{a^2}_{\parallel}} h_{r r} {\cal{Y}} + \frac{h \; \bar {\cal{Y}}_{\theta \theta}}{{a_{\perp}}^2 r^2}+ \frac{h \; \bar {\cal{Y}}_{\phi \phi}}{{a_{\perp}}^2 r^2 \sin^2 \theta}\\ 
\nonumber &+& 2 K {\cal{Y}} + G {\cal{Y}}_{:\theta \theta} + \frac{G {\cal{Y}}_{:\phi \phi}}{\sin^2 \theta} \Bigg) + \frac{\delta \rho}{3\bar{\rho}} \\ 
\nonumber
&+& \frac{1}{3}\Bigg\{ \partial_{\theta} \int \frac{1}{{a_{\perp}}^2 r^2} \Bigg( {\bar{\rm{v}}} \; \bar {\cal{Y}}_\theta + \tilde{\rm{v}} \; {\cal{Y}}_\theta -  h_t^{\text{axial}} \bar {\cal{Y}}_\theta - h_t^{\text{polar}} {\cal{Y}}_{\theta}  \Bigg) dt \\ \nonumber &+& \partial_{\phi} \int \frac{1}{{a_{\perp}}^2 r^2 \sin^2 \theta} \Bigg( {\bar{\rm{v}}} \; \bar {\cal{Y}}_\phi + \tilde{\rm{v}} \; {\cal{Y}}_\phi -  h_t^{\text{axial}} \bar {\cal{Y}}_\phi - h_t^{\text{polar}} {\cal{Y}}_{\phi}  \Bigg) dt \\ 
\nonumber &+& \cot \theta \int \frac{1}{{a_{\perp}}^2 r^2} \Bigg( {\bar{\rm{v}}} \; \bar {\cal{Y}}_\theta + \tilde{\rm{v}} \; {\cal{Y}}_\theta -  h_t^{\text{axial}} \bar {\cal{Y}}_\theta - h_t^{\text{polar}} {\cal{Y}}_{\theta}  \Bigg) dt \\ 
\nonumber &-& \partial_r \int \frac{{\cal{Y}}(1-\kappa r^2)}{{a_{\parallel}}^2} \Bigg(\frac{1}{2} h_{t r} + \frac{a_{\parallel}}{\sqrt{(1-\kappa r^2)}} \tilde{w} \Bigg) dt \\ 
\nonumber &-& \Bigg(\Bigg(\frac{a_{\parallel}}{\sqrt{(1-\kappa r^2)}}\Bigg)^\dagger + 2 \frac{(a_{\perp} r)^\dagger a_{\parallel}}{a_{\perp} r \sqrt{(1-\kappa r^2)}}
\\ 
\nonumber &+& \frac{{\bar{\rho}}^\dagger a_{\parallel} }{\bar{\rho} \sqrt{(1-\kappa r^2)}} \Bigg) \int \frac{(1-\kappa r^2)}{{a^2}_{\parallel}} {\cal{Y}} \Bigg(\frac{1}{2} h_{t r} + \frac{a_{\parallel}}{\sqrt{(1-\kappa r^2)}} \tilde{w} \Bigg) dt \Bigg\} \,,
\eea
which is our gauge invariant quantity, conserved on all scales with only adiabatic pressure perturbations, but expressed in terms of the perturbation functions used in \cite{Tim1}. This relates directly to the density perturbation on constant curvature hypersurfaces through \eq{GIdenspertZeta2}
\begin{equation*}
\delta \tilde{\rho} \Big|_{\psi=0} = - 3 \bar{\rho} \SMTP\, .
\end{equation*}
Equation \eqref{GSzetafull} is clearly more complicated than \eq{zetafull4}.



\chapter{Gauge Transformations, Relations and Alternative Gauges}
\label{Gauge Transformations}

\section{General Gauge Transformations}
\label{General Gauge Transformations}

We now give the gauge transformations for the perturbed quantities
used in Chapters 3 and 4 and in Subsection~\ref{Flat to
  Long Gauge Relations} below for easy reference. Following the notation of
Ref.~\cite{Malik:2004tf}, quantities in the new coordinate system are
  denoted by a tilde. We use the active approach throughout.

The matter variables, the velocity and the density perturbations,
transform as
\bea
\label{GGTV}
\tilde{\hat v}_\alpha &=& \hat v_\alpha + \frac{\delta t}{a} \,,\\
\label{GGTrho}
\tilde{\delta \rho}_\alpha 
&=& \delta \rho_\alpha - \dot{\bar{\rho}}_\alpha \delta t \,,
\eea
where $\hat v_\alpha$ is defined in \eq{FlatNewv}.

The perturbations of the metric transform as
\bea
\label{GGTphi2}
\tilde{\Phi} &=& \Phi - \dot{\delta t} \,,\\
\label{GGTpsi2}
\tilde{\psi} &=& \psi + H \delta t \,,\\
\tilde{B} &=& B - a\dot{\delta x}+ \delta t \,,\\
\tilde{E} &=& E - \delta x \,.
\eea

\section{Flat to Longitudinal Gauge Relations}
\label{Flat to Long Gauge Relations}

The relation between the velocity in flat gauge ($\tilde{\psi}=\tilde{E}=0$) and in longitudinal gauge ($\tilde{B} = \tilde{E} = 0$) is given by
\be
\label{FLGTV}
\hat{v}_{\alpha(\rm flat)} = v_{\alpha(\rm long)} + {B}_{(\rm flat)}\,.
\ee
The relation for the  density perturbations is
\be
\label{FLGTrho}
{\delta \rho}_{\alpha(\rm flat)} = \delta \rho_{\alpha(\rm long)} - a \dot{\bar{\rho}}_\alpha {B}_{(\rm flat)} \,.
\ee
The transformation behaviour of the metric perturbations and the fact that
$\Phi = \psi$ in longitudinal gauge (from the trace free part of the $i-j$ component of the Einstein field equations, \eq{TraceFreeGijDEEFE2}) in the absence of anisotropic
stress gives 
\be
\label{FLGTpsi}
{B}_{(\rm flat)} = - \frac{\Phi_{(\rm long)}}{Ha} \,.
\ee

\section{Longitudinal Gauge with with Arbitrary Numbers of Fields and DM Fluids}
\label{Long2f2dm}

As mentioned in Section~\ref{Fixing} the \PY~code was originally written in longitudinal gauge as this is the one commonly used in the literature in the field, see e.g.~\cite{Amendola:2014kwa}. However due to numerical instabilities caused by the constraint \eq{Phiconstraint} for $\Phi$ below, this version was abandoned. We include the equations below for reference and completeness.

For a given DM species, $\alpha$, the evolution equation for the perturbation is
\bea
\label{IntDEPertEcons4}
\nonumber \dot{\delta \rho_\alpha} &+& 3 H (\delta \rho_\alpha + \delta P_\alpha) - \left(3 \dot{\Phi} + \frac{k^2 v_\alpha}{a}\right)({\bar{\rho}}_\alpha + {\bar{P}}_\alpha) = - \sum\limits_{I} \kappa \C_{I \alpha} ({\bar{\rho}}_\alpha - 3 {\bar{P}}_\alpha) {\dot{\delta \ph}}_I \\ &-& \sum\limits_{I} \kappa \C_{I \alpha} (\delta \rho_\alpha - 3 \delta P_\alpha) {\dot{\bar{\ph}}}_I .
\eea
Momentum conservation is given by

\be
\label{IntDEPertMomconsLong}
\dot{v}_\alpha =  \kappa \sum\limits_{I} \C_{I \alpha} (\bar{\rho}_\alpha - 3 \bar{P}_\alpha) \frac{\delta \ph_I}{a} + 3H \frac{\dot{\bar{P}}_\alpha}{\dot{\bar{\rho}}_\alpha} (v_\alpha) - H(v_\alpha) - \frac{\Phi}{a} - \frac{\delta P_\alpha}{a({\bar{\rho}_\alpha} + \bar{P}_\alpha)} .
\ee

The evolution equation for the fields, labelled $I$, $J$, is
\bea
\label{IntDEPertEconsSF2}
\nonumber {\ddot{\delta \ph}}_I &+& 3 H {\dot{\delta \ph}}_I + \sum\limits_{J} V,_{\ph_I \ph_J} \delta \ph_J - 4 \dot{\Phi} {\dot{\bar{\ph}}}_I + \frac{k^2}{a^2} \delta \ph_I + 2 V,_{\ph_I} \Phi - 2 \sum\limits_{\alpha} \kappa \C_{I \alpha} ({\bar{\rho}}_\alpha - 3 {\bar{P}}_\alpha) \Phi \\ &-& \sum\limits_{\alpha} \kappa \C_{I \alpha} (\delta \rho_\alpha - 3 \delta P_\alpha) = 0 . 
\eea
The Einstein Field Equations are as follows. From the $0-0$ component we get
\be
\label{G00DEEFE2}
3 H (\dot{\Phi} + H \Phi) + \frac{k^2}{a^2}\Phi = - \frac{\kappa^2}{2} \left[\sum\limits_{\alpha} \delta \rho_\alpha + \sum\limits_{I}(-\Phi \dot{\bar{\ph}}^2_I + {\dot{\delta \ph}}_I {\dot{\bar{\ph}}}_I + V,_{\ph_I} \delta \ph_I)\right] .
\ee
From the $0-i$ component we get
\be
\label{Gi0DEEFE2}
\dot{\Phi} + H \Phi = - \frac{\kappa^2}{2} \left[\sum\limits_{\alpha} a v_\alpha(\bar{\rho}_\alpha + \bar{P}_\alpha) - \sum\limits_{I} {\dot{\bar{\ph}}}_I \delta \ph_I \right] .
\ee
From the trace of $i-j$ component we get
\be
\label{TraceGijDEEFE2}
\ddot{\Phi} + 4 H \dot{\Phi} + (3 H^2 + 2 \dot{H}) \Phi = \frac{\kappa^2}{2} \left[\sum\limits_{\alpha} \delta P_\alpha - \sum\limits_{I} \left(\Phi \dot{\bar{\ph}}^2_I - {\dot{\delta \ph}}_I {\dot{\bar{\ph}}}_I + V,_{\ph_I} \delta \ph_I\right) \right] .
\ee
From the trace-free part of the $i-j$ component we get
\be
\label{TraceFreeGijDEEFE2}
\psi = \Phi ,
\ee
since $\sigma_s = 0$.\\
From \eq{G00DEEFE2} and \eq{Gi0DEEFE2} we get

\bea
\label{Phiconstraint}
\nonumber \Phi &=& \left( \sum\limits_{I} \dot{\bar{\ph}}^2_I - \frac{2 k^2}{(\kappa a)^2} \right)^{-1} \Bigg[ \sum\limits_{\alpha} \left( \delta \rho_\alpha - 3 H a v_\alpha(\bar{\rho}_\alpha + \bar{P}_\alpha) \right) \\ &+& \sum\limits_{I} \left( {\delta \dot{\ph}_I \dot{\bar{\ph}}_I} + {V,_{\ph_I} \delta \ph_I} + 3 H \dot{\bar{\ph}}_I \delta \ph_I \right) \Bigg]
\eea

\section{Synchronous Comoving Gauge with Arbitrary Numbers of Fields and DM Fluids}
\label{Synch2f2dm}
Synchronous gauge had been considered for use in the \PY~code. This was partly because it has been used in codes such as CAMB and CLASS~\cite{Lewis:1999bs,Blas:2011rf}. The equations from Section~\ref{IntDEperts} are presented here in synchronous co-moving gauge ($\tilde{\Phi} = \tilde{B} = \tilde{v} = 0$), but otherwise in full generality, allowing for multiple fields and fluids. This is done for reference and completeness. For a given DM species, $\alpha$, the evolution equation for the perturbation is

\bea
\label{IntDEPertEconsSynch}
\nonumber \dot{\delta \rho_\alpha} &+& 3 H (\delta \rho_\alpha + \delta P_\alpha) - \left(3 \dot{\psi} + k^2 \dot{E}\right)({\bar{\rho}}_\alpha + {\bar{P}}_\alpha) = - \sum\limits_{I} \kappa \C_{I \alpha} ({\bar{\rho}}_\alpha - 3 {\bar{P}}_\alpha) {\dot{\delta \ph}}_I \\ &-& \sum\limits_{I} \kappa \C_{I \alpha} (\delta \rho_\alpha - 3 \delta P_\alpha) {\dot{\bar{\ph}}}_I .
\eea
Momentum conservation is given by
\be
\label{IntDEPertMomconsSynch}
 \kappa \sum\limits_{I} \C_{I \alpha} ({\bar{\rho}}_\alpha - 3 \bar{P_\alpha}) \delta \ph_I  = \frac{\delta P_\alpha}{{\bar{\rho}}_\alpha + \bar{P_\alpha}} .
\ee
The evolution equation for the fields, labelled $I$, $J$, is
\bea
\label{IntDEPertEconsSFSynch}
\nonumber {\ddot{\delta \ph}}_I &+& 3 H {\dot{\delta \ph}}_I + \sum\limits_{J} V,_{\ph_I \ph_J} \delta \ph_J - \left(3 \dot{\psi} + k^2 \dot{E}\right) {\dot{\bar{\ph}}}_I + \frac{k^2}{a^2} \delta \ph_I \\ &-& \sum\limits_{\alpha} \kappa \C_{I \alpha} (\delta \rho_\alpha - 3 \delta P_\alpha) - 2 \kappa \sum\limits_{\alpha} \C_{I \alpha} (\bar{\rho}_{\alpha} = 0 . 
\eea
The Einstein Field Equations are as follows. From the $0-0$ component we get
\be
\label{G00DEEFESynch}
3 H (\dot{\psi})+ \frac{k^2}{a^2} (\psi + H a^2 \dot{E} ) = - \frac{\kappa^2}{2} \left[\sum\limits_{\alpha} \delta \rho_\alpha + \sum\limits_{I}({\dot{\delta \ph}}_I {\dot{\bar{\ph}}}_I + V,_{\ph_I} \delta \ph_I)\right] .
\ee
From the $0-i$ component we get
\be
\label{Gi0DEEFESynch}
\dot{\psi} =  \frac{\kappa^2}{2} \sum\limits_{I} {\dot{\bar{\ph}}}_I \delta \ph_I .
\ee
From the trace of $i-j$  component we get
\be
\label{TraceGijDEEFESynch}
\ddot{\psi} + 3 H \dot{\psi} = \frac{\kappa^2}{2} \left[\sum\limits_{\alpha} \delta P_\alpha + \sum\limits_{I} \left({\dot{\delta \ph}}_I {\dot{\bar{\ph}}}_I - V,_{\ph_I} \delta \ph_I\right) \right] .
\ee
From the trace-free part of the $i-j$  component we get
\be
\label{TraceFreeGijDEEFESynch}
{\dot{\sigma}}_s + H \sigma_s + \psi = 0 ,
\ee
where $\sigma_s$ is the scalar shear and $\sigma_s = a^2 \dot{E}$.

\begin{singlespace}



\end{singlespace}



\begin{thebibliography}{99}



\bibitem{Adam:2015rua} 
  R.~Adam {\it et al.} [Planck Collaboration],
  arXiv:1502.01582 [astro-ph.CO].

\bibitem{ACT} 
  J.~Dunkley, R.~Hlozek, J.~Sievers, V.~Acquaviva, P.~A.~R.~Ade, P.~Aguirre, M.~Amiri and J.~W.~Appel {\it et al.},
  Astrophys.\ J.\  {\bf 739}, 52 (2011)
  [arXiv:1009.0866 [astro-ph.CO]].
 
\bibitem{SPT}  N.~R.~Hall, L.~Knox, C.~L.~Reichardt, P.~A.~R.~Ade, K.~A.~Aird, B.~A.~Benson, L.~E.~Bleem and J.~E.~Carlstrom {\it et al.},
  Astrophys.\ J.\  {\bf 718}, 632 (2010)
  [arXiv:0912.4315 [astro-ph.CO]].

\bibitem{Perlmutter:1998np} 
  S.~Perlmutter {\it et al.}  [Supernova Cosmology Project Collaboration],
  Astrophys.\ J.\  {\bf 517}, 565 (1999)
  [astro-ph/9812133].
  
\bibitem{Riess:1998cb} 
  A.~G.~Riess {\it et al.} [Supernova Search Team Collaboration],
  Astron.\ J.\  {\bf 116}, 1009 (1998)
  [astro-ph/9805201].

\bibitem{Kowalski:2008ez} 
  M.~Kowalski {\it et al.} [Supernova Cosmology Project Collaboration],
  Astrophys.\ J.\  {\bf 686}, 749 (2008)
  doi:10.1086/589937
  [arXiv:0804.4142 [astro-ph]].

\bibitem{Anderson:2013zyy} 
  L.~Anderson {\it et al.}  [BOSS Collaboration],
  arXiv:1312.4877 [astro-ph.CO].

\bibitem{Bonnett:2015pww} 
  C.~Bonnett {\it et al.} [DES Collaboration],
  arXiv:1507.05909 [astro-ph.CO].

\bibitem{Dawson:2015wdb} 
  K.~S.~Dawson {\it et al.},
  arXiv:1508.04473 [astro-ph.CO].

\bibitem{celerier}   M.~-N.~Celerier and J.~Schneider,
  Phys.\ Lett.\ A {\bf 249}, 37 (1998)
  [astro-ph/9809134].
  
\bibitem{Moresco:2012jh} 
  M.~Moresco, A.~Cimatti, R.~Jimenez, L.~Pozzetti, G.~Zamorani, M.~Bolzonella, J.~Dunlop and F.~Lamareille {\it et al.},
  JCAP {\bf 1208}, 006 (2012)
  [arXiv:1201.3609 [astro-ph.CO]].
  

\bibitem{Ade:2013ktc} 
  P.~A.~R.~Ade {\it et al.}  [Planck Collaboration],
  arXiv:1303.5062 [astro-ph.CO].

\bibitem{Clarkson:2012bg} 
  C.~Clarkson,
  Comptes Rendus Physique {\bf 13}, 682 (2012)
  [arXiv:1204.5505 [astro-ph.CO]].
  
\bibitem{Vargas:2015ctw} 
  C.~Z.~Vargas, F.~T.~Falciano and R.~R.~R.~Reis,
  arXiv:1512.02571 [astro-ph.CO].


\bibitem{Meyer:2014qla} 
  S.~Meyer, M.~Redlich and M.~Bartelmann,
  JCAP {\bf 1503}, no. 03, 053 (2015)
  doi:10.1088/1475-7516/2015/03/053
  [arXiv:1412.3012 [astro-ph.CO]].
  

\bibitem{Sussman:2015bea} 
  R.~A.~Sussman and J.~Larena,
  Class.\ Quant.\ Grav.\  {\bf 32}, no. 16, 165012 (2015)
  doi:10.1088/0264-9381/32/16/165012
  [arXiv:1503.04589 [gr-qc]].

\bibitem{Iribarrem:2014dta} 
  A.~Iribarrem, P.~Andreani, S.~February, C.~Gruppioni, A.~R.~Lopes, M.~B.~Ribeiro and W.~R.~Stoeger,
  Astron.\ Astrophys.\  {\bf 563}, A20 (2014)
  [arXiv:1401.6572 [astro-ph.CO]].

\bibitem{Lim:2013rra} 
  W.~C.~Lim, M.~Regis and C.~Clarkson,
  JCAP {\bf 1310}, 010 (2013)
  [arXiv:1308.0902 [astro-ph.CO]].

\bibitem{Biswas}   T.~Biswas, A.~Notari and W.~Valkenburg,
  JCAP {\bf 1011}, 030 (2010)
  [arXiv:1007.3065 [astro-ph.CO]].

\bibitem{Bolejko1}   K.~Bolejko, M.~-N.~Celerier and A.~Krasinski,
  Class.\ Quant.\ Grav.\  {\bf 28}, 164002 (2011)
  [arXiv:1102.1449 [astro-ph.CO]].

\bibitem{Bolejko3}   K.~Bolejko and J.~S.~B.~Wyithe,
  JCAP {\bf 0902}, 020 (2009)
  [arXiv:0807.2891 [astro-ph]].

\bibitem{CFL}   T.~Clifton, P.~G.~Ferreira and K.~Land,
  Phys.\ Rev.\ Lett.\  {\bf 101}, 131302 (2008)
  [arXiv:0807.1443 [astro-ph]].

\bibitem{GBH2}   J.~Garcia-Bellido and T.~Haugboelle,
  JCAP {\bf 0804}, 003 (2008)
  [arXiv:0802.1523 [astro-ph]].
  
\bibitem{MZS10}   A.~Moss, J.~P.~Zibin and D.~Scott,
  Phys.\ Rev.\ D {\bf 83}, 103515 (2011)
  [arXiv:1007.3725 [astro-ph.CO]].
  
\bibitem{ZMS}   J.~P.~Zibin, A.~Moss and D.~Scott,
  Phys.\ Rev.\ Lett.\  {\bf 101}, 251303 (2008)
  [arXiv:0809.3761 [astro-ph]].
  
\bibitem{goodman}   J.~Goodman,
  Phys.\ Rev.\ D {\bf 52}, 1821 (1995)
  [astro-ph/9506068].


\bibitem{Sussman1}
 R.~A.~Sussman,
  Class.\ Quant.\ Grav.\  {\bf 30}, 235001 (2013)
  [arXiv:1305.3683 [gr-qc]].

\bibitem{Zumalacarregui:2012pq} 
  M.~Zumalacarregui, J.~Garcia-Bellido and P.~Ruiz-Lapuente,
  JCAP {\bf 1210}, 009 (2012)
  [arXiv:1201.2790 [astro-ph.CO]].
  

\bibitem{YNS}   C.~-M.~Yoo, K.~-i.~Nakao and M.~Sasaki,
  JCAP {\bf 1007}, 012 (2010)
  [arXiv:1005.0048 [astro-ph.CO]].

\bibitem{Alnes}   H.~Alnes and M.~Amarzguioui,
  Phys.\ Rev.\ D {\bf 74}, 103520 (2006)
  [astro-ph/0607334].
  
\bibitem{CFZ}   T.~Clifton, P.~G.~Ferreira and J.~Zuntz,
  JCAP {\bf 0907}, 029 (2009)
  [arXiv:0902.1313 [astro-ph.CO]].

\bibitem{Bull:2011wi} 
  P.~Bull, T.~Clifton and P.~G.~Ferreira,
  Phys.\ Rev.\ D {\bf 85}, 024002 (2012)
  [arXiv:1108.2222 [astro-ph.CO]].

\bibitem{ZS}  P.~Zhang and A.~Stebbins,
  Phys.\ Rev.\ Lett.\  {\bf 107}, 041301 (2011)
  [arXiv:1009.3967 [astro-ph.CO]].

\bibitem{MZ11}   J.~P.~Zibin and A.~Moss,
  Class.\ Quant.\ Grav.\  {\bf 28}, 164005 (2011)
  [arXiv:1105.0909 [astro-ph.CO]].

\bibitem{YNS2}   C.~-M.~Yoo, K.~-i.~Nakao and M.~Sasaki,
  JCAP {\bf 1010}, 011 (2010)
  [arXiv:1008.0469 [astro-ph.CO]].

\bibitem{SZ80}   R.~A.~Sunyaev and Y.~.B.~Zeldovich,
  Mon.\ Not.\ Roy.\ Astron.\ Soc.\  {\bf 190}, 413 (1980).

\bibitem{kSZobs1}   W.~L.~Holzapfel, P.~A.~R.~Ade, S.~E.~Church, P.~D.~Mauskopf, Y.~Rephaeli, T.~M.~Wilbanks and A.~E.~Lange,
  [astro-ph/9702223].

\bibitem{kSZobs2}    B.~A.~Benson, S.~E.~Church, P.~A.~R.~Ade, J.~J.~Bock, K.~M.~Ganga, J.~R.~Hinderks, P.~D.~Mauskopf and B.~Philhour {\it et al.},
  Astrophys.\ J.\  {\bf 592}, 674 (2003)
  [astro-ph/0303510].

\bibitem{kSZobs3}    T.~Kitayama, E.~Komatsu, N.~Ota, T.~Kuwabara, Y.~Suto, K.~Yoshikawa, M.~Hattori and H.~Matsuo,
  Publ.\ Astron.\ Soc.\ Jap.\  {\bf 56}, 17 (2004)
  [astro-ph/0311574].

\bibitem{GBH}   J.~Garcia-Bellido and T.~Haugboelle,
  JCAP {\bf 0809}, 016 (2008)
  [arXiv:0807.1326 [astro-ph]].

\bibitem{Finelli:2014yha} 
  F.~Finelli, J.~Garcia-Bellido, A.~Kovacs, F.~Paci and I.~Szapudi,
  arXiv:1405.1555 [astro-ph.CO].
  
\bibitem{Alonso:2010zv} 
  D.~Alonso, J.~Garcia-Bellido, T.~Haugbolle and J.~Vicente,
  Phys.\ Rev.\ D {\bf 82}, 123530 (2010)
  [arXiv:1010.3453 [astro-ph.CO]].

\bibitem{Alonso:2012ds} 
  D.~Alonso, J.~Garcia-Bellido, T.~Haugboelle and A.~Knebe,
  Phys.\ Dark Univ.\  {\bf 1}, 24 (2012)
  [arXiv:1204.3532 [astro-ph.CO]].

\bibitem{Bardeen80}
J.~M.~Bardeen,
Phys.\ Rev.\ D {\bf 22}, 1882 (1980).

\bibitem{KS}
H.~Kodama and M.~Sasaki,
Prog.\ Theor.\ Phys.\ Suppl.\  {\bf 78}, 1 (1984).

\bibitem{Lyth85}
D.~H.~Lyth,
Phys.\ Rev.\ D {\bf 31}, 1792 (1985).

\bibitem{WMLL} 
  D.~Wands, K.~A.~Malik, D.~H.~Lyth and A.~R.~Liddle,
  Phys.\ Rev.\ D {\bf 62}, 043527 (2000)
  [astro-ph/0003278].

\bibitem{MW2003} 
  K.~A.~Malik and D.~Wands,
  Class.\ Quant.\ Grav.\  {\bf 21}, L65 (2004)
  [astro-ph/0307055].

\bibitem{LMS}
  D.~H.~Lyth, K.~A.~Malik and M.~Sasaki,
  JCAP {\bf 0505}, 004 (2005)
  [astro-ph/0411220].

\bibitem{Salopek:1990jq}
D.~S.~Salopek and J.~R.~Bond,
Phys.\ Rev.\ D {\bf 42}, 3936 (1990).

\bibitem{Rigopoulos03}
G.~I.~Rigopoulos and E.~P.~S.~Shellard,
Phys.\ Rev.\  D {\bf 68}, 123518 (2003)
[arXiv:astro-ph/0306620].

\bibitem{Langlois:2005qp}
  D.~Langlois and F.~Vernizzi,
  Phys.\ Rev.\  D {\bf 72}, 103501 (2005)
  [arXiv:astro-ph/0509078].

\bibitem{Bridgman:2001mc}
  H.~A.~Bridgman, K.~A.~Malik and D.~Wands,
  Phys.\ Rev.\ D {\bf 65}, 043502 (2002)
  [arXiv:astro-ph/0107245].


\bibitem{Abolhasani:2013zya} 
A.~A.~Abolhasani, R.~Emami,  J.~T.~Firouzjaee and H.~Firouzjahi,
  JCAP {\bf 1308}, 016 (2013)
  [arXiv:1302.6986 [astro-ph.CO]].

\bibitem{Bondi} 
  H.~Bondi,
  Mon.\ Not.\ Roy.\ Astron.\ Soc.\  {\bf 107}, 410 (1947).
  doi:10.1093/mnras/107.5-6.410

\bibitem{Ellis}
  G.~F.~R.~Ellis and H.~van Elst,
  NATO Adv.\ Study Inst.\ Ser.\ C.\ Math.\ Phys.\ Sci.\  {\bf 541}, 1 (1999)
  [gr-qc/9812046].

\bibitem{Gerlach:1979rw} 
  U.~H.~Gerlach and U.~K.~Sengupta,
  Phys.\ Rev.\ D {\bf 19}, 2268 (1979).

\bibitem{Gerlach:1980tx} 
  U.~H.~Gerlach and U.~K.~Sengupta,
  Phys.\ Rev.\ D {\bf 22}, 1300 (1980).

\bibitem{Tim1}
  C.~Clarkson, T.~Clifton and S.~February,
  JCAP {\bf 0906}, 025 (2009)
  [arXiv:0903.5040 [astro-ph.CO]].

\bibitem{February:2012fp} 
  S.~February, C.~Clarkson and R.~Maartens,
  JCAP {\bf 1303}, 023 (2013)
  [arXiv:1206.1602 [astro-ph.CO]].

\bibitem{February:2013qza} 
  S.~February, J.~Larena, C.~Clarkson and D.~Pollney,
  arXiv:1311.5241 [astro-ph.CO].
  
\bibitem{Amendola:1999dr} 
  L.~Amendola,
  Mon.\ Not.\ Roy.\ Astron.\ Soc.\  {\bf 312}, 521 (2000)
  [astro-ph/9906073].

\bibitem{Holden:1999hm} 
  D.~J.~Holden and D.~Wands,
  Phys.\ Rev.\ D {\bf 61}, 043506 (2000)
  [gr-qc/9908026].

\bibitem{Amendola:1999er} 
  L.~Amendola,
  Phys.\ Rev.\ D {\bf 62}, 043511 (2000)
  [astro-ph/9908023].

\bibitem{Koivisto:2005nr} 
  T.~Koivisto,
  Phys.\ Rev.\ D {\bf 72}, 043516 (2005)
  [astro-ph/0504571].

\bibitem{Gonzalez:2006cj} 
  T.~Gonzalez, G.~Leon and I.~Quiros,
  Class.\ Quant.\ Grav.\  {\bf 23}, 3165 (2006)
  [astro-ph/0702227].

\bibitem{Valiviita:2008iv} 
  J.~Valiviita, E.~Majerotto and R.~Maartens,
  JCAP {\bf 0807}, 020 (2008)
  [arXiv:0804.0232 [astro-ph]].

\bibitem{Amendola:2014kwa} 
  L.~Amendola, T.~Barreiro and N.~J.~Nunes,
  arXiv:1407.2156 [astro-ph.CO].

\bibitem{Farrar:2003uw} 
  G.~R.~Farrar and P.~J.~E.~Peebles,
  Astrophys.\ J.\  {\bf 604}, 1 (2004)
  [astro-ph/0307316].
  
\bibitem{Copeland:2003cv} 
  E.~J.~Copeland, N.~J.~Nunes and M.~Pospelov,
  Phys.\ Rev.\ D {\bf 69}, 023501 (2004)
  [hep-ph/0307299].
  
\bibitem{Brookfield:2007au} 
  A.~W.~Brookfield, C.~van de Bruck and L.~M.~H.~Hall,
  Phys.\ Rev.\ D {\bf 77}, 043006 (2008)
  [arXiv:0709.2297 [astro-ph]].

\bibitem{Baldi:2012kt} 
  M.~Baldi,
  Annalen Phys.\  {\bf 524}, 602 (2012)
  [arXiv:1204.0514 [astro-ph.CO]].

\bibitem{Piloyan:2013mla} 
  A.~Piloyan, V.~Marra, M.~Baldi and L.~Amendola,
  JCAP {\bf 1307}, 042 (2013)
  [arXiv:1305.3106 [astro-ph.CO]].

\bibitem{AmenTsuji} 
  L.~Amendola, S.~Tsujikawa
  Cambridge, UK: Univ. Pr. (2010) 503 p

\bibitem{Koivisto:2015qua} 
  T.~S.~Koivisto, E.~N.~Saridakis and N.~Tamanini,
  JCAP {\bf 1509}, 047 (2015)
  doi:10.1088/1475-7516/2015/09/047
  [arXiv:1505.07556 [astro-ph.CO]].

\bibitem{Piloyan:2014gta} 
  A.~Piloyan, V.~Marra, M.~Baldi and L.~Amendola,
  JCAP {\bf 1402}, 045 (2014)
  [arXiv:1401.2656 [astro-ph.CO]].


\bibitem{Liddle:1998jc} 
  A.~R.~Liddle, A.~Mazumdar and F.~E.~Schunck,
  Phys.\ Rev.\ D {\bf 58}, 061301 (1998)
  [astro-ph/9804177].
  
\bibitem{Malik:1998gy} 
  K.~A.~Malik and D.~Wands,
  Phys.\ Rev.\ D {\bf 59}, 123501 (1999)
  doi:10.1103/PhysRevD.59.123501
  [astro-ph/9812204].
    
\bibitem{Kanti:1999vt} 
  P.~Kanti and K.~A.~Olive,
  Phys.\ Rev.\ D {\bf 60}, 043502 (1999)
  doi:10.1103/PhysRevD.60.043502
  [hep-ph/9903524].

\bibitem{Kitching:2015fra} 
  T.~D.~Kitching {\it et al.},
  arXiv:1501.03978 [astro-ph.CO].

\bibitem{Raccanelli:2015qqa} 
  A.~Raccanelli {\it et al.},
  arXiv:1501.03821 [astro-ph.CO].
       

\bibitem{Hubble}
  E.~Hubble,
  Proc.\ Nat.\ Acad.\ Sci.\  {\bf 15}, 168 (1929).
  doi:10.1073/pnas.15.3.168

\bibitem{PenziasWilson}
  A.~A.~Penzias and R.~W.~Wilson,
  Astrophys.\ J.\  {\bf 142}, 419 (1965).
  doi:10.1086/148307
  
\bibitem{Guth}
  A.~H.~Guth,
  Phys.\ Rev.\ D {\bf 23}, 347 (1981).
  doi:10.1103/PhysRevD.23.347

\bibitem{2df}
  M.~Colless,
  Phil.\ Trans.\ Roy.\ Soc.\ Lond.\ A {\bf 357}, 105 (1999)
  doi:10.1098/rsta.1999.0317
  [astro-ph/9804079].

\bibitem{6df}
  D.~H.~Jones {\it et al.},
  Mon.\ Not.\ Roy.\ Astron.\ Soc.\  {\bf 399}, 683 (2009)
  doi:10.1111/j.1365-2966.2009.15338.x
  [arXiv:0903.5451 [astro-ph.CO]].
  
\bibitem{SDSS}
  K.~N.~Abazajian {\it et al.} [SDSS Collaboration],
  Astrophys.\ J.\ Suppl.\  {\bf 182}, 543 (2009)
  doi:10.1088/0067-0049/182/2/543
  [arXiv:0812.0649 [astro-ph]].

\bibitem{COBE}
	  NASA,
	  \emph{COBE Data Products}.
	  \begin{tt}http:\slash\slash lambda.gsfc.nasa.gov\slash product\slash cobe\slash c\_products\_table.cfm. \end{tt}

\bibitem{WMAP}
  D.~Larson {\it et al.},
  Astrophys.\ J.\ Suppl.\  {\bf 192}, 16 (2011)
  doi:10.1088/0067-0049/192/2/16
  [arXiv:1001.4635 [astro-ph.CO]].

\bibitem{PLANCK}
  J.~A.~Tauber {\it et al.},
  Astron.\ Astrophys.\  {\bf 520}, A1 (2010).
  doi:10.1051/0004-6361/200912983
        
        
\bibitem{Wald84}
  R.~M.~Wald,
  Chicago, Usa: Univ. Pr. ( 1984) 491p

\bibitem{Friedmann}
  A.~Friedman,
  Z.\ Phys.\  {\bf 10}, 377 (1922)
  [Gen.\ Rel.\ Grav.\  {\bf 31}, 1991 (1999)].
  doi:10.1007/BF01332580

\bibitem{Liddle:2000cg} 
  A.~R.~Liddle and D.~H.~Lyth,
  Cambridge, UK: Univ. Pr. (2000) 400 p

\bibitem{CMB2}
  R.~H.~Dicke, P.~J.~E.~Peebles, P.~G.~Roll and D.~T.~Wilkinson,
  Astrophys.\ J.\  {\bf 142}, 414 (1965).
  doi:10.1086/148306

\bibitem{Ade:2015xua} 
  P.~A.~R.~Ade {\it et al.} [Planck Collaboration],
  Astron.\ Astrophys.\  {\bf 594}, A13 (2016)
  doi:10.1051/0004-6361/201525830
  [arXiv:1502.01589 [astro-ph.CO]].

\bibitem{Dodelson}
  S.~Dodelson,
  ``Modern cosmology,''
  Amsterdam, Netherlands: Academic Pr. (2003) 440p.

\bibitem{Hou:2017uap} 
  S.~Q.~Hou, J.~J.~He, A.~Parikh, D.~Kahl, C.~A.~Bertulani, T.~Kajino, G.~J.~Mathews and G.~Zhao,
  Astrophys.\ J.\  {\bf 834}, no. 2, 165 (2017)
  doi:10.3847/1538-4357/834/2/165
  [arXiv:1701.04149 [astro-ph.CO]].

\bibitem{Rindler}
   W.~Rindler,
  Gen.\ Rel.\ Grav.\  {\bf 34}, 133 (2002)
  [Mon.\ Not.\ Roy.\ Astron.\ Soc.\  {\bf 116}, 662 (1956)].
  doi:10.1023/A:1015347106729
  
\bibitem{Dicke}
  R.~H.~Dicke,
  ``Gravitation and the Universe,''
  American Philosophical Society\  {\bf 78}, 1-82 (1970).

\bibitem{Peacock}
  J.~A.~Peacock,
  ``Cosmological physics,''
  Cambridge, UK: Univ. Pr. (1999) 682p.

\bibitem{Liddle}
  A.~R.~Liddle,
  astro-ph/9901124.
  
\bibitem{Giacomelli}
  G.~Giacomelli, L.~Patrizii and Z.~Sahnoun,
  doi:10.1142/9789814340861.0039
  arXiv:1105.2724 [hep-ex].

\bibitem{Press}
  W.~H.~Press, B.~S.~Ryden and D.~N.~Spergel,
  Astrophys.\ J.\  {\bf 347}, 590 (1989).
  doi:10.1086/168151

\bibitem{Taylor}
  A.~N.~Taylor and A.~R.~Liddle,
  Phys.\ Rev.\ D {\bf 64}, 023513 (2001)
  doi:10.1103/PhysRevD.64.023513
  [astro-ph/0011365].

\bibitem{Asaka}
  T.~Asaka, M.~Kawasaki and T.~Yanagida,
  Phys.\ Rev.\ D {\bf 60}, 103518 (1999)
  doi:10.1103/PhysRevD.60.103518
  [hep-ph/9904438].

\bibitem{Sitter}
  W.~de Sitter,
  ``On the Relativity of Inertia: Remarks Concerning Einstein's Latest Hypothesis,''
  Proc. Kon. Ned. Akad. Wet.\  {\bf 19}, 1217-1225 (1917).


\bibitem{Zeldovich}
Y.~B.~Zeldovich,
``Cosmological field theory for observational astronomers,''
Sov. Sci. Rev. E Astrophys. Space Phys.\ {\bf 5}, 1-37 (1986).

\bibitem{Kinney}
  W.~H.~Kinney,
  arXiv:0902.1529 [astro-ph.CO].

\bibitem{Morandi:2016cet} 
  A.~Morandi and M.~Sun,
  Mon.\ Not.\ Roy.\ Astron.\ Soc.\  {\bf 457}, no. 3, 3266 (2016)
  doi:10.1093/mnras/stw143
  [arXiv:1601.03741 [astro-ph.CO]].





\bibitem{Cadabra}
K.~Peeters,
Comput.\ Phys.\ Commun.\  {\bf 176}, 550 (2007)
[arXiv:cs/0608005].
K.~Peeters,
arXiv:hep-th/0701238.

\bibitem{din} \emph{Introducing Einstein's Relativity} by R. D'inverno Clarendon Press, 1992

\bibitem{MW2008}
  K.~A.~Malik and D.~Wands,
  Phys.\ Rept.\  {\bf 475}, 1 (2009)
  [arXiv:0809.4944 [astro-ph]].

\bibitem{kmdrmgi}
  K.~A.~Malik and D.~R.~Matravers,
  Gen.\ Rel.\ Grav.\  {\bf 45}, 1989 (2013)
  doi:10.1007/s10714-013-1573-2
  [arXiv:1206.1478 [astro-ph.CO]].

\bibitem{Bonnor}
	W.~B.~Bonnor 
	MNRAS, 159, 261 (1972)


\bibitem{Christopherson:2008ry} 
  A.~J.~Christopherson and K.~A.~Malik,
  Phys.\ Lett.\ B {\bf 675}, 159 (2009)
  [arXiv:0809.3518 [astro-ph]].

\bibitem{Hellaby}
  A.~A.~H.~Alfedeel and C.~Hellaby,
  Gen.\ Rel.\ Grav.\  {\bf 42}, 1935 (2010)
  [arXiv:0906.2343 [gr-qc]].


\bibitem{Faraoni:2015uma} 
  V.~Faraoni,
  Gen.\ Rel.\ Grav.\  {\bf 47}, no. 7, 84 (2015)
  doi:10.1007/s10714-015-1926-0
  [arXiv:1506.06358 [gr-qc]].




\bibitem{Bitbucket} 
  {https://bitbucket.org/pyessence/pyessence}

\bibitem{Pyweb} 
  {http://pyessence.leithes.co.uk/}

\bibitem{PYDOCREF} 
  A.~Leithes,
  arXiv:1608.00910 [astro-ph.CO].
  
\bibitem{Planck:2013jfk} 
  P.~A.~R.~Ade {\it et al.} [Planck Collaboration],
  Astron.\ Astrophys.\  {\bf 571}, A22 (2014)
  doi:10.1051/0004-6361/201321569
  [arXiv:1303.5082 [astro-ph.CO]].


\bibitem{Malik:2004tf} 
  K.~A.~Malik and D.~Wands,
  JCAP {\bf 0502}, 007 (2005)
  [astro-ph/0411703].

\bibitem{Bull:2015lja} 
  P.~Bull,
  Astrophys.\ J.\  {\bf 817}, no. 1, 26 (2016)
  doi:10.3847/0004-637X/817/1/26
  [arXiv:1509.07562 [astro-ph.CO]].

\bibitem{Macaulay:2013swa} 
  E.~Macaulay, I.~K.~Wehus and H.~K.~Eriksen,
  Phys.\ Rev.\ Lett.\  {\bf 111}, no. 16, 161301 (2013)
  [arXiv:1303.6583 [astro-ph.CO]].
  
\bibitem{Abbott:2015swa} 
  T.~Abbott {\it et al.} [DES Collaboration],
  arXiv:1507.05552 [astro-ph.CO].


\bibitem{Samushia:2011cs} 
  L.~Samushia, W.~J.~Percival and A.~Raccanelli,
  Mon.\ Not.\ Roy.\ Astron.\ Soc.\  {\bf 420}, 2102 (2012)
  doi:10.1111/j.1365-2966.2011.20169.x
  [arXiv:1102.1014 [astro-ph.CO]].
  

\bibitem{ACQ} 
  A.~Leithes, K.~A.~Malik, D.~J.~Mulryne and N.~J.~Nunes,
  arXiv:1608.00908 [astro-ph.CO].

\bibitem{Valiviita:2015dfa} 
  J.~Väliviita and E.~Palmgren,
  JCAP {\bf 1507}, no. 07, 015 (2015)
  doi:10.1088/1475-7516/2015/07/015
  [arXiv:1504.02464 [astro-ph.CO]].

\bibitem{Jennings:2009qg} 
  E.~Jennings, C.~M.~Baugh, R.~E.~Angulo and S.~Pascoli,
  Mon.\ Not.\ Roy.\ Astron.\ Soc.\  {\bf 401}, 2181 (2010)
  doi:10.1111/j.1365-2966.2009.15819.x
  [arXiv:0908.1394 [astro-ph.CO]].


\bibitem{Pagels:1981ke} 
  H.~Pagels and J.~R.~Primack,
  Phys.\ Rev.\ Lett.\  {\bf 48}, 223 (1982).
  doi:10.1103/PhysRevLett.48.223

\bibitem{Padmanabhan:2006kz} 
  T.~Padmanabhan,
  AIP Conf.\ Proc.\  {\bf 843}, 111 (2006)
  [astro-ph/0602117].

\bibitem{Blas:2011rf} 
  D.~Blas, J.~Lesgourgues and T.~Tram,
  JCAP {\bf 1107}, 034 (2011)
  [arXiv:1104.2933 [astro-ph.CO]].
  

\bibitem{Lewis:1999bs} 
  A.~Lewis, A.~Challinor and A.~Lasenby,
  Astrophys.\ J.\  {\bf 538}, 473 (2000)
  [astro-ph/9911177].

\bibitem{Carrilho:2015cma} 
  P.~Carrilho and K.~A.~Malik,
  JCAP {\bf 1602}, no. 02, 021 (2016)
  doi:10.1088/1475-7516/2016/02/021
  [arXiv:1507.06922 [astro-ph.CO]].


\bibitem{I.:2016gam} 
  E.~G.~Chirinos Isidro, C.~Zuñiga Vargas and W.~Zimdahl,
  JCAP {\bf 1605}, no. 05, 003 (2016)
  doi:10.1088/1475-7516/2016/05/003
  [arXiv:1602.08583 [gr-qc]].


\bibitem{Das:2015vda} 
  M.~Das, T.~Saito, D.~Iono, M.~Honey and S.~Ramya,
  Astrophys.\ J.\  {\bf 815}, no. 1, 40 (2015)
  doi:10.1088/0004-637X/815/1/40
  [arXiv:1510.07411 [astro-ph.GA]].

\bibitem{Aragon-Calvo:2016vye} 
  M.~A.~Aragon-Calvo, M.~C.~Neyrinck and J.~Silk,
  arXiv:1607.07881 [astro-ph.GA].

\bibitem{Sussman:2015wna} 
  R.~A.~Sussman, I.~Delgado Gaspar and J.~C.~Hidalgo,
  JCAP {\bf 1603}, no. 03, 012 (2016)
  Erratum: [JCAP {\bf 1606}, no. 06, E03 (2016)]
  doi:10.1088/1475-7516/2016/06/E03, 10.1088/1475-7516/2016/03/012
  [arXiv:1507.02306 [gr-qc]].

\bibitem{Musoke:2015kql} 
  N.~K.~Musoke, D.~D.~McNutt, A.~A.~Coley and D.~A.~Brooks,
  Gen.\ Rel.\ Grav.\  {\bf 48}, no. 3, 27 (2016)
  doi:10.1007/s10714-016-2022-9
  [arXiv:1511.01435 [gr-qc]].

\bibitem{Reddy:2016wlp} 
  D.~R.~K.~Reddy, G.~Ramesh and S.~Umadevi,
  arXiv:1601.02648 [physics.gen-ph].

\bibitem{Camci:2016yed} 
  U.~Camci, A.~Yildirim and I.~Basaran Oz,
  Astropart.\ Phys.\  {\bf 76}, 29 (2016)
  doi:10.1016/j.astropartphys.2015.12.006
  [arXiv:1605.00864 [gr-qc]].
  
\bibitem{Keresztes:2015pxa} 
  Z.~Keresztes, M.~Forsberg, M.~Bradley, P.~K.~S.~Dunsby and L.~Gergely,
  JCAP {\bf 1511}, no. 11, 042 (2015)
  doi:10.1088/1475-7516/2015/11/042
  [arXiv:1507.08300 [gr-qc]].


\bibitem{Elder:2016yxm} 
  B.~Elder, J.~Khoury, P.~Haslinger, M.~Jaffe, H.~Müller and P.~Hamilton,
  arXiv:1603.06587 [astro-ph.CO].
  
\bibitem{Tamanini:2016klr} 
  N.~Tamanini and M.~Wright,
  JCAP {\bf 1604}, no. 04, 032 (2016)
  doi:10.1088/1475-7516/2016/04/032
  [arXiv:1602.06903 [gr-qc]].

\bibitem{Bouhmadi-Lopez:2016cja} 
  M.~Bouhmadi-López, K.~S.~Kumar, J.~Marto, J.~Morais and A.~Zhuk,
  JCAP {\bf 1607}, no. 07, 050 (2016)
  doi:10.1088/1475-7516/2016/07/050
  [arXiv:1605.03212 [gr-qc]].

\bibitem{Guendelman:2015jii} 
  E.~Guendelman, E.~Nissimov and S.~Pacheva,
  Eur.\ Phys.\ J.\ C {\bf 76}, no. 2, 90 (2016)
  doi:10.1140/epjc/s10052-016-3938-7
  [arXiv:1511.07071 [gr-qc]].

\bibitem{Maunakea} 
  A.~McConnachie,  {\it et al.},
  [arXiv:1606.00043 [astro-ph.IM]].

\bibitem{Bacon:2015dqe} 
  D.~Bacon {\it et al.},
  PoS AASKA {\bf 14}, 145 (2015)
  [arXiv:1501.03977 [astro-ph.CO]].

\end{thebibliography}
\end{document}